\documentclass[aps,pra,reprint,nofootinbib,amsmath,amssymb,showpacs,floatfix]{revtex4-1}

\usepackage{graphicx,units}

\newcommand{\ket}[1]{\ensuremath{\vert{#1\rangle}}} 

\newcommand{\braket}[2]{\ensuremath{{\langle #1}\vert{#2 \rangle}}}

\newcommand{\I}{\text{i}}
\newcommand{\E}{\text{e}}
\providecommand{\abs}[1]{\left\lvert#1\right\rvert}

\usepackage[colorlinks,allcolors=blue,hyperindex,breaklinks]{hyperref}

\begin{document}

\title{Decoherence: From Interpretation to Experiment}

\author{Maximilian Schlosshauer}
\email{schlossh@up.edu}

\affiliation{Department of Physics, University of Portland, 5000 North Willamette Boulevard, Portland, Oregon 97203, USA}

\begin{abstract}
  I offer a few selected reflections on the decoherence program, with an emphasis on Zeh's role and views. First, I discuss Zeh's commitment to a realistic interpretation of the quantum state, which he saw as necessary for a consistent understanding of the decoherence process. I suggest that this commitment has been more fundamental than, and prior to, his support of an Everett-style interpretation of quantum mechanics. Seen through this lens, both his defense of Everett and the genesis of his ideas on decoherence emerge as consequences of his realistic view of the quantum state. Second, I give an overview of experiments on decoherence and describe, using the study of collisional decoherence as an example, the close interplay between experimental advances and theoretical modeling in decoherence research.\\[.2cm]
  Published in: \emph{From Quantum to Classical}, Fundamental Theories of Physics, vol.~204, edited by C.\ Kiefer
  (Springer, Cham, 2022), pp.~45--64, doi: \href{https://doi.org/10.1007/978-3-030-88781-0_3}{\texttt{10.1007/978-3-030-88781-0\_3}}
\end{abstract}

\maketitle

\section{Introduction}

I never had the opportunity to meet Dieter Zeh in person, but over the years we maintained a fairly regular correspondence. The last email I received from him dates from April 1, 2015, by which time our exchanges had become rather infrequent.  Because this email happens to be an excellent representation of his inimitable style and thoughtfully articulated views, and because it touches on some points to which I shall return later in this essay, I will take the liberty to quote it in full below. Zeh had sent it to me in response to a paper \cite{Camilleri:2015:oo} that Kristian Camilleri and I had written on the relation between decoherence and the views of Bohr and Heisenberg. There, we had suggested that Bohr's doctrine of ``classical concepts'' should be understood epistemologically. As we put it, 
\begin{quote} 
Bohr's fundamental point was that any interpretation of quantum mechanics must in the end fall back on the use of classical concepts, because such concepts play an indispensible role in experimental contexts in which we acquire empirical knowledge of the world.
\end{quote}
Thus, we saw Bohr's insistence on the primacy of classical concepts as motivated by his emphasis on the functional role of experiments, and we therefore suggested that this understanding of classical concepts should be distinguished from the kind of classicality obtained from dynamical solutions to the quantum--classical problem, such as the account given by decoherence. 

``This argument may be acceptable,'' Zeh wrote to me, ``if it is to be understood merely from a historical point of view'':
\begin{quote}
In this case, Bohr's classical concepts would be akin to Wittgenstein's famous ladder that one may afterward discard. But I'm sure that this is not what Bohr had in mind. I say this because, on such a reading, the axiomatic view of quantum theory would amount to nothing more than, for example, Sommerfeld's axiomatic introduction of Maxwell's equations (as presented in his textbook)---that is to say, without taking a historical or operational route. (Of course, one must not forget the empirical justification, in contrast with modern cosmological theories!) Bohr, however, has always categorically rejected a dynamical analysis of the measurement process, and certainly he would have never applied the wave function to the ``environment.'' 

Personally, I do not care much for the historical development, only for the outcome of this development. But one should require that this outcome be conceptually complete. In particular, usually I have explicitly used the term ``quasiclassical'' concepts, although later on I realized that Gell-Mann and Hartle had used this term to refer to something completely different from wave packets. Needless to say, in this context the wave function, or a superposition, are to be understood ontically (what else?); the use of ``observables'' appears to be entirely out of context in such a framework.\footnote{My translation.}
\end{quote}

In just these two paragraphs, Zeh takes us on a tour de force, giving not only his own assessment of Bohr's views, but also finding the opportunity to mention Wittgenstein's ladder, Maxwell's equations, Sommerfeld's textbook, cosmological theories, and the distinction between axiomatic, operational, and empirical approaches to theories. It is a feat of compression, and a wonderful example of the breadth and depth of his thinking that was always on display. Most of his emails (and papers) read like this. 

In this essay, I will share a few observations on the decoherence program and the role played by Zeh in its development. I will do so through two different lenses. First, in Sec.~\ref{sec:interpretation}, I will recount Zeh's interpretive stance toward quantum mechanics and how it shaped his conception and understanding of decoherence. Second, in Sec.~\ref{sec:experiments}, I will discuss some of the recent experiments on decoherence and their interplay with theoretical models, using, in particular, the example of collisional decoherence to illustrate an effort that started with Zeh's formative ideas and continues to this date.  I will offer concluding remarks in Sec.~\ref{sec:concluding-remarks}.

\section{\label{sec:interpretation}Interpretation}

Zeh's discovery \cite{Zeh:1970:yt} of the crucial importance of the environment for the description of quantum systems was made in virtual isolation and dismissed for years afterward \cite{Freitas:2008:yy,Camilleri:2009:aq,Freire:2009:aa}. Zeh concluded at the time that his early work on decoherence had all but destroyed his mainstream academic career \cite{Freitas:2008:yy,Freire:2009:aa}, and the 1970s and early 1980s constituted what he has called the ``dark ages of decoherence'' \cite{Zeh:2006:na}. In his 2008 interview with Freitas \cite{Freitas:2008:yy}, he recounted: 
\begin{quote}
Except for Wigner, nobody seemed to be interested in my suggestion, and so it was just neglected. This changed not until 1981, when Zurek's paper appeared. (\dots) I was very happy that it had appeared in the \emph{Physical Review}. I expected that the idea would now soon become better known, which turned out to be true---slowly. 
\end{quote}
Camilleri \cite{Camilleri:2009:aq} (see also his contribution to this volume) has given an analysis of the early history of decoherence, showing how it was shaped by Zeh's and Zurek's contrasting interpretive commitments. He traces one reason for the slow reception to Zeh's association of decoherence with an Everett-style ``many-minds'' interpretation, which Zeh had independently formulated \cite{Zeh:1970:yt,Zeh:1973:wq} when he first conceived of the ideas that later became the theory of decoherence. Zeh continued to argue in favor of this interpretation for the rest of this life \cite{Zeh:2016:ii}, writing, for example, in 2006 \cite{Zeh:2006:na}:
\begin{quote}
A consistent interpretation of [quantum] theory in accordance with the observed world requires a \emph{novel and nontrivial identification of observers} with appropriate quantum states of local systems which exist only in certain, dynamically autonomous \emph{components} of the global wave function. Accordingly, it is the observer who ``splits'' indeterministically---not the (quantum) world. [Emphasis in the original.]
\end{quote}
He called this interpretation ``consistent'' (one of his favorite terms) because, in his view, it required no separation into quantum and classical, be it at the level of terminology or physics. He suggested that by ``consistently using this global unitary description, all those much discussed `absurdities' of quantum theory can be explained'' \cite{Zeh:2016:ii}, and that \cite{Zeh:2000:rr}
\begin{quote}
the Heisenberg--Bohr picture of quantum mechanics can now be claimed dead. Neither classical concepts, nor any uncertainty relations, complementarity, observables, quantum logic, quantum statistics, or quantum jumps have to be introduced on a fundamental level.
\end{quote}
Elsewhere, he wrote that ``this interpretation is an attempt to replace the `pragmatic irrationalism' that is common in quantum theory textbooks (complementarity, dualism, fundamental uncertainty etc.) by a consistent application of just those concepts which are actually, and without exception \emph{successfully}, used when the theory is applied'' \cite{Zeh:2006:na}. 

Yet it seems to me that Zeh's attachment to the Everett view was never primary. Rather, it appears to have been a \emph{consequence} of his much deeper conviction that, as he put it in his email, ``the wave function, or a superposition, are to be understood ontically (what else?),'' together with the purely empirically grounded assumption of the universal validity of the Schr\"odinger evolution. In his interview with Freitas \cite{Freitas:2008:yy}, he explained his interpretation of the wave function:
\begin{quote}
I do believe that the wave function describes the real world. I don't mean that we may not find something entirely new some day, but from all we know now, it's the only consistent description. (\dots) I think it is untenable to claim that the wave function is only a tool to calculate probabilities. There are many individual (real) physical properties that have to be described by nonlocal superpositions. For example, you have to describe total spin eigenstates of a couple of particles as superpositions of product states by means of Clebsch--Gordan coefficients. Even if you take the particles apart, they have to stay in this superposition in order to conserve angular momentum. This total spin can be confirmed individually if you bring the particles back together again, or statistically by means of Bell type correlations. So the observable (hence real) total spin is defined by the nonlocal quantum state. I don't see any other way to formulate this. That's why I believe that the wave function describes reality. 
\end{quote}
He went on to elaborate:
\begin{quote} 
What I mean by saying that a concept is real (or describes reality) is that it can be used consistently to describe our observations. So I would never use this word in a framework where you need complementary concepts, or a conceptual dualism, that is, where you have to switch between different concepts. So Bohr had to conclude that there is no microscopic reality in his view (and he had even to assume uncertainty relations to apply to the classical properties of macroscopic objects). I think this question came up in the beginning of quantum theory when people started asking: ``If the electron behaves sometimes like a particle, and sometimes like a wave, then what is it really?'' It's a question for consistency. But I think that you can consistently derive all apparent particle aspects from a universal wave function. Consider Mott's description of particle tracks, for example. So in this respect, I think that the wave function can be regarded as a concept that represents reality. 
\end{quote}
Indeed, he appeared to be open to any interpretation or version of quantum theory in which the quantum state was considered real. In this vein, his main objection to dynamical collapse theories \cite{Bassi:2003:yb} was simply that there was no experimental evidence of collapse-induced effects. As he noted in his interview with Freitas \cite{Freitas:2008:yy}:
\begin{quote}
Dynamical collapse theories are very honest and serious suggestions, but they must lead to observable consequences. One has to demonstrate experimentally where and how the global Schr\"odinger equation breaks down, and as long as that has not be done, it's very speculative to suggest such things. So I prefer to assume the opposite, which necessarily leads to Everett. 
\end{quote}
He was open to Bohmian mechanics, too, since it was realistic (a physical pilot wave guiding physical particles), though he found the role of particles as distinguishing the ``relevant'' branch of the wave function, while leaving an infinity of ``empty'' branches, to contribute unnecessary clutter \cite{Zeh:1999:rr}. Everett, then, seemed to him the most appealing interpretation arising from the assumption of the physical reality of the wave function, because, in his view, it was minimalistic, in the sense that it required only one physical entity (the wave function), one kinematical concept (the superposition principle), and one dynamical law (the Schr\"odinger equation). Faced with the ontological extravagance of such an interpretation, with its myriad parallel (but, thanks to decoherence, dynamically decoupled) branches, he argued that \cite{Zeh:2000:rr}
\begin{quote}
the existence of ``other'' components (with their separate conscious versions of ourselves) is a heuristic fiction, based on the assumption of a general validity of dynamical laws that have always been confirmed when tested.
\end{quote}
His concept of a ``heuristic fiction,'' a term he frequently used when discussing the question of the reality of the branches, was in reference to the philosophy of Vaihinger \cite{Vaihinger:1935:jj}, who had suggested that we may treat concepts and entities ``as if'' they were real if they can be extrapolated from consistently confirmed physical laws.\footnote{When Freitas, in his interview with Zeh \cite{Freitas:2008:yy}, posed the question of the reality of the ``unobserved branches of the wave function,'' Zeh did not mention Vaihinger's concept. Instead, his response was simply that this reality 
``is the consequence, but I think it's also a matter of definition what you call `real.' In any normal definition of the word, I would call [the branches] real.''} Note that in Zeh's statement, the assumption of the reality of the wave function is implicit, for he makes reference only to ``dynamical laws''  (namely, the Schr\"odinger equation) as giving rise to the ``heuristic fiction'' of coexisting branches. 

Interestingly, he never warmed to attempts to derive the Born rule from within the Everett interpretation, even though this problem---and, more generally, the meaning of probability---are often thought to be the true Achilles heel of the Everett interpretation \cite{Hemmo:2007:uu,Saunders:2010:im}. He was content to say that the probability weights had to be postulated \cite{Zeh:2016:ii}, emphasizing that \cite{Freitas:2008:yy}
\begin{quote}
the only consequence you can derive from decoherence is that the branches become dynamically autonomous: separate ``worlds.'' What remains open is only the precise probability weight, which means, for example, that not all branches are equally probable, so we do not end up in one of these so-called ``maverick'' components, where frequencies of measurement results do not agree with the Born rule. 
\end{quote}

Just as the Everett view of quantum mechanics was to Zeh a consequence of a global, universal wave function governed by universal unitary dynamics, so was ubiquitous entanglement and the resulting decoherence. This is why he never stopped insisting that, as was noted above, decoherence must not be separated from the interpretation of quantum mechanics. It also meant that decoherence, by itself, was unsurprising---or, as he put it \cite{Zeh:1996:gy}, that it is 
\begin{quote}
a normal consequence of interacting quantum mechanical systems. It can hardly be denied to occur---but it cannot explain anything that could not have been explained before. Remarkable is only its quantitative (realistic) aspect that seems to have been overlooked for long. 
\end{quote}
This is not to imply that Zeh considered his discovery of decoherence trivial or obvious, though Joos may well be correct when he considers the comparably late recognition of the consequences of environmental entanglement a ``historical accident'' \cite[p.~13]{Joos:1999:po}.\footnote{As for the prehistory of decoherence, there are some remarks one can find, for example, in the writings of Heisenberg that hint at the role of the environment, albeit without the all-important reference to entanglement \cite{Camilleri:2015:oo}. For instance, Heisenberg suggested that ``the interference terms are \dots\ removed by the partly undefined interactions of the measuring apparatus, with the system and with the rest of the world'' \cite[p.~23]{Heisenberg:1955:lm}.} Rather, it is to suggest that what Zeh might have ultimately found more relevant was the notion of the reality of the wave function that had led him to his discovery in the first place. As he explained in 2006 \cite{Zeh:2006:na}, the concept of decoherence 
\begin{quote}
arose as a by-product of arguments favoring either a collapse of the wave function as part of its dynamics, or an Everett-type interpretation. In contrast to the Copenhagen interpretation, which insists on fundamental classical concepts, both these interpretations regard the wave function as a complete and universal representation of reality. 
\end{quote}
Perhaps this may also help explain why Zeh did become little involved in the intense experimental and theoretical research efforts on decoherence we have seen over the past two or three decades (though he was always a close and critical observer). He may have felt that, in the way he had originally formulated decoherence, what mattered was already fundamentally in place. I was under the impression that he thought that many of the recent developments---be it experimental confirmation, realistic modeling, fresh terminology, or applications in quantum information science---did not add much that was genuinely new, and that, to the contrary, they often diluted or even misrepresented the purity and universality of his conception of decoherence. One of his concerns in this context was the dissociation of decoherence from the kind of realistic, Everettian view he had been championing. Thus, when the first cavity QED experiments \cite{Brune:1996:om,Maitre:1997:tv,Raimond:1997:um} allowed for the observation of a controlled decoherence process and provided an impressive confirmation of his insights about the role of the environment, he was surprised to read the authors' interpretation \cite{Raimond:1997:um} of their observations:
\begin{quote}
At first I expected that these results would be seen as a confirmation of universal entanglement, and hence Everett, or at least as a severe argument against Copenhagen. But it turned out to be just the opposite---they later regarded entanglement as a form of complementarity!  (Quoted in Ref.~\cite{Camilleri:2009:aq}.)
\end{quote}
Comments such as this suggest that Zeh saw the insights brought about by decoherence theory not only as improving upon the kind of interpretive framework constructed by the Copenhagen view, but as actually  being at odds with it \cite{Camilleri:2009:aq}. (Of course, given the variety of viewpoints that have been lumped together under the heading of the Copenhagen interpretation \cite{Howard:2004:mh}, one needs to assess claims of this kind in the context of the specific views of individuals such as Bohr and Heisenberg \cite{Camilleri:2015:oo}.) As early as in his first paper on decoherence \cite{Zeh:1970:yt}, Zeh was also always careful (and correct) to stress that decoherence does not solve the measurement problem, because the reduced density matrices obtained from the trace over the environment are improper and therefore not ignorance-interpretable (see also his Refs.~\cite{Zeh:2004:zm,Zeh:1999:qr,Zeh:2006:na}). This, indeed, is now the universally acknowledged position. 

Another one of Zeh's concerns was with the proper definition of the term ``decoherence.''  Zeh emphasized that decoherence should refer only to a process of uncontrollable entanglement with a large number of degrees of freedom \cite{Zeh:2006:na}. In particular, it should be carefully distinguished from the phenomenon of ``ensemble dephasing'' obtained by averaging over many noisy realizations of the evolution for an ensemble of systems \cite{Zeh:2003:mq,Zeh:2004:zm}. In the literature, this distinction is often blurred, with the term ``decoherence'' quite loosely applied to any process that leads to a diminishing of coherence. While at the level of the reduced density matrix of the system, loss of coherence due to environmental entanglement and loss of coherence due to classical noise may manifest themselves in a similar manner, the two processes are physically distinguished in the sense that the effects of noise could in principle be reversed by local operations on the system (akin to the spin-echo technique in NMR), while undoing the effect of environmental entanglement requires measurements on the environment to recover information that has been transferred from the system \cite{Myatt:2000:yy,Zurek:2002:ii}. Because one does not usually have full control over the environment, the entanglement and resulting leakage of information from system to environment become effectively irreversible, and it is this irreversibility that Zeh emphasized as an essential property of the decoherence process \cite{Zeh:2006:na}. (Incidentally, one of his other major research interests was the arrow of time, a topic on which he published a book \cite{Zeh:2001:tt}.) Such uncontrollable entanglement is in contrast with the controlled entanglement we may use as a resource, for example, in a Bell experiment or in quantum information processing, and decoherence is typically detrimental to the presence and exploitation of such controlled entanglement \cite{Zyczkowski:2001:ii,Lee:2004:uu,Barreiro:2010:aa}.\footnote{Under very specific, carefully designed conditions, however, decoherence may sometimes act as a generator of controlled entanglement \cite{Braun:2002:aa,Benatti:2003:aa,Kim:2002:oo,Jakobczyk:2002:oo}.}  

Despite Zeh's efforts to tie decoherence theory to a realistic interpretation of the quantum state, its wider reception by the physics community has been largely characterized by indifference to the attachment of a particular interpretation of quantum mechanics. This may be no surprise, as decoherence is simply an application of the standard quantum formalism to open quantum systems, and therefore its consequences can be couched in the language of any interpretation of quantum mechanics \cite{Bacciagaluppi:2003:yz,Schlosshauer:2003:tv}. In some of these interpretations, however, the concepts and insights of decoherence have come to play an essential role. For example, decoherence is a crucial ingredient in making the Everett interpretation work, since it leads to a dynamical definition of the branches and therefore addresses the preferred-basis problem that had posed a serious challenge to this type of interpretation \cite{Zurek:1998:re,Butterfield:2001:ua,Wallace:2003:iq,Wallace:2003:iz,Wallace:2010:im}. A relative-state view thoroughly informed by decoherence is at the heart of Zurek's ``existential interpretation'' \cite{Zurek:1993:pu,Zurek:1998:re,Zurek:2004:yb}, which later came to include further results derived from system--environment entanglement, such as quantum Darwinism \cite{Zurek:2009:om,Zurek:2014:xx,Zwolak:2016:zz} and a proposed derivation of Born's rule on the basis of what Zurek has termed environment-assisted invariance \cite{Zurek:2003:rv,Zurek:2003:pl,Zurek:2004:yb,Zurek:2018:on}. In modal interpretations \cite{Clifton:1996:op}, decoherence can specify the values to be made definite \cite{Bacciagaluppi:1996:po,Bene:2001:po} and identify problematic rules for value assignments \cite{Bacciagaluppi:2000:yz}. In the consistent-histories interpretation \cite{Griffiths:1984:tr,Omnes:1994:pz,Griffiths:2002:tr}, decoherence is used to dynamically define consistent, quasiclassical histories \cite{Zurek:1993:pu,Paz:1993:ww,Albrecht:1992:rz,Albrecht:1993:pq,Twamley:1993:bz} and records \cite{Albrecht:1992:rz,Albrecht:1993:pq,Paz:1993:ww,Zurek:1993:pu,Zurek:2002:ii,GellMann:1998:xy}, with the stability of such records explained by their redundant environmental encoding \cite{Zurek:1993:pu,Paz:1993:ww,Zurek:2002:ii,Zurek:2003:pl,Riedel:2016:oo}. Ideas associated with the views of Bohr and his followers, such as the importance of amplification, irreversibility, and communication, have been illuminated and supported by reference to the role of the environment \cite{Zurek:2003:pl,Ollivier:2003:za,Ollivier:2004:im,Blume:2004:oo,Blume:2005:oo,Zurek:2009:om,Riedel:2010:un,Riedel:2011:un,Riedel:2012:un,Streltsov:2013:oo,Zurek:2013:xx,Zurek:2018:om,Zurek:2018:on}. In some interpretations, on the other hand, decoherence appears to play no important role. For example, in QBism \cite{Fuchs:2014:pp}, an interpretation in which quantum states encode an observer's personal probabilistic expectations associated with his future measurement interactions, decoherence would simply reflect the observer's adjustment of these expectations in light of the presence of an environment.\footnote{Unsurprisingly, Zeh's own assessment of QBism was negative \cite{Zeh:2016:ii}. He saw it as an interpretation that ``replaces the whole physical world by a black box, representing an abstract concept of `information' about inconsistent classical variables, and assumed to be available to vaguely defined `agents' rather than to observers who may be consistent parts of the physical world to be described.''}

At its most basic, interpretation-neutral level, decoherence simply addresses a consistency problem, by giving a dynamical account of how and when the quantum probability distributions approach the classically expected distributions. (This, surely, is a view of decoherence Zeh would have considered incomplete at best.) It seems that decoherence has proven plainly too useful and relevant a concept to experimenters and theoreticians alike to be claimed by any particular interpretive stance. Likewise, it has been found to fit too well the pragmatic, information-theoretic view of quantum mechanics suggested by modern quantum information science to be left to any particular ontological or metaphysical camp.

\section{\label{sec:experiments}Experiments}

The field of decoherence research has seen astonishing growth since the 1980s, and today decoherence has become one of the central paradigms of modern quantum mechanics  \cite{Zurek:2002:ii,Paz:2001:aa,Breuer:2002:oq, Joos:2003:jh,Schlosshauer:2007:un,Hornberger:2009:aq,Schlosshauer:2019:qd}. There have been at least two significant catalysts for this development. The first is the dramatic progress that has been made in the experimental realization and control of quantum systems and quantum phenomena. The other catalyst has been the influential role played by quantum information science. Both are areas for which a deep and detailed understanding of the physical sources and dynamics of decoherence is essential, and, accordingly, they have been a driving factor for research into decoherence. This has led to ever-more realistic decoherence models (below we will discuss the example of models for collisional decoherence). It has also resulted in the development of a variety of methods for the effective control and mitigation of decoherence, a task of acute importance in any implementation of quantum computing devices. Such methods include decoherence-free subspaces \cite{Lidar:2003:aa,Lidar:2014:pp}, reservoir engineering \cite{Dalvit:2000:bb,Poyatos:1996:um,Myatt:2000:yy,Turchette:2000:aa,Carvalho:2001:ua}, dynamical decoupling \cite{Viola:1998:uu,Viola:1999:zp,Zanardi:1999:oo,Viola:2000:pp,Wu:2002:aa,Wu:2002:bb}, and quantum error correction \cite{Steane:1996:cd,Shor:1995:rx,Steane:2001:dx,Gaitan:2008:uu,Lidar:2013:pp}. In turn, these methods have nourished further experimental advances. Decoherence-free subspaces, for example, have been experimentally implemented in quantum systems such as photons \cite{Kwiat:2000:kv, Altepeter:2004:ll}, trapped ions \cite{Kielpinski:2001:uu,Roos:204:pp,Haffner:2005:zz,Langer:2005:uu}, and NMR qubits \cite{Viola:2001:ra}. They have been experimentally shown to improve the performance of quantum algorithms \cite{Mohseni:2003:pp} and fault-tolerant quantum key distribution \cite{Zhang:2006:zz}, and have even found use in applications beyond quantum information, for example, in protecting a neutron interferometer from mechanical vibrations \cite{Pushin:2011:zz}. 

More generally, the information-theoretic view of quantum mechanics has proved helpful in understanding and formalizing decoherence as a process of environmental monitoring that results in a transfer of information from the system into quantum correlations between system and environment \cite{Hornberger:2009:aq,Zurek:2009:om,Schlosshauer:2019:qd}. This perspective has also inspired new ways of applying the ideas of decoherence, for example, by studying the redundant storage of information in the environment, as explored in the program of quantum Darwinism \cite{Zurek:2009:om,Zurek:2014:xx,Zwolak:2016:zz}. 

There are numerous experiments in which the observation and control of decoherence processes are important components. Examples include atom--photon interactions in a cavity \cite{Brune:1996:om,Maitre:1997:tv,Raimond:2001:aa,Kaiser:2001:tm,Haroche:2006:hh,Deleglise:2008:oo}, matter-wave interferometry \cite{Arndt:1999:rc,Hornberger:2003:tv,Hackermuller:2003:uu,Hackermuller:2004:rd,Fein:2019:aa}, 
superconducting systems \cite{Chiorescu:2003:ta,Vion:2002:oo,Yu:2002:yb,Martinis:2002:qq}, ion traps \cite{Schneider:1998:yz,Miquel:1997:zz, Myatt:2000:yy,Turchette:2000:aa,Steane:2000:ii,Haffner:2003:oo, Leibfried:2003:mm,SchmidtKaler:2003:pp,Brouard:2004:in,Grotz:2006:km, Haffner:2008:pp}, quantum dots \cite{Kuhlmann:2013:aa,Arnold:2014:oo,Fischer:2009:ii,Kuhlmann:2013:aa,Urbaszek:2013:pp,Delteil:2014:aa,Tighineanu:2018:ii}, and micromechanical resonators \cite{Fong:2012:aa,Zhang:2014:oo,Miao:2014:ii,Moser:2014:uu,Maillet:2016:zz,Maillet:2016:zz}. Some of these experiments chiefly aim to minimize decoherence so that the desired superposition states of mesoscopically or macroscopically distinguishable states can be prepared and maintained, often with an eye toward the implementation of devices for quantum information processing. Other experiments are expressedly designed to study the dynamics of decoherence itself, by making possible the observation of its gradual action, as well as by  exerting appropriate control over the strength and structure of the interaction with the environment such that different decoherence timescales and preferred bases can be realized. 

\begin{figure}
\centering
\includegraphics[scale=1]{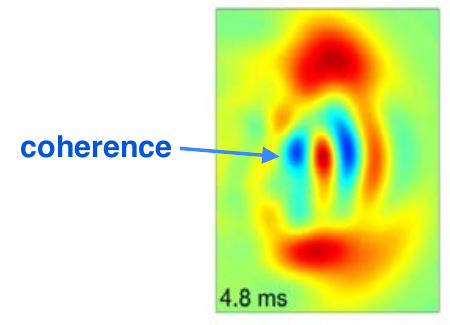} \,\,\, \includegraphics[scale=1]{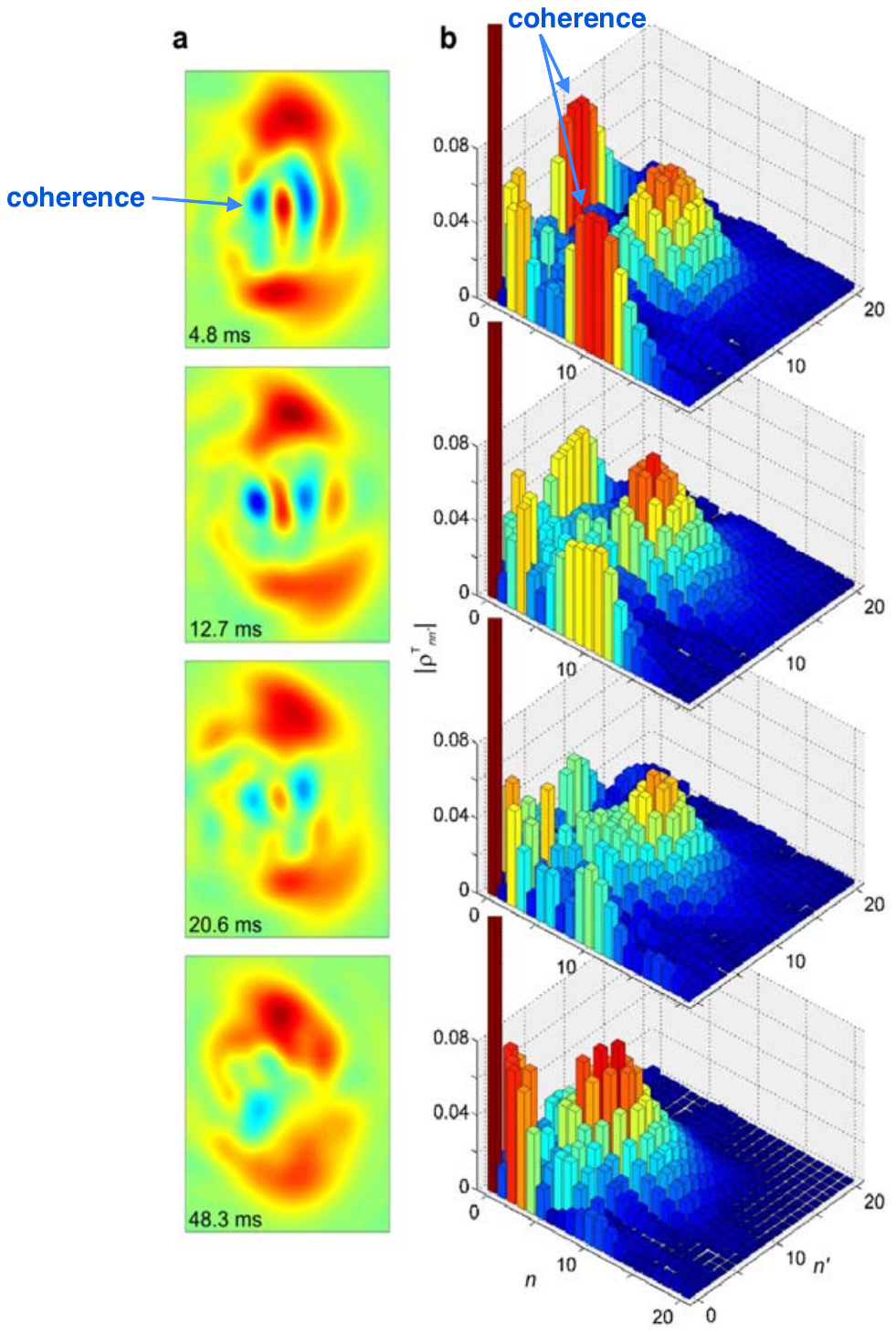}

\hspace*{2cm}\includegraphics[scale=1]{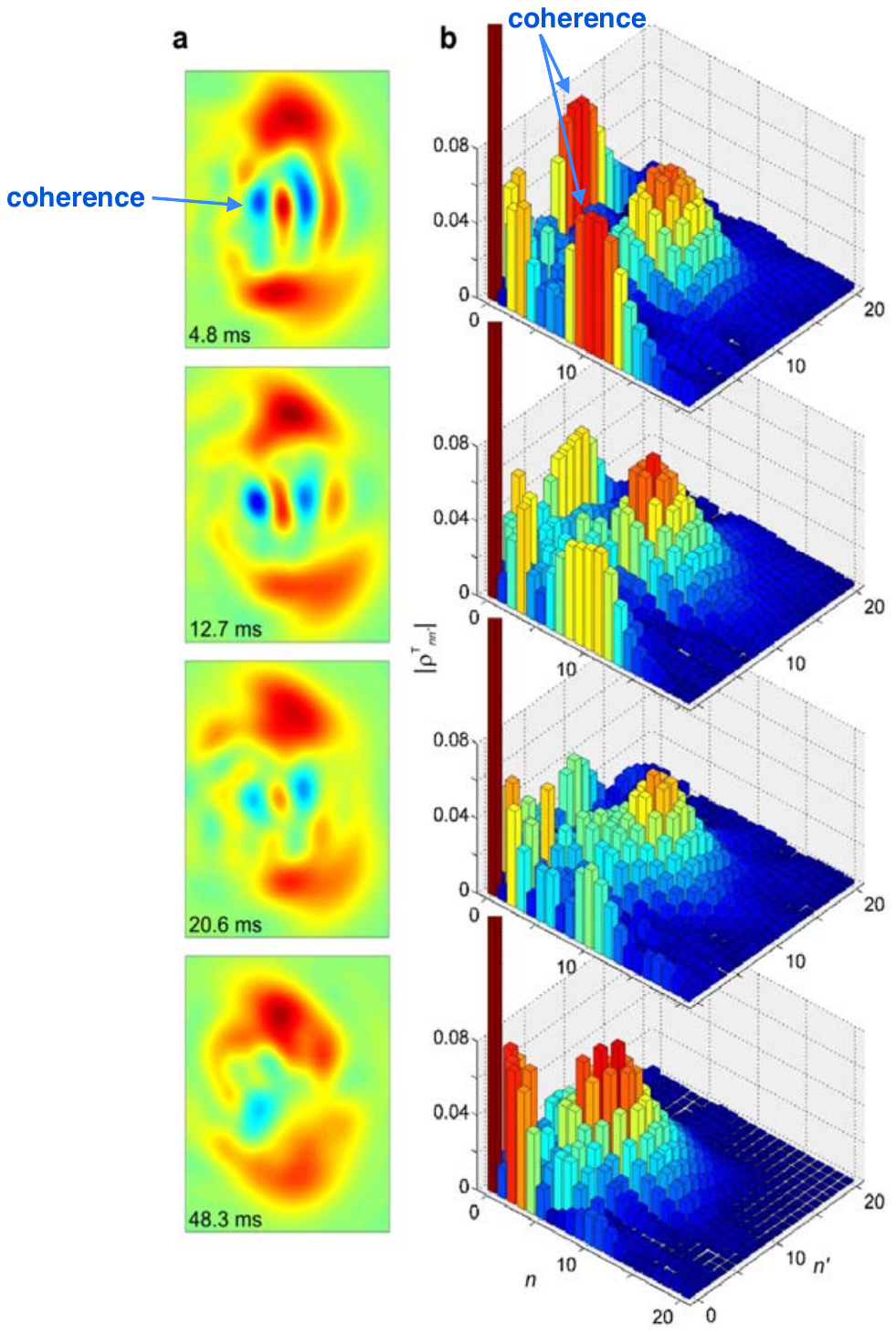} \,\,\, \includegraphics[scale=1]{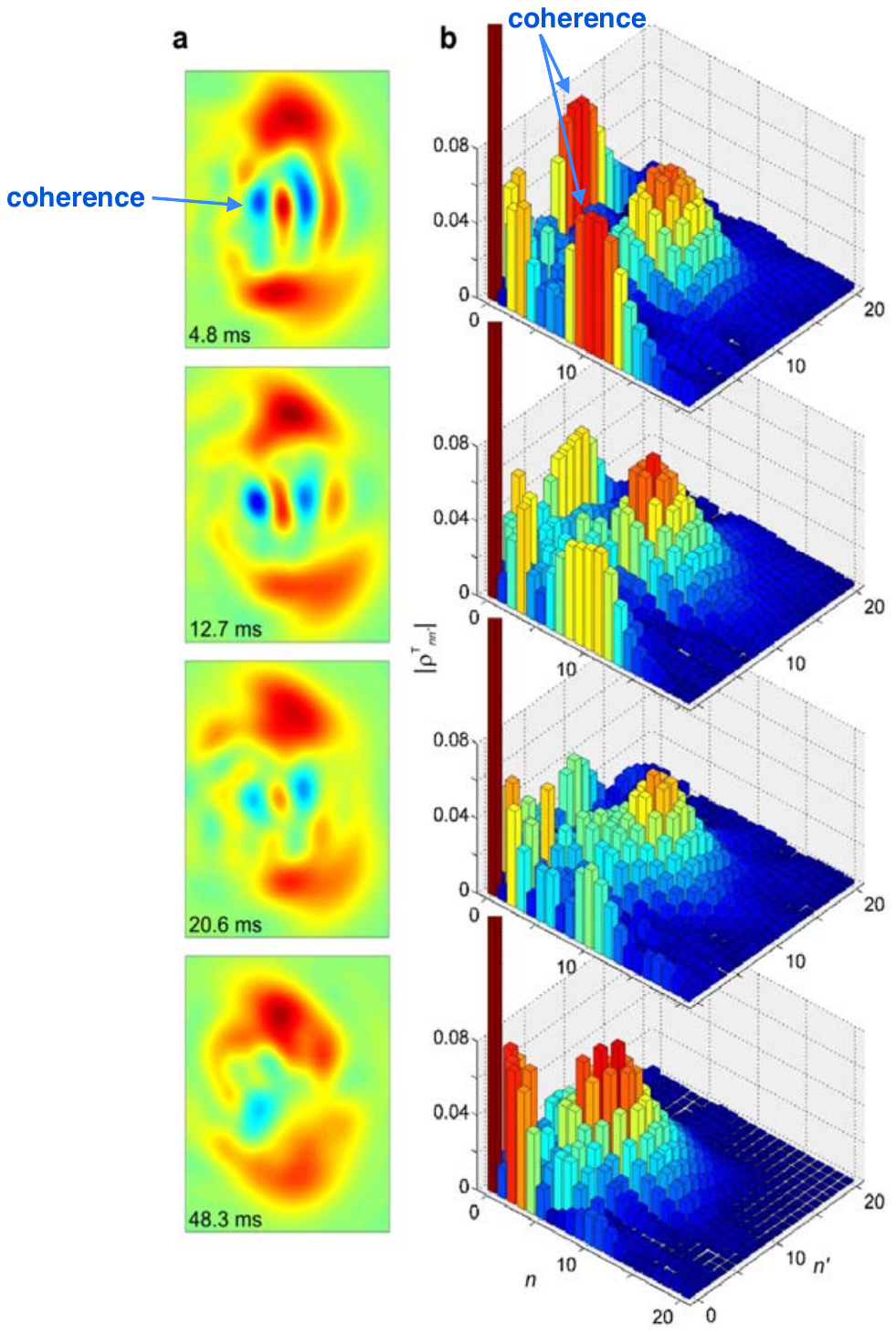}
\caption{Observation of decoherence in the cavity QED experiment of Ref.~\cite{Deleglise:2008:oo}. A superposition of two coherent photon fields with mean photon number $\bar{n}=3.5$ is prepared and its gradual decoherence is monitored. The figure shows the 2D Wigner representation of the photon state as a function of the time. Oscillations in the region between the two peaks represent coherence between the components of the superposition. Figure adapted with permission from Ref.~\cite{Deleglise:2008:oo}.}
\label{fig:qeddecoh}
\end{figure}
 
\begin{figure}
{\footnotesize  \hspace{0cm} (a) }

\vspace{.2cm}

\centering
\includegraphics[scale=1]{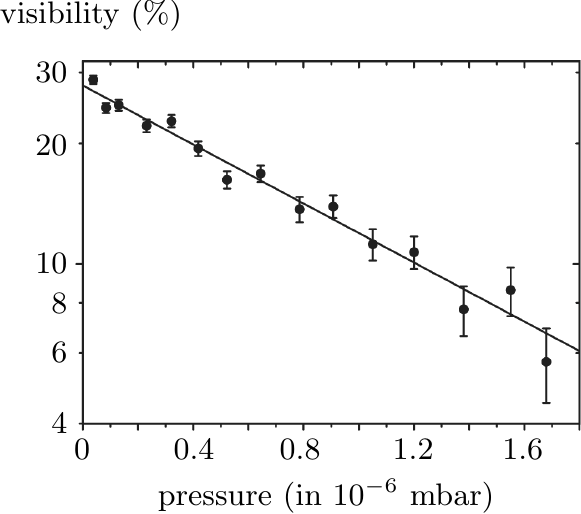}

\vspace{.3cm}

{\footnotesize  \hspace{0cm} (b) }

\vspace{.2cm}

\includegraphics [scale=.7]{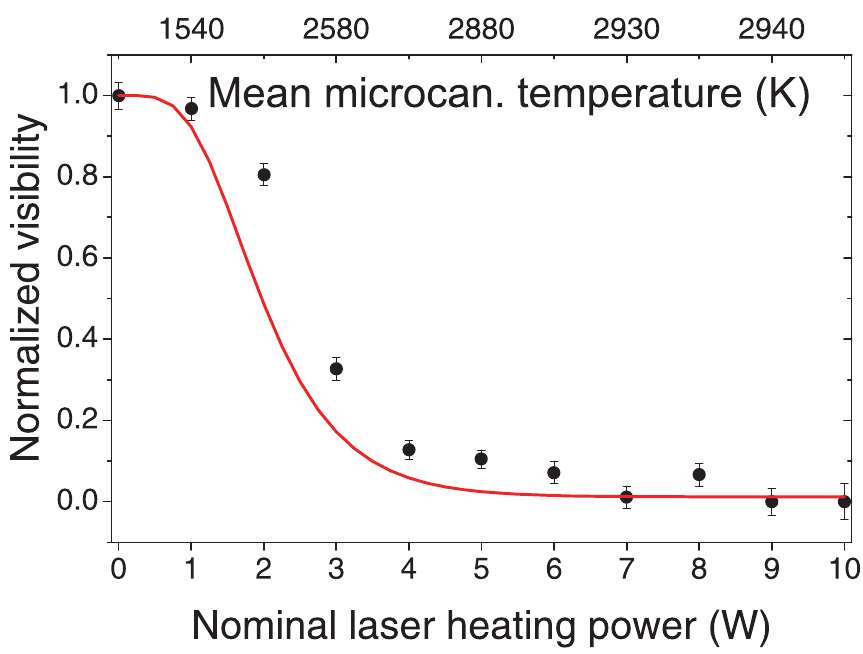}
\caption{Reduction of the visibility of interference fringes for C$_{70}$ fullerenes in the presence of decoherence, as measured in the experiments of Refs.~\cite{Hornberger:2003:tv,Hackermuller:2003:uu, Hackermuller:2004:rd}. (a) Decoherence due to collisions with a background gas, shown as a function of gas pressure \cite{Hornberger:2003:tv,Hackermuller:2003:uu}. The solid line represents theoretical predictions \cite{Hornberger:2003:un,Hornberger:2004:bb,Hornberger:2005:mo}. Figure adapted with permission from  Ref.~\cite{Hackermuller:2003:uu}. (b) Decoherence due to emission of thermal radiation from heated molecules \cite{Hackermuller:2004:rd}. The solid line shows the theoretical prediction \cite{Hackermuller:2004:rd,Hornberger:2005:mo,Hornberger:2006:tx}. Figure reproduced with permission from Ref.~\cite{Hornberger:2012:ii}. }
\label{fig:c70-vis}
\end{figure}

Two classic classes of experiments in this latter category are the cavity QED experiments of Refs.~\cite{Brune:1996:om,Maitre:1997:tv,Raimond:2001:aa,Kaiser:2001:tm,Haroche:2006:hh,Deleglise:2008:oo} and the matter-wave interferometry experiments with large molecules of Refs.~\cite{Arndt:1999:rc,Hornberger:2003:tv,Hackermuller:2003:uu,Hackermuller:2004:rd,Fein:2019:aa}. In the cavity QED experiments, nonclassical superpositions of two coherent-field states $\ket{\alpha \E^{\I \chi}}$ and $\ket{\alpha \E^{-\I \chi}}$ were generated inside the cavity, followed by the observation of their controlled decoherence. Specifically, in the experiment of Ref.~\cite{Deleglise:2008:oo}, the mean photon number was $\bar{n}=\abs{\alpha}^2=3.5$ and the phase $\chi=0.37\pi$, resulting in a very small overlap of the two state components (given by $\abs{\braket{\alpha \E^{\I \chi}}{\alpha \E^{-\I \chi}}}^2 = \E^{- 4\abs{\alpha}^2\sin^2\chi} \approx 8 \times 10^{-6}$). The gradual decoherence of the superposition was measured by reconstructing the photon state at different moments in time (see Fig.~\ref{fig:qeddecoh}). The theoretically predicted decoherence time \cite{Walls:1985:pp,Brune:1992:zz,Haroche:2006:hh,Deleglise:2008:oo} was found to be in good agreement with the experimentally measured value. In the matter-wave interferometry experiments \cite{Arndt:1999:rc,Hornberger:2003:tv,Hackermuller:2003:uu,Hackermuller:2004:rd,Fein:2019:aa}, interference fringes were observed for large molecules in a Talbot--Lau interferometer, and controlled decay of the visibility of  interference fringes due to collisions with background gas molecules \cite{Hornberger:2003:tv,Hackermuller:2003:uu}, as well as due to thermal decoherence \cite{Hackermuller:2004:rd}, was demonstrated (see Fig.~\ref{fig:c70-vis}). For both collisional and thermal decoherence, which-path information carried away by environmental particles---in the form of gas molecules scattered by the fullerenes, and photons emitted from the heated fullerenes---led to a decrease in spatial coherence between wave-function components corresponding to passage through different slits in the diffraction grating. By adjusting the pressure of the background gas, the strength of the collisional decoherence could be tuned. Likewise, the amount of thermal decoherence was varied by heating the fullerenes to different temperatures. 

A detailed understanding of the sources and dynamics of decoherence has also helped realize and observe increasingly large nonclassical superposition states \cite{Arndt:2014:oo}. For example, interference fringes for molecules consisting of up to 2,000 atoms and with masses in excess of 25,000~Da have been reported \cite{Fein:2019:aa}. In cavity QED, mesoscopic superposition states involving $29$ photons have been realized \cite{Auffeves:2003:za}, and superpositions of coherent states involving 111 photons have been reported for a superconducting qubit coupled to a waveguide cavity resonator \cite{Vlastakis:2013:pp}. Superconducting qubits such as quantronium, transmon, and fluxonium qubits \cite{Devoret:2013:pp} have achieved coherence times on the order of $\unit[100]{\mu s}$ \cite{Rigetti:2012:aa,Sears:2012:ee}. A superposition of two hyperfine levels in a trapped $^9$Be$^+$ ion has shown coherence times of \unit[10]{s} \cite{Langer:2005:uu}, and superpositions of two vibrational states for a trapped $^{40}$Ca$^+$ ion have achieved motional center-of-mass-mode coherence times around \unit[100]{ms}. For an experimental realization of a decoherence-free subspace using trapped pairs of ions, coherence times of up to \unit[34]{s} have been observed \cite{Haffner:2005:zz,Langer:2005:uu}. 

Such experiments, with their ability to realize highly nonclassical superposition states and exert exquisite control over the dominant decoherence processes, also play important roles in testing the universal validity of quantum mechanics \cite{Leggett:2002:uy,Marshall:2003:om,Bassi:2005:om,Pikovski:2012:aa,Arndt:2014:oo,Wan:2016:oo,Kaltenbaek:2016:pp,Stickler:2018:ii}. The goal here is to look for (or rule out) hypothetical physical mechanisms \cite{Bassi:2003:yb} that may induce a nonunitary evolution of the wave function for a closed system. The challenge in testing such collapse models is to distinguish any observation of a collapse from what may phenomenologically be explained in terms of decoherence \cite{Tegmark:1993:uz,Nimmrichter:2013:aa}. The pervasiveness of decoherence, and the fact that the effectiveness of the collapse mechanism increases with the size of the system (just as decoherence typically becomes more effective for larger systems), make the unambiguous detection of a novel collapse mechanism difficult. Many experiments have been proposed for probing the mass regimes relevant to collapse models. Examples include
matter-wave interferometry \cite{Nimmrichter:2011:pr,Hornberger:2012:ii,Arndt:2014:oo}, micromechanical oscillators \cite{Marshall:2003:om,Pikovski:2012:aa}, interference of free nanoparticles \cite{Romero:2011:aa,Wan:2016:oo}, molecular nanorotors \cite{Stickler:2018:ii}, and even space-based experiments involving systems on the order of $10^{10}$ atoms \cite{Kaltenbaek:2016:pp}. The motivation for situating experiments in space comes from considerations of decoherence, since such an environment would offer very low background gas pressures (thus minimizing collisional decoherence), very low temperatures (minimizing thermal decoherence), and microgravity (minimizing decoherence due to gravitational time dilation \cite{Pikovski:2015:oo}); the last property may even allow for tests of quantum gravity models \cite{Kaltenbaek:2016:pp}. None of these experiments have yet been realized, although matter-wave interferometry with molecular clusters between $10^6$ and $\unit[10^8]{amu}$ (the regime relevant to testing the collapse theories proposed in Refs.~\cite{Adler:2007:um,Bassi:2010:aa}) may soon be within reach of experimental feasibility \cite{Hornberger:2012:ii,Arndt:2014:oo}.

It should be noted that not all reported experimental observations of a loss of coherence are due to a proper decoherence process as defined by an entanglement-mediated transfer of information to an environment. Frequently, such losses are caused by fluctuations (noise) acting locally on the system, manifesting itself in the form of dephasing for an ensemble of systems (see the brief discussion in Sec.~\ref{sec:interpretation} regarding the distinction between these types of coherence loss). For instance, many instances of an observed loss of coherence in superconducting quantum systems fall into this latter category, such as the dephasing of superpositions of supercurrents in SQUIDs \cite{Chiorescu:2003:ta}, dephasing of superpositions of charge states in charge qubits \cite{Vion:2002:oo}, and dephasing of superpositions of macroscopically distinct phase states in phase qubits \cite{Yu:2002:yb,Martinis:2002:qq}. The same is true for a number of experiments on ion traps. Here, dephasing between the internal qubit states of the ion has been observed due to sources such as the magnetic trapping field \cite{SchmidtKaler:2003:pp,Brouard:2004:in,Grotz:2006:km, Haffner:2008:pp}, laser fluctuations \cite{Schneider:1998:yz,Miquel:1997:zz}, ac-Stark shifts \cite{Haffner:2003:oo}, off-resonant excitations \cite{Steane:2000:ii}, and detuning errors \cite{Leibfried:2003:mm}. Loss of coherence in ion traps has also been studied in terms of the dephasing of superpositions of motional states induced by a random noise process simulated by introducing fluctuations of the trap frequency \cite{Myatt:2000:yy,Turchette:2000:aa}. 

Above, we have alluded to the connection between experimental advances and the development of realistic decoherence models. This connection may be nicely illustrated with the example of collisional decoherence. The importance of collisional decoherence in accounting for preferred macroscopic states was already recognized by Zeh in his 1970 paper \cite{Zeh:1970:yt}. There, he discussed the example of the two spatially distinct configurations of a sugar molecule (right-handed and left-handed) that ``interact in different ways with their environment.'' A dominant form of such interactions are environmental scattering processes that resolve differences between the spatial configurations, leading to the dynamical superselection of ``handedness'' as the stable state. This, Zeh suggested, explained why superpositions of the right-handed and left-handed configurations are not observed, even though such superpositions also form energy eigenstates, which are usually the preferred states for microscopic systems (see also Refs.~\cite{Harris:1981:rc,Zeh:1999:qr,Trost:2009:ll,Bahrami:2012:oo}). 

One of the first models of collisional decoherence was studied in the classic paper by Joos and Zeh \cite{Joos:1985:iu}. This work was an extension of research on the quantum Zeno effect Joos had done as a doctoral student under Zeh's supervision \cite{Joos:1984:km}. In his interview with Freitas \cite{Freitas:2008:yy}, Zeh later recalled that he had instructed Joos to studiously steer clear of any reference to matters of interpretation, in order to avoid the kind of criticism and rejection from the academic community Zeh had experienced \cite{Camilleri:2009:aq,Freire:2009:aa} with his papers from the 1970s:
\begin{quote}
I suggested that he might think about the role of the environment, but without ever talking about Everett. So he concentrated on consequences on the reduced density matrix. This was something nobody could deny. 
\end{quote}
Joos and Zeh's paper presented a master-equation treatment of the reduced density matrix and derived some classic order-of-magnitude estimates of decoherence rates. These estimates indicated the extreme efficiency with which superpositions of macroscopially distinct positions would be decohered. Even an environment as subtle, weak, and inescapable as the cosmic background radiation was found to be capable of having an appreciable effect: Joos and Zeh used the example of a dust grain of size $\unit[10^{-3}]{cm}$ delocalized over a distance roughly equal to its size, and calculated a decoherence timescale on the order of one second.

In 1985, when Joos and Zeh's paper was published, such order-of-magnitude estimates were sufficient, as experiments that would actually resolve these collisional-decoherence dynamics in a gradual, controlled, and measurable fashion were not available. This changed around 2000 when the matter-wave interferometry experiments enabled an experimental test of models of collisional decoherence \cite{Hornberger:2003:tv,Hackermuller:2003:uu,Hornberger:2003:un,Hornberger:2004:bb,Hornberger:2005:mo}. Now precise quantitative results were needed to compare with experimental values. Previous work by Gallis and Fleming \cite{Gallis:1990:un} and Di{\'o}si \cite{Diosi:1995:um} had already brought improvements and generalizations to Joos and Zeh's master equation, by relaxing the no-recoil assumption for the central particle, relaxing an assumption regarding the relative wavelengths of system and environmental particles, and addressing a mathematical issue that had resulted in decoherence rates that were too large by a factor of $2\pi$ (see also Ref.~\cite{Adler:2006:yb}). In 2003, Hornberger and Sipe \cite{Hornberger:2003:un} gave a rigorous derivation of the master equation for collisional decoherence. It used an approach quite different from Di{\'o}si's, and its predictions were found to agree well with the decrease in fringe visibility observed in the matter-wave experiments with fullerenes  shown in Fig.~\ref{fig:c70-vis} \cite{Hornberger:2003:un,Hornberger:2004:bb,Hornberger:2005:mo}. 

But this was not the final word on modeling. Hornberger and coworkers also gave a general, nonperturbative treatment based on the quantum linear Boltzmann equation \cite{Hornberger:2006:tb,Hornberger:2006:tc,Hornberger:2008:ii,Busse:2009:aa,Vacchini:2009:pp,Busse:2010:aa,Busse:2010:oo}, which provided a unified account of the dynamical interplay between phase-space decoherence and dissipation. Previous results were recovered as limiting cases, and it was shown that the pointer states dynamically selected by the environmental interaction were exponentially localized solitonic wave functions obeying the classical equations of motion \cite{Busse:2009:aa,Busse:2010:aa}. Subsequently, insights gleaned from molecular interferometry experiments have led to further refinements of these models. For example, near-field interferometry with massive molecules was found to be highly sensitive to molecular rotations \cite{Gring:2010:aa,Stickler:2015:zz}. This observation helped motivate the generalization of previous models of collisional decoherence, in which the central particle was treated as a pointlike particle with no orientational degrees of freedom, to the derivation of master equations describing the spatio-orientational decoherence of rotating, anisotropic, nonspherical molecules immersed into an environment of gas particles \cite{Walter:2016:zz,Stickler:2016:yy,Papendell:2017:yy, Stickler:2018:oo,Stickler:2018:uu}. These models were found to make predictions in good agreement with experimental data \cite{Stickler:2018:oo}.

\section{\label{sec:concluding-remarks}Concluding remarks}

An early version of what would become Zeh's 1970 paper \cite{Zeh:1970:yt} was rejected with the argument that ``quantum theory does not apply to macroscopic objects''  \cite{Camilleri:2009:aq}. Even if such an application was not necessarily considered explicitly forbidden, a common view seems to have been that such an application would be either too difficult in practice (given the complexity of larger systems) or unnecessary (since the systems are known to behave classically, i.e., not to exhibit quantum effects). Indeed, for a long time the distinction between quantum and classical was often synonymous with the distinction between microsopic and macrosopic. In such light, Zeh's early work may be seen as motivated by his sense that this conflation was in error, leading him to his suggestion that everything, no matter its size, not only \emph{could}, but \emph{should} be described by a (realistically interpreted) quantum-mechanical wave function. This approach he referred to as a ``universally valid quantum theory,'' which he defined in his 1970 paper as a ``universe [described] by a wave function that obeys the Schr\"odinger equation'' \cite[p.~73]{Zeh:1970:yt}. 

To Schr\"odinger himself, the consideration of an amplifying chain of entanglement from microscopic to macroscopic scales had served  merely as a \emph{reductio ad absurdum}, meant to illustrate the ``ridiculous'' consequences of applying quantum mechanics in this way. Zeh not only viewed these consequences in a fresh and unprejudiced way, but, crucially, he also realized that the very quantum property of entanglement, ``\emph{the} characteristic trait of quantum mechanics, the one that enforces its entire departure from classical lines of thought'' (Schr\"odinger's famous words \cite[p.~555]{Schrodinger:1935:jn}), could provide a dynamical explanation of the emergence of classicality. Thus, in a twist of irony, the justification for why macroscopic systems may be treated classically came from Zeh's radically quantum-mechanical treatment of such systems. In this way, decoherence theory showed why the quantum--classical and microscopic--macroscopic boundaries, although fundamentally independent of each other, \emph{appear} to coincide in practice. This became especially clear with Joos and Zeh's 1985 paper \cite{Joos:1985:iu} and Zurek's estimate of a decoherence timescale \cite{Zurek:1986:uz},\footnote{Among these early studies, one should also mention the papers by Walls and Milburn \cite{Walls:1985:pp} and by Caldeira and Leggett \cite{Caldeira:1985:tt}, which investigated the influence of damping on the coherence of superpositions. In particular, the master equation derived by Caldeira and Leggett \cite{Caldeira:1983:on} has been widely used in modeling decoherence and dissipation \cite{Gallis:1990:un,Gallis:1992:im,Anglin:1997:za}, and Camilleri \cite{Camilleri:2009:aq} has suggested that Caldeira and Leggett's work ``was of decisive importance not only for the mathematical treatment and conceptual development of environment-induced decoherence, but also for the interpretational context.'' It is important to bear in mind, however, that while damping (dissipation) is always accompanied by a loss of coherence, decoherence may be substantial even when there is no loss of energy from the system, as already demonstrated by one of the first models of decoherence \cite{Zurek:1982:tv}.} both of which demonstrated that, in general, the larger the system, the stronger its decoherence, and, furthermore, that on mesoscopic and macroscopic scales such decoherence is virtually instantaneous compared to the timescales for dissipation and classical noise effects. 

The fundamental idea of decoherence theory---that ubiquitous, uncontrollable entanglement leads to local nonobservability of interference and dynamical superselection, as already spelled out in Zeh's seminal 1970 paper \cite{Zeh:1970:yt}---has proven uniquely fruitful, inspiring and making possible new experiments, as well as revolutionizing our theoretical understanding and experimental control of the quantum-to-classical transition. Given the current interest in the construction of devices for quantum computing, it is clear that decoherence will continue to play a central role in quantum science for the foreseeable future. If anything, its role will only amplify, as ever-larger multiqubit systems are realized and quantum phenomena involving coherence and entanglement are explored on increasingly macroscopic scales. The search for deviations from standard quantum mechanics, while not as prominent a research effort as quantum information science, will surely provide additional impetus for theoretical and experimental studies of decoherence. 

We must be grateful to Zeh not only for his original insight, but also for the tenacity with which he pursued his ideas. He was a pioneer, and a bold and independent thinker. His voice will be missed.

% \bibliography{../Bib/References} 

\begin{thebibliography}{196}%
\makeatletter
\providecommand \@ifxundefined [1]{%
 \@ifx{#1\undefined}
}%
\providecommand \@ifnum [1]{%
 \ifnum #1\expandafter \@firstoftwo
 \else \expandafter \@secondoftwo
 \fi
}%
\providecommand \@ifx [1]{%
 \ifx #1\expandafter \@firstoftwo
 \else \expandafter \@secondoftwo
 \fi
}%
\providecommand \natexlab [1]{#1}%
\providecommand \enquote  [1]{``#1''}%
\providecommand \bibnamefont  [1]{#1}%
\providecommand \bibfnamefont [1]{#1}%
\providecommand \citenamefont [1]{#1}%
\providecommand \href@noop [0]{\@secondoftwo}%
\providecommand \href [0]{\begingroup \@sanitize@url \@href}%
\providecommand \@href[1]{\@@startlink{#1}\@@href}%
\providecommand \@@href[1]{\endgroup#1\@@endlink}%
\providecommand \@sanitize@url [0]{\catcode `\\12\catcode `\$12\catcode
  `\&12\catcode `\#12\catcode `\^12\catcode `\_12\catcode `\%12\relax}%
\providecommand \@@startlink[1]{}%
\providecommand \@@endlink[0]{}%
\providecommand \url  [0]{\begingroup\@sanitize@url \@url }%
\providecommand \@url [1]{\endgroup\@href {#1}{\urlprefix }}%
\providecommand \urlprefix  [0]{URL }%
\providecommand \Eprint [0]{\href }%
\providecommand \doibase [0]{http://dx.doi.org/}%
\providecommand \selectlanguage [0]{\@gobble}%
\providecommand \bibinfo  [0]{\@secondoftwo}%
\providecommand \bibfield  [0]{\@secondoftwo}%
\providecommand \translation [1]{[#1]}%
\providecommand \BibitemOpen [0]{}%
\providecommand \bibitemStop [0]{}%
\providecommand \bibitemNoStop [0]{.\EOS\space}%
\providecommand \EOS [0]{\spacefactor3000\relax}%
\providecommand \BibitemShut  [1]{\csname bibitem#1\endcsname}%
\let\auto@bib@innerbib\@empty
%</preamble>
\bibitem [{\citenamefont {Camilleri}\ and\ \citenamefont
  {Schlosshauer}(2015)}]{Camilleri:2015:oo}%
  \BibitemOpen
  \bibfield  {author} {\bibinfo {author} {\bibfnamefont {K.}~\bibnamefont
  {Camilleri}}\ and\ \bibinfo {author} {\bibfnamefont {M.}~\bibnamefont
  {Schlosshauer}},\ }\href@noop {} {\bibfield  {journal} {\bibinfo  {journal}
  {Stud. Hist. Phil. Mod. Phys.}\ }\textbf {\bibinfo {volume} {49}},\ \bibinfo
  {pages} {73} (\bibinfo {year} {2015})}\BibitemShut {NoStop}%
\bibitem [{\citenamefont {Zeh}(1970)}]{Zeh:1970:yt}%
  \BibitemOpen
  \bibfield  {author} {\bibinfo {author} {\bibfnamefont {H.~D.}\ \bibnamefont
  {Zeh}},\ }\href@noop {} {\bibfield  {journal} {\bibinfo  {journal} {Found.
  Phys.}\ }\textbf {\bibinfo {volume} {1}},\ \bibinfo {pages} {69} (\bibinfo
  {year} {1970})}\BibitemShut {NoStop}%
\bibitem [{Fre()}]{Freitas:2008:yy}%
  \BibitemOpen
  \href@noop {} {\bibinfo {title} {Interview with H.\ Dieter {Z}eh, conducted at Waldhilsbach, Germany, by
  {F}{\'a}bio {F}reitas, {J}uly 25--26, 2008}.\ }\bibinfo {howpublished}
  {Audio recording and transcript deposited at the American Institute of
  Physics}\BibitemShut {NoStop}%
\bibitem [{\citenamefont {Camilleri}(2009)}]{Camilleri:2009:aq}%
  \BibitemOpen
  \bibfield  {author} {\bibinfo {author} {\bibfnamefont {K.}~\bibnamefont
  {Camilleri}},\ }\href@noop {} {\bibfield  {journal} {\bibinfo  {journal}
  {Stud. Hist. Phil. Mod. Phys.}\ }\textbf {\bibinfo {volume} {40}},\ \bibinfo
  {pages} {290} (\bibinfo {year} {2009})}\BibitemShut {NoStop}%
\bibitem [{\citenamefont {{Freire Jr.}}(2009)}]{Freire:2009:aa}%
  \BibitemOpen
  \bibfield  {author} {\bibinfo {author} {\bibfnamefont {O.}~\bibnamefont
  {{Freire Jr.}}},\ }\href {\doibase 10.1016/j.shpsb.2009.09.002} {\bibfield
  {journal} {\bibinfo  {journal} {Stud. Hist. Phil. Mod. Phys.}\ }\textbf
  {\bibinfo {volume} {40}},\ \bibinfo {pages} {280} (\bibinfo {year}
  {2009})}\BibitemShut {NoStop}%
\bibitem [{\citenamefont {Zeh}(2006)}]{Zeh:2006:na}%
  \BibitemOpen
  \bibfield  {author} {\bibinfo {author} {\bibfnamefont {H.~D.}\ \bibnamefont
  {Zeh}},\ }in\ \href@noop {} {\emph {\bibinfo {booktitle} {Quantum
  Decoherence}}},\ \bibinfo {editor} {edited by\ \bibinfo {editor}
  {\bibfnamefont {B.}~\bibnamefont {Duplantier}}, \bibinfo {editor}
  {\bibfnamefont {J.-M.}\ \bibnamefont {Raimond}},  and\ \bibinfo {editor}
  {\bibfnamefont {V.}~\bibnamefont {Rivasseau}}}\ (\bibinfo  {publisher}
  {Birkh{\"a}user, Basel},\ \bibinfo {year} {2006}), pp.\ \bibinfo {pages}
  {151--175}\BibitemShut {NoStop}%
\bibitem [{\citenamefont {Zeh}(1973)}]{Zeh:1973:wq}%
  \BibitemOpen
  \bibfield  {author} {\bibinfo {author} {\bibfnamefont {H.~D.}\ \bibnamefont
  {Zeh}},\ }\href@noop {} {\bibfield  {journal} {\bibinfo  {journal} {Found.
  Phys.}\ }\textbf {\bibinfo {volume} {3}},\ \bibinfo {pages} {109} (\bibinfo
  {year} {1973})}\BibitemShut {NoStop}%
\bibitem [{\citenamefont {Zeh}(2016)}]{Zeh:2016:ii}%
  \BibitemOpen
  \bibfield  {author} {\bibinfo {author} {\bibfnamefont {H.~D.}\ \bibnamefont
  {Zeh}},\ }\href@noop {} {\bibfield  {journal} {\bibinfo  {journal} {Z.
  Naturf.}\ }\textbf {\bibinfo {volume} {71a}},\ \bibinfo {pages} {195}
  (\bibinfo {year} {2016})}\BibitemShut {NoStop}%
\bibitem [{\citenamefont {Zeh}(2000{\natexlab{a}})}]{Zeh:2000:rr}%
  \BibitemOpen
  \bibfield  {author} {\bibinfo {author} {\bibfnamefont {H.~D.}\ \bibnamefont
  {Zeh}},\ }\href@noop {} {\bibfield  {journal} {\bibinfo  {journal} {Found.
  Phys. Lett.}\ }\textbf {\bibinfo {volume} {13}},\ \bibinfo {pages} {221}
  (\bibinfo {year} {2000}{\natexlab{a}})}\BibitemShut {NoStop}%
\bibitem [{\citenamefont {Bassi}\ and\ \citenamefont
  {Ghirardi}(2003)}]{Bassi:2003:yb}%
  \BibitemOpen
  \bibfield  {author} {\bibinfo {author} {\bibfnamefont {A.}~\bibnamefont
  {Bassi}}\ and\ \bibinfo {author} {\bibfnamefont {G.~C.}\ \bibnamefont
  {Ghirardi}},\ }\href@noop {} {\bibfield  {journal} {\bibinfo  {journal}
  {Phys. Rep.}\ }\textbf {\bibinfo {volume} {379}},\ \bibinfo {pages} {257}
  (\bibinfo {year} {2003})}\BibitemShut {NoStop}%
\bibitem [{\citenamefont {Zeh}(1999)}]{Zeh:1999:rr}%
  \BibitemOpen
  \bibfield  {author} {\bibinfo {author} {\bibfnamefont {H.~D.}\ \bibnamefont
  {Zeh}},\ }\href@noop {} {\bibfield  {journal} {\bibinfo  {journal} {Found.
  Phys. Lett.}\ }\textbf {\bibinfo {volume} {12}},\ \bibinfo {pages} {197}
  (\bibinfo {year} {1999})}\BibitemShut {NoStop}%
\bibitem [{\citenamefont {Vaihinger}(1935)}]{Vaihinger:1935:jj}%
  \BibitemOpen
  \bibfield  {author} {\bibinfo {author} {\bibfnamefont {H.}~\bibnamefont
  {Vaihinger}},\ }\href@noop {} {\emph {\bibinfo {title} {The Philosophy of
  ``As if'': A System of the Theoretical, Practical and Religious Fictions of
  Mankind}}},\ \bibinfo {edition} {2nd}\ ed.\ (\bibinfo  {publisher} {Routledge
  \& Kegan Paul},\ \bibinfo {address} {London},\ \bibinfo {year} {1935}),
  \bibinfo {note} {translated by C. K. Ogden}\BibitemShut {NoStop}%
\bibitem [{\citenamefont {Hemmo}\ and\ \citenamefont
  {Pitowsky}(2007)}]{Hemmo:2007:uu}%
  \BibitemOpen
  \bibfield  {author} {\bibinfo {author} {\bibfnamefont {M.}~\bibnamefont
  {Hemmo}}\ and\ \bibinfo {author} {\bibfnamefont {I.}~\bibnamefont
  {Pitowsky}},\ }\href {\doibase 10.1016/j.shpsb.2006.04.005} {\bibfield
  {journal} {\bibinfo  {journal} {Stud. Hist. Phil. Mod. Phys.}\ }\textbf
  {\bibinfo {volume} {38}},\ \bibinfo {pages} {333} (\bibinfo {year}
  {2007})}\BibitemShut {NoStop}%
\bibitem [{\citenamefont {Saunders}\ \emph {et~al.}(2010)\citenamefont
  {Saunders}, \citenamefont {Barrett}, \citenamefont {Kent}, and\
  \citenamefont {Wallace}}]{Saunders:2010:im}%
  \BibitemOpen
  \bibinfo {editor} {\bibfnamefont {S.}~\bibnamefont {Saunders}}, \bibinfo
  {editor} {\bibfnamefont {J.}~\bibnamefont {Barrett}}, \bibinfo {editor}
  {\bibfnamefont {A.}~\bibnamefont {Kent}},  and\ \bibinfo {editor}
  {\bibfnamefont {D.}~\bibnamefont {Wallace}},\ eds.,\ \href@noop {} {\emph
  {\bibinfo {title} {Many Worlds? {E}verett, Quantum Theory and Reality}}}\
  (\bibinfo  {publisher} {Oxford University Press},\ \bibinfo {address}
  {Oxford},\ \bibinfo {year} {2010})\BibitemShut {NoStop}%
\bibitem [{\citenamefont {Zeh}(1997)}]{Zeh:1996:gy}%
  \BibitemOpen
  \bibfield  {author} {\bibinfo {author} {\bibfnamefont {H.~D.}\ \bibnamefont
  {Zeh}},\ }in\ \href@noop {} {\emph {\bibinfo {booktitle} {New Developments on
  Fundamental Problems in Quantum Physics (Oviedo II)}}},\ \bibinfo {editor}
  {edited by\ \bibinfo {editor} {\bibfnamefont {M.}~\bibnamefont {Ferrero}}\
  and\ \bibinfo {editor} {\bibfnamefont {A.}~\bibnamefont {van~der Merwe}}}\
  (\bibinfo  {publisher} {Kluwer},\ \bibinfo {address} {Dordrecht},\ \bibinfo
  {year} {1997}), pp.\ \bibinfo {pages} {441--452}\BibitemShut {NoStop}%
\bibitem [{\citenamefont {Joos}(2000)}]{Joos:1999:po}%
  \BibitemOpen
  \bibfield  {author} {\bibinfo {author} {\bibfnamefont {E.}~\bibnamefont
  {Joos}},\ }in\ \href@noop {} {\emph {\bibinfo {booktitle} {Decoherence:
  Theoretical, Experimental, and Conceptual Problems}}},\ \bibinfo {editor}
  {edited by\ \bibinfo {editor} {\bibfnamefont {P.}~\bibnamefont {Blanchard}},
  \bibinfo {editor} {\bibfnamefont {D.}~\bibnamefont {Giulini}}, \bibinfo
  {editor} {\bibfnamefont {E.}~\bibnamefont {Joos}}, \bibinfo {editor}
  {\bibfnamefont {C.}~\bibnamefont {Kiefer}}, and\ \bibinfo {editor}
  {\bibfnamefont {I.-O.}\ \bibnamefont {Stamatescu}}}\ (\bibinfo  {publisher}
  {Springer},\ \bibinfo {address} {Berlin},\ \bibinfo {year} {2000}), pp.\
  \bibinfo {pages} {1--17}\BibitemShut {NoStop}%
\bibitem [{\citenamefont {Heisenberg}(1955)}]{Heisenberg:1955:lm}%
  \BibitemOpen
  \bibfield  {author} {\bibinfo {author} {\bibfnamefont {W.}~\bibnamefont
  {Heisenberg}},\ }in\ \href@noop {} {\emph {\bibinfo {booktitle} {Niels Bohr
  and the Development of Physics: Essays Dedicated to Niels Bohr on the
  Occasion of his Seventieth Birthday}}},\ \bibinfo {editor} {edited by\
  \bibinfo {editor} {\bibfnamefont {W.}~\bibnamefont {Pauli}}, \bibinfo
  {editor} {\bibfnamefont {L.}~\bibnamefont {Rosenfeld}}, and\ \bibinfo
  {editor} {\bibfnamefont {V.}~\bibnamefont {Weisskopf}}}\ (\bibinfo
  {publisher} {McGraw Hill},\ \bibinfo {address} {New York},\ \bibinfo {year}
  {1955}), pp.\ \bibinfo {pages} {12--29}\BibitemShut {NoStop}%
\bibitem [{\citenamefont {Brune}\ \emph {et~al.}(1996)\citenamefont {Brune},
  \citenamefont {Hagley}, \citenamefont {Dreyer}, \citenamefont {Ma{\^i}tre},
  \citenamefont {Maali}, \citenamefont {Wunderlich}, \citenamefont {Raimond},\
  and\ \citenamefont {Haroche}}]{Brune:1996:om}%
  \BibitemOpen
  \bibfield  {author} {\bibinfo {author} {\bibfnamefont {M.}~\bibnamefont
  {Brune}}, \bibinfo {author} {\bibfnamefont {E.}~\bibnamefont {Hagley}},
  \bibinfo {author} {\bibfnamefont {J.}~\bibnamefont {Dreyer}}, \bibinfo
  {author} {\bibfnamefont {X.}~\bibnamefont {Ma{\^i}tre}}, \bibinfo {author}
  {\bibfnamefont {A.}~\bibnamefont {Maali}}, \bibinfo {author} {\bibfnamefont
  {C.}~\bibnamefont {Wunderlich}}, \bibinfo {author} {\bibfnamefont {J.~M.}\
  \bibnamefont {Raimond}}, and\ \bibinfo {author} {\bibfnamefont
  {S.}~\bibnamefont {Haroche}},\ }\href@noop {} {\bibfield  {journal} {\bibinfo
   {journal} {Phys. Rev. Lett.}\ }\textbf {\bibinfo {volume} {77}},\ \bibinfo
  {pages} {4887} (\bibinfo {year} {1996})}\BibitemShut {NoStop}%
\bibitem [{\citenamefont {Ma{\^i}tre}\ \emph {et~al.}(1997)\citenamefont
  {Ma{\^i}tre}, \citenamefont {Hagley}, \citenamefont {Dreyer}, \citenamefont
  {Maali}, \citenamefont {Brune}, \citenamefont {Raimond},\ and\ \citenamefont
  {Haroche}}]{Maitre:1997:tv}%
  \BibitemOpen
  \bibfield  {author} {\bibinfo {author} {\bibfnamefont {X.}~\bibnamefont
  {Ma{\^i}tre}}, \bibinfo {author} {\bibfnamefont {E.}~\bibnamefont {Hagley}},
  \bibinfo {author} {\bibfnamefont {J.}~\bibnamefont {Dreyer}}, \bibinfo
  {author} {\bibfnamefont {A.}~\bibnamefont {Maali}}, \bibinfo {author}
  {\bibfnamefont {C.~W.~M.}\ \bibnamefont {Brune}}, \bibinfo {author}
  {\bibfnamefont {J.~M.}\ \bibnamefont {Raimond}}, and\ \bibinfo {author}
  {\bibfnamefont {S.}~\bibnamefont {Haroche}},\ }\href@noop {} {\bibfield
  {journal} {\bibinfo  {journal} {J. Mod. Opt.}\ }\textbf {\bibinfo {volume}
  {44}},\ \bibinfo {pages} {2023} (\bibinfo {year} {1997})}\BibitemShut
  {NoStop}%
\bibitem [{\citenamefont {Raimond}\ \emph {et~al.}(1997)\citenamefont
  {Raimond}, \citenamefont {Brune},\ and\ \citenamefont
  {Haroche}}]{Raimond:1997:um}%
  \BibitemOpen
  \bibfield  {author} {\bibinfo {author} {\bibfnamefont {J.~M.}\ \bibnamefont
  {Raimond}}, \bibinfo {author} {\bibfnamefont {M.}~\bibnamefont {Brune}}, 
  and\ \bibinfo {author} {\bibfnamefont {S.}~\bibnamefont {Haroche}},\
  }\href@noop {} {\bibfield  {journal} {\bibinfo  {journal} {Phys. Rev. Lett.}\
  }\textbf {\bibinfo {volume} {79}},\ \bibinfo {pages} {1964} (\bibinfo {year}
  {1997})}\BibitemShut {NoStop}%
\bibitem [{\citenamefont {Howard}(2004)}]{Howard:2004:mh}%
  \BibitemOpen
  \bibfield  {author} {\bibinfo {author} {\bibfnamefont {D.}~\bibnamefont
  {Howard}},\ }\href@noop {} {\bibfield  {journal} {\bibinfo  {journal} {Phil.
  Sci.}\ }\textbf {\bibinfo {volume} {71}},\ \bibinfo {pages} {669} (\bibinfo
  {year} {2004})}\BibitemShut {NoStop}%
\bibitem [{\citenamefont {Zeh}(2004)}]{Zeh:2004:zm}%
  \BibitemOpen
  \bibfield  {author} {\bibinfo {author} {\bibfnamefont {H.~D.}\ \bibnamefont
  {Zeh}},\ }in\ \href@noop {} {\emph {\bibinfo {booktitle} {Science and
  Ultimate Reality: Quantum Theory, Cosmology and Complexity}}},\ \bibinfo
  {editor} {edited by\ \bibinfo {editor} {\bibfnamefont {J.~D.}\ \bibnamefont
  {Barrow}}, \bibinfo {editor} {\bibfnamefont {P.~C.~W.}\ \bibnamefont
  {Davies}}, and\ \bibinfo {editor} {\bibfnamefont {C.~L.}\ \bibnamefont
  {{Harper Jr.}}}}\ (\bibinfo  {publisher} {Cambridge University Press, Cambridge},\
  \bibinfo {year} {2004}), pp.\ \bibinfo {pages} {103--120}\BibitemShut
  {NoStop}%
\bibitem [{\citenamefont {Zeh}(2000{\natexlab{b}})}]{Zeh:1999:qr}%
  \BibitemOpen
  \bibfield  {author} {\bibinfo {author} {\bibfnamefont {H.~D.}\ \bibnamefont
  {Zeh}},\ }in\ \href@noop {} {\emph {\bibinfo {booktitle} {Decoherence:
  {T}heoretical, Experimental, and Conceptual Problems}}},\ \bibinfo {series
  and number} {Lecture Notes in Physics {No.\ 538}},\ \bibinfo {editor} {edited
  by\ \bibinfo {editor} {\bibfnamefont {P.}~\bibnamefont {Blanchard}}, \bibinfo
  {editor} {\bibfnamefont {D.}~\bibnamefont {Giulini}}, \bibinfo {editor}
  {\bibfnamefont {E.}~\bibnamefont {Joos}}, \bibinfo {editor} {\bibfnamefont
  {C.}~\bibnamefont {Kiefer}}, and\ \bibinfo {editor} {\bibfnamefont
  {I.}~\bibnamefont {Stamatescu}}}\ (\bibinfo  {publisher} {Springer},\
  \bibinfo {address} {Berlin},\ \bibinfo {year} {2000}), pp.\ \bibinfo {pages}
  {19--42}\BibitemShut {NoStop}%
\bibitem [{\citenamefont {Zeh}(2003)}]{Zeh:2003:mq}%
  \BibitemOpen
  \bibfield  {author} {\bibinfo {author} {\bibfnamefont {H.~D.}\ \bibnamefont
  {Zeh}},\ }in\ \href@noop {} {\emph {\bibinfo {booktitle} {Decoherence and the
  Appearance of a Classical World in Quantum Theory}}},\ \bibinfo {editor}
  {edited by\ \bibinfo {editor} {\bibfnamefont {E.}~\bibnamefont {Joos}},
  \bibinfo {editor} {\bibfnamefont {H.~D.}\ \bibnamefont {Zeh}}, \bibinfo
  {editor} {\bibfnamefont {C.}~\bibnamefont {Kiefer}}, \bibinfo {editor}
  {\bibfnamefont {D.}~\bibnamefont {Giulini}}, \bibinfo {editor} {\bibfnamefont
  {J.}~\bibnamefont {Kupsch}}, and\ \bibinfo {editor} {\bibfnamefont {I.-O.}\
  \bibnamefont {Stamatescu}}}\ (\bibinfo  {publisher} {Springer},\ \bibinfo
  {address} {New York},\ \bibinfo {year} {2003}), \bibinfo {edition} {2nd}\
  ed.,\ Chap.~\bibinfo {chapter} {2}\BibitemShut {NoStop}%
\bibitem [{\citenamefont {Myatt}\ \emph {et~al.}(2000)\citenamefont {Myatt},
  \citenamefont {King}, \citenamefont {Turchette}, \citenamefont {Sackett},
  \citenamefont {Kielpinski}, \citenamefont {Itano}, \citenamefont {Monroe},\
  and\ \citenamefont {Wineland}}]{Myatt:2000:yy}%
  \BibitemOpen
  \bibfield  {author} {\bibinfo {author} {\bibfnamefont {C.~J.}\ \bibnamefont
  {Myatt}}, \bibinfo {author} {\bibfnamefont {B.~E.}\ \bibnamefont {King}},
  \bibinfo {author} {\bibfnamefont {Q.~A.}\ \bibnamefont {Turchette}}, \bibinfo
  {author} {\bibfnamefont {C.~A.}\ \bibnamefont {Sackett}}, \bibinfo {author}
  {\bibfnamefont {D.}~\bibnamefont {Kielpinski}}, \bibinfo {author}
  {\bibfnamefont {W.~M.}\ \bibnamefont {Itano}}, \bibinfo {author}
  {\bibfnamefont {C.}~\bibnamefont {Monroe}}, and\ \bibinfo {author}
  {\bibfnamefont {D.~J.}\ \bibnamefont {Wineland}},\ }\href@noop {} {\bibfield
  {journal} {\bibinfo  {journal} {Nature}\ }\textbf {\bibinfo {volume} {403}},\
  \bibinfo {pages} {269} (\bibinfo {year} {2000})}\BibitemShut {NoStop}%
\bibitem [{\citenamefont {Zurek}(2003{\natexlab{a}})}]{Zurek:2002:ii}%
  \BibitemOpen
  \bibfield  {author} {\bibinfo {author} {\bibfnamefont {W.~H.}\ \bibnamefont
  {Zurek}},\ }\href {\doibase 10.1103/RevModPhys.75.715} {\bibfield  {journal}
  {\bibinfo  {journal} {Rev. Mod. Phys.}\ }\textbf {\bibinfo {volume} {75}},\
  \bibinfo {pages} {715} (\bibinfo {year} {2003}{\natexlab{a}})}\BibitemShut
  {NoStop}%
\bibitem [{\citenamefont {Zeh}(2007)}]{Zeh:2001:tt}%
  \BibitemOpen
  \bibfield  {author} {\bibinfo {author} {\bibfnamefont {H.~D.}\ \bibnamefont
  {Zeh}},\ }\href@noop {} {\emph {\bibinfo {title} {The Physical Basis of the
  Direction of Time}}},\ \bibinfo {edition} {5th}\ ed.\ (\bibinfo  {publisher}
  {Springer},\ \bibinfo {address} {Berlin},\ \bibinfo {year}
  {2007})\BibitemShut {NoStop}%
\bibitem [{\citenamefont {\ifmmode~\dot{Z}\else \.{Z}\fi{}yczkowski}\ \emph
  {et~al.}(2001)\citenamefont {\ifmmode~\dot{Z}\else \.{Z}\fi{}yczkowski},
  \citenamefont {Horodecki}, \citenamefont {Horodecki},\ and\ \citenamefont
  {Horodecki}}]{Zyczkowski:2001:ii}%
  \BibitemOpen
  \bibfield  {author} {\bibinfo {author} {\bibfnamefont {K.}~\bibnamefont
  {\ifmmode~\dot{Z}\else \.{Z}\fi{}yczkowski}}, \bibinfo {author}
  {\bibfnamefont {P.}~\bibnamefont {Horodecki}}, \bibinfo {author}
  {\bibfnamefont {M.}~\bibnamefont {Horodecki}}, and\ \bibinfo {author}
  {\bibfnamefont {R.}~\bibnamefont {Horodecki}},\ }\href {\doibase
  10.1103/PhysRevA.65.012101} {\bibfield  {journal} {\bibinfo  {journal} {Phys.
  Rev. A}\ }\textbf {\bibinfo {volume} {65}},\ \bibinfo {pages} {012101}
  (\bibinfo {year} {2001})}\BibitemShut {NoStop}%
\bibitem [{\citenamefont {Lee}\ \emph {et~al.}(2004)\citenamefont {Lee},
  \citenamefont {Kim}, \citenamefont {Ahn}, \citenamefont {McAneney},\ and\
  \citenamefont {Kim}}]{Lee:2004:uu}%
  \BibitemOpen
  \bibfield  {author} {\bibinfo {author} {\bibfnamefont {J.}~\bibnamefont
  {Lee}}, \bibinfo {author} {\bibfnamefont {I.}~\bibnamefont {Kim}}, \bibinfo
  {author} {\bibfnamefont {D.}~\bibnamefont {Ahn}}, \bibinfo {author}
  {\bibfnamefont {H.}~\bibnamefont {McAneney}}, and\ \bibinfo {author}
  {\bibfnamefont {M.~S.}\ \bibnamefont {Kim}},\ }\href {\doibase
  10.1103/PhysRevA.70.024301} {\bibfield  {journal} {\bibinfo  {journal} {Phys.
  Rev. A}\ }\textbf {\bibinfo {volume} {70}},\ \bibinfo {pages} {024301}
  (\bibinfo {year} {2004})}\BibitemShut {NoStop}%
\bibitem [{\citenamefont {Barreiro}\ \emph {et~al.}(2010)\citenamefont
  {Barreiro}, \citenamefont {Schindler}, \citenamefont {G{\"u}hne},
  \citenamefont {Monz}, \citenamefont {Chwalla}, \citenamefont {Roos},
  \citenamefont {Hennrich},\ and\ \citenamefont {Blatt}}]{Barreiro:2010:aa}%
  \BibitemOpen
  \bibfield  {author} {\bibinfo {author} {\bibfnamefont {J.~T.}\ \bibnamefont
  {Barreiro}}, \bibinfo {author} {\bibfnamefont {P.}~\bibnamefont {Schindler}},
  \bibinfo {author} {\bibfnamefont {O.}~\bibnamefont {G{\"u}hne}}, \bibinfo
  {author} {\bibfnamefont {T.}~\bibnamefont {Monz}}, \bibinfo {author}
  {\bibfnamefont {M.}~\bibnamefont {Chwalla}}, \bibinfo {author} {\bibfnamefont
  {C.~F.}\ \bibnamefont {Roos}}, \bibinfo {author} {\bibfnamefont
  {M.}~\bibnamefont {Hennrich}}, and\ \bibinfo {author} {\bibfnamefont
  {R.}~\bibnamefont {Blatt}},\ }\href@noop {} {\bibfield  {journal} {\bibinfo
  {journal} {Nature Phys.}\ }\textbf {\bibinfo {volume} {6}},\ \bibinfo {pages}
  {943} (\bibinfo {year} {2010})}\BibitemShut {NoStop}%
\bibitem [{\citenamefont {Braun}(2002)}]{Braun:2002:aa}%
  \BibitemOpen
  \bibfield  {author} {\bibinfo {author} {\bibfnamefont {D.}~\bibnamefont
  {Braun}},\ }\href {\doibase 10.1103/PhysRevLett.89.277901} {\bibfield
  {journal} {\bibinfo  {journal} {Phys. Rev. Lett.}\ }\textbf {\bibinfo
  {volume} {89}},\ \bibinfo {pages} {277901} (\bibinfo {year}
  {2002})}\BibitemShut {NoStop}%
\bibitem [{\citenamefont {Benatti}\ \emph {et~al.}(2003)\citenamefont
  {Benatti}, \citenamefont {Floreanini},\ and\ \citenamefont
  {Piani}}]{Benatti:2003:aa}%
  \BibitemOpen
  \bibfield  {author} {\bibinfo {author} {\bibfnamefont {F.}~\bibnamefont
  {Benatti}}, \bibinfo {author} {\bibfnamefont {R.}~\bibnamefont {Floreanini}},
   and\ \bibinfo {author} {\bibfnamefont {M.}~\bibnamefont {Piani}},\ }\href
  {\doibase 10.1103/PhysRevLett.91.070402} {\bibfield  {journal} {\bibinfo
  {journal} {Phys. Rev. Lett.}\ }\textbf {\bibinfo {volume} {91}},\ \bibinfo
  {pages} {070402} (\bibinfo {year} {2003})}\BibitemShut {NoStop}%
\bibitem [{\citenamefont {Kim}\ \emph {et~al.}(2002)\citenamefont {Kim},
  \citenamefont {Lee}, \citenamefont {Ahn},\ and\ \citenamefont
  {Knight}}]{Kim:2002:oo}%
  \BibitemOpen
  \bibfield  {author} {\bibinfo {author} {\bibfnamefont {M.~S.}\ \bibnamefont
  {Kim}}, \bibinfo {author} {\bibfnamefont {J.}~\bibnamefont {Lee}}, \bibinfo
  {author} {\bibfnamefont {D.}~\bibnamefont {Ahn}}, and\ \bibinfo {author}
  {\bibfnamefont {P.~L.}\ \bibnamefont {Knight}},\ }\href {\doibase
  10.1103/PhysRevA.65.040101} {\bibfield  {journal} {\bibinfo  {journal} {Phys.
  Rev. A}\ }\textbf {\bibinfo {volume} {65}},\ \bibinfo {pages} {040101}
  (\bibinfo {year} {2002})}\BibitemShut {NoStop}%
\bibitem [{\citenamefont {Jak{\'o}bczyk}(2002)}]{Jakobczyk:2002:oo}%
  \BibitemOpen
  \bibfield  {author} {\bibinfo {author} {\bibfnamefont {L.}~\bibnamefont
  {Jak{\'o}bczyk}},\ }\href {\doibase 10.1088/0305-4470/35/30/313} {\bibfield
  {journal} {\bibinfo  {journal} {J. Phys. A: Math. Theor.}\ }\textbf {\bibinfo
  {volume} {35}},\ \bibinfo {pages} {6383} (\bibinfo {year}
  {2002})}\BibitemShut {NoStop}%
\bibitem [{\citenamefont {Bacciagaluppi}(2012)}]{Bacciagaluppi:2003:yz}%
  \BibitemOpen
  \bibfield  {author} {\bibinfo {author} {\bibfnamefont {G.}~\bibnamefont
  {Bacciagaluppi}},\ }in\ \href@noop {} {\emph {\bibinfo {booktitle} {The
  Stanford Encyclopedia of Philosophy}}},\ \bibinfo {editor} {edited by\
  \bibinfo {editor} {\bibfnamefont {E.~N.}\ \bibnamefont {Zalta}}}\ (\bibinfo
  {year} {2012}), \bibinfo {note} {online at
  http://plato.stanford.edu/archives/win2012/entries/qm-decoherence}\BibitemShut
  {NoStop}%
\bibitem [{\citenamefont {Schlosshauer}(2004)}]{Schlosshauer:2003:tv}%
  \BibitemOpen
  \bibfield  {author} {\bibinfo {author} {\bibfnamefont {M.}~\bibnamefont
  {Schlosshauer}},\ }\href@noop {} {\bibfield  {journal} {\bibinfo  {journal}
  {Rev. Mod. Phys.}\ }\textbf {\bibinfo {volume} {76}},\ \bibinfo {pages}
  {1267} (\bibinfo {year} {2004})}\BibitemShut {NoStop}%
\bibitem [{\citenamefont {Zurek}(1998)}]{Zurek:1998:re}%
  \BibitemOpen
  \bibfield  {author} {\bibinfo {author} {\bibfnamefont {W.~H.}\ \bibnamefont
  {Zurek}},\ }\href@noop {} {\bibfield  {journal} {\bibinfo  {journal} {Philos.
  Trans. R. Soc. London, Ser. A}\ }\textbf {\bibinfo {volume} {356}},\ \bibinfo
  {pages} {1793} (\bibinfo {year} {1998})}\BibitemShut {NoStop}%
\bibitem [{\citenamefont {Butterfield}(2001)}]{Butterfield:2001:ua}%
  \BibitemOpen
  \bibfield  {author} {\bibinfo {author} {\bibfnamefont {J.~N.}\ \bibnamefont
  {Butterfield}},\ }in\ \href@noop {} {\emph {\bibinfo {booktitle} {Quantum
  Mechanics: Scientific Perspectives on Divine Action}}},\ \bibinfo {editor}
  {edited by\ \bibinfo {editor} {\bibfnamefont {R.~J.}\ \bibnamefont
  {Russell}}, \bibinfo {editor} {\bibfnamefont {P.}~\bibnamefont {Clayton}},
  \bibinfo {editor} {\bibfnamefont {K.}~\bibnamefont {Wegter-McNelly}}, and\
  \bibinfo {editor} {\bibfnamefont {J.}~\bibnamefont {Polkinghorne}}}\
  (\bibinfo  {publisher} {Vatican Observatory and The Center for Theology and
  the Natural Sciences},\ \bibinfo {address} {Vatican City State},\ \bibinfo
  {year} {2001}), pp.\ \bibinfo {pages} {111--140},\ \bibinfo {note} {also
  available as an eprint from the Pittsburgh Philosophy of Science Archive at
  http://philsci-archive.pitt.edu/archive/00000203}\BibitemShut {NoStop}%
\bibitem [{\citenamefont {Wallace}(2002)}]{Wallace:2003:iq}%
  \BibitemOpen
  \bibfield  {author} {\bibinfo {author} {\bibfnamefont {D.}~\bibnamefont
  {Wallace}},\ }\href@noop {} {\bibfield  {journal} {\bibinfo  {journal} {Stud.
  Hist. Philos. Mod. Phys.}\ }\textbf {\bibinfo {volume} {33}},\ \bibinfo
  {pages} {637} (\bibinfo {year} {2002})}\BibitemShut {NoStop}%
\bibitem [{\citenamefont {Wallace}(2003)}]{Wallace:2003:iz}%
  \BibitemOpen
  \bibfield  {author} {\bibinfo {author} {\bibfnamefont {D.}~\bibnamefont
  {Wallace}},\ }\href@noop {} {\bibfield  {journal} {\bibinfo  {journal} {Stud.
  Hist. Philos. Mod. Phys.}\ }\textbf {\bibinfo {volume} {34}},\ \bibinfo
  {pages} {87} (\bibinfo {year} {2003})}\BibitemShut {NoStop}%
\bibitem [{\citenamefont {Wallace}(2010)}]{Wallace:2010:im}%
  \BibitemOpen
  \bibfield  {author} {\bibinfo {author} {\bibfnamefont {D.}~\bibnamefont
  {Wallace}},\ }in\ \href@noop {} {\emph {\bibinfo {booktitle} {Many Worlds?
  {E}verett, Quantum Theory and Reality}}},\ \bibinfo {editor} {edited by\
  \bibinfo {editor} {\bibfnamefont {S.}~\bibnamefont {Saunders}}, \bibinfo
  {editor} {\bibfnamefont {J.}~\bibnamefont {Barrett}}, \bibinfo {editor}
  {\bibfnamefont {A.}~\bibnamefont {Kent}}, and\ \bibinfo {editor}
  {\bibfnamefont {D.}~\bibnamefont {Wallace}}}\ (\bibinfo  {publisher} {Oxford
  University Press},\ \bibinfo {address} {Oxford},\ \bibinfo {year} {2010}),
  pp.\ \bibinfo {pages} {53--72}\BibitemShut {NoStop}%
\bibitem [{\citenamefont {Zurek}(1993)}]{Zurek:1993:pu}%
  \BibitemOpen
  \bibfield  {author} {\bibinfo {author} {\bibfnamefont {W.~H.}\ \bibnamefont
  {Zurek}},\ }\href@noop {} {\bibfield  {journal} {\bibinfo  {journal} {Prog.
  Theor. Phys.}\ }\textbf {\bibinfo {volume} {89}},\ \bibinfo {pages} {281}
  (\bibinfo {year} {1993})}\BibitemShut {NoStop}%
\bibitem [{\citenamefont {Zurek}(2005)}]{Zurek:2004:yb}%
  \BibitemOpen
  \bibfield  {author} {\bibinfo {author} {\bibfnamefont {W.~H.}\ \bibnamefont
  {Zurek}},\ }\href@noop {} {\bibfield  {journal} {\bibinfo  {journal} {Phys.
  Rev. A}\ }\textbf {\bibinfo {volume} {71}},\ \bibinfo {pages} {052105}
  (\bibinfo {year} {2005})}\BibitemShut {NoStop}%
\bibitem [{\citenamefont {Zurek}(2009)}]{Zurek:2009:om}%
  \BibitemOpen
  \bibfield  {author} {\bibinfo {author} {\bibfnamefont {W.~H.}\ \bibnamefont
  {Zurek}},\ }\href@noop {} {\bibfield  {journal} {\bibinfo  {journal} {Nature
  Phys.}\ }\textbf {\bibinfo {volume} {5}},\ \bibinfo {pages} {181} (\bibinfo
  {year} {2009})}\BibitemShut {NoStop}%
\bibitem [{\citenamefont {Zurek}(2014)}]{Zurek:2014:xx}%
  \BibitemOpen
  \bibfield  {author} {\bibinfo {author} {\bibfnamefont {W.~H.}\ \bibnamefont
  {Zurek}},\ }\href@noop {} {\bibfield  {journal} {\bibinfo  {journal} {Phys.
  Today}\ }\textbf {\bibinfo {volume} {67}},\ \bibinfo {pages} {44} (\bibinfo
  {year} {2014})}\BibitemShut {NoStop}%
\bibitem [{\citenamefont {Zwolak}\ \emph {et~al.}(2016)\citenamefont {Zwolak},
  \citenamefont {Riedel},\ and\ \citenamefont {Zurek}}]{Zwolak:2016:zz}%
  \BibitemOpen
  \bibfield  {author} {\bibinfo {author} {\bibfnamefont {M.}~\bibnamefont
  {Zwolak}}, \bibinfo {author} {\bibfnamefont {C.~J.}\ \bibnamefont {Riedel}},
  \ and\ \bibinfo {author} {\bibfnamefont {W.~H.}\ \bibnamefont {Zurek}},\
  }\href@noop {} {\bibfield  {journal} {\bibinfo  {journal} {Sci. Rep.}\
  }\textbf {\bibinfo {volume} {6}},\ \bibinfo {pages} {25277} (\bibinfo {year}
  {2016})}\BibitemShut {NoStop}%
\bibitem [{\citenamefont {Zurek}(2003{\natexlab{b}})}]{Zurek:2003:rv}%
  \BibitemOpen
  \bibfield  {author} {\bibinfo {author} {\bibfnamefont {W.~H.}\ \bibnamefont
  {Zurek}},\ }\href@noop {} {\bibfield  {journal} {\bibinfo  {journal} {Phys.
  Rev. Lett.}\ }\textbf {\bibinfo {volume} {90}},\ \bibinfo {pages} {120404}
  (\bibinfo {year} {2003}{\natexlab{b}})}\BibitemShut {NoStop}%
\bibitem [{\citenamefont {Zurek}(2004)}]{Zurek:2003:pl}%
  \BibitemOpen
  \bibfield  {author} {\bibinfo {author} {\bibfnamefont {W.~H.}\ \bibnamefont
  {Zurek}},\ }in\ \href@noop {} {\emph {\bibinfo {booktitle} {Science and
  Ultimate Reality}}},\ \bibinfo {editor} {edited by\ \bibinfo {editor}
  {\bibfnamefont {J.~D.}\ \bibnamefont {Barrow}}, \bibinfo {editor}
  {\bibfnamefont {P.~C.~W.}\ \bibnamefont {Davies}}, and\ \bibinfo {editor}
  {\bibfnamefont {C.~H.}\ \bibnamefont {Harper}}}\ (\bibinfo  {publisher}
  {Cambridge University Press},\ \bibinfo {address} {Cambridge, England},\
  \bibinfo {year} {2004}), pp.\ \bibinfo {pages} {121--137}\BibitemShut
  {NoStop}%
\bibitem [{\citenamefont {Zurek}(2018{\natexlab{a}})}]{Zurek:2018:on}%
  \BibitemOpen
  \bibfield  {author} {\bibinfo {author} {\bibfnamefont {W.~H.}\ \bibnamefont
  {Zurek}},\ }\href@noop {} {\bibfield  {journal} {\bibinfo  {journal} {Phil.
  Trans. R. Soc. A}\ }\textbf {\bibinfo {volume} {376}},\ \bibinfo {pages}
  {20180107} (\bibinfo {year} {2018}{\natexlab{a}})}\BibitemShut {NoStop}%
\bibitem [{\citenamefont {Clifton}(1996)}]{Clifton:1996:op}%
  \BibitemOpen
  \bibfield  {author} {\bibinfo {author} {\bibfnamefont {R.}~\bibnamefont
  {Clifton}},\ }\href@noop {} {\bibfield  {journal} {\bibinfo  {journal} {Br.
  J. Philos. Sci.}\ }\textbf {\bibinfo {volume} {47}},\ \bibinfo {pages} {371}
  (\bibinfo {year} {1996})}\BibitemShut {NoStop}%
\bibitem [{\citenamefont {Bacciagaluppi}\ and\ \citenamefont
  {Hemmo}(1996)}]{Bacciagaluppi:1996:po}%
  \BibitemOpen
  \bibfield  {author} {\bibinfo {author} {\bibfnamefont {G.}~\bibnamefont
  {Bacciagaluppi}}\ and\ \bibinfo {author} {\bibfnamefont {M.}~\bibnamefont
  {Hemmo}},\ }\href@noop {} {\bibfield  {journal} {\bibinfo  {journal} {Stud.
  Hist. Philos. Mod. Phys.}\ }\textbf {\bibinfo {volume} {27}},\ \bibinfo
  {pages} {239} (\bibinfo {year} {1996})}\BibitemShut {NoStop}%
\bibitem [{\citenamefont {Bene}(2001)}]{Bene:2001:po}%
  \BibitemOpen
  \bibfield  {author} {\bibinfo {author} {\bibfnamefont {G.}~\bibnamefont
  {Bene}},\ }\href@noop {} {\  (\bibinfo {year} {2001})},\ \Eprint
  {http://arxiv.org/abs/quant-ph/0104112} {quant-ph/0104112} \BibitemShut
  {NoStop}%
\bibitem [{\citenamefont {Bacciagaluppi}(2000)}]{Bacciagaluppi:2000:yz}%
  \BibitemOpen
  \bibfield  {author} {\bibinfo {author} {\bibfnamefont {G.}~\bibnamefont
  {Bacciagaluppi}},\ }\href@noop {} {\bibfield  {journal} {\bibinfo  {journal}
  {Found. Phys.}\ }\textbf {\bibinfo {volume} {30}},\ \bibinfo {pages} {1431}
  (\bibinfo {year} {2000})}\BibitemShut {NoStop}%
\bibitem [{\citenamefont {Griffiths}(1984)}]{Griffiths:1984:tr}%
  \BibitemOpen
  \bibfield  {author} {\bibinfo {author} {\bibfnamefont {R.~B.}\ \bibnamefont
  {Griffiths}},\ }\href@noop {} {\bibfield  {journal} {\bibinfo  {journal} {J.
  Stat. Phys.}\ }\textbf {\bibinfo {volume} {36}},\ \bibinfo {pages} {219}
  (\bibinfo {year} {1984})}\BibitemShut {NoStop}%
\bibitem [{\citenamefont {Omn{\`e}s}(1994)}]{Omnes:1994:pz}%
  \BibitemOpen
  \bibfield  {author} {\bibinfo {author} {\bibfnamefont {R.}~\bibnamefont
  {Omn{\`e}s}},\ }\href@noop {} {\emph {\bibinfo {title} {The Interpretation of
  Quantum Mechanics}}}\ (\bibinfo  {publisher} {Princeton University Press},\
  \bibinfo {address} {Princeton},\ \bibinfo {year} {1994})\BibitemShut
  {NoStop}%
\bibitem [{\citenamefont {Griffiths}(2002)}]{Griffiths:2002:tr}%
  \BibitemOpen
  \bibfield  {author} {\bibinfo {author} {\bibfnamefont {R.~B.}\ \bibnamefont
  {Griffiths}},\ }\href@noop {} {\emph {\bibinfo {title} {Consistent Quantum
  Theory}}}\ (\bibinfo  {publisher} {Cambridge University Press},\ \bibinfo
  {address} {Cambridge},\ \bibinfo {year} {2002})\BibitemShut {NoStop}%
\bibitem [{\citenamefont {Paz}\ and\ \citenamefont
  {Zurek}(1993)}]{Paz:1993:ww}%
  \BibitemOpen
  \bibfield  {author} {\bibinfo {author} {\bibfnamefont {J.~P.}\ \bibnamefont
  {Paz}}\ and\ \bibinfo {author} {\bibfnamefont {W.~H.}\ \bibnamefont
  {Zurek}},\ }\href@noop {} {\bibfield  {journal} {\bibinfo  {journal} {Phys.
  Rev. D}\ }\textbf {\bibinfo {volume} {48}},\ \bibinfo {pages} {2728}
  (\bibinfo {year} {1993})}\BibitemShut {NoStop}%
\bibitem [{\citenamefont {Albrecht}(1992)}]{Albrecht:1992:rz}%
  \BibitemOpen
  \bibfield  {author} {\bibinfo {author} {\bibfnamefont {A.}~\bibnamefont
  {Albrecht}},\ }\href@noop {} {\bibfield  {journal} {\bibinfo  {journal}
  {Phys. Rev. D}\ }\textbf {\bibinfo {volume} {46}},\ \bibinfo {pages} {5504}
  (\bibinfo {year} {1992})}\BibitemShut {NoStop}%
\bibitem [{\citenamefont {Albrecht}(1993)}]{Albrecht:1993:pq}%
  \BibitemOpen
  \bibfield  {author} {\bibinfo {author} {\bibfnamefont {A.}~\bibnamefont
  {Albrecht}},\ }\href@noop {} {\bibfield  {journal} {\bibinfo  {journal}
  {Phys. Rev. D}\ }\textbf {\bibinfo {volume} {48}},\ \bibinfo {pages} {3768}
  (\bibinfo {year} {1993})}\BibitemShut {NoStop}%
\bibitem [{\citenamefont {Twamley}(1993)}]{Twamley:1993:bz}%
  \BibitemOpen
  \bibfield  {author} {\bibinfo {author} {\bibfnamefont {J.}~\bibnamefont
  {Twamley}},\ }\href@noop {} {\bibfield  {journal} {\bibinfo  {journal} {Phys.
  Rev. D}\ }\textbf {\bibinfo {volume} {48}},\ \bibinfo {pages} {5730}
  (\bibinfo {year} {1993})}\BibitemShut {NoStop}%
\bibitem [{\citenamefont {Gell-Mann}\ and\ \citenamefont
  {Hartle}(1998)}]{GellMann:1998:xy}%
  \BibitemOpen
  \bibfield  {author} {\bibinfo {author} {\bibfnamefont {M.}~\bibnamefont
  {Gell-Mann}}\ and\ \bibinfo {author} {\bibfnamefont {J.~B.}\ \bibnamefont
  {Hartle}},\ }in\ \href@noop {} {\emph {\bibinfo {booktitle} {Quantum
  Classical Correspondence: The 4th Drexel Symposium on Quantum
  Nonintegrability}}},\ \bibinfo {editor} {edited by\ \bibinfo {editor}
  {\bibfnamefont {D.~H.}\ \bibnamefont {Feng}}\ and\ \bibinfo {editor}
  {\bibfnamefont {B.~L.}\ \bibnamefont {Hu}}}\ (\bibinfo  {publisher}
  {International Press},\ \bibinfo {address} {Cambridge, Massachussetts},\
  \bibinfo {year} {1998}), pp.\ \bibinfo {pages} {3--35}\BibitemShut {NoStop}%
\bibitem [{\citenamefont {Riedel}\ \emph {et~al.}(2016)\citenamefont {Riedel},
  \citenamefont {Zurek},\ and\ \citenamefont {Zwolak}}]{Riedel:2016:oo}%
  \BibitemOpen
  \bibfield  {author} {\bibinfo {author} {\bibfnamefont {C.~J.}\ \bibnamefont
  {Riedel}}, \bibinfo {author} {\bibfnamefont {W.~H.}\ \bibnamefont {Zurek}}, 
  and\ \bibinfo {author} {\bibfnamefont {M.}~\bibnamefont {Zwolak}},\
  }\href@noop {} {\bibfield  {journal} {\bibinfo  {journal} {Phys. Rev. A}\
  }\textbf {\bibinfo {volume} {93}},\ \bibinfo {pages} {032126} (\bibinfo
  {year} {2016})}\BibitemShut {NoStop}%
\bibitem [{\citenamefont {Ollivier}\ \emph {et~al.}(2004)\citenamefont
  {Ollivier}, \citenamefont {Poulin},\ and\ \citenamefont
  {Zurek}}]{Ollivier:2003:za}%
  \BibitemOpen
  \bibfield  {author} {\bibinfo {author} {\bibfnamefont {H.}~\bibnamefont
  {Ollivier}}, \bibinfo {author} {\bibfnamefont {D.}~\bibnamefont {Poulin}}, 
  and\ \bibinfo {author} {\bibfnamefont {W.~H.}\ \bibnamefont {Zurek}},\
  }\href@noop {} {\bibfield  {journal} {\bibinfo  {journal} {Phys. Rev. Lett.}\
  }\textbf {\bibinfo {volume} {93}},\ \bibinfo {pages} {220401} (\bibinfo
  {year} {2004})}\BibitemShut {NoStop}%
\bibitem [{\citenamefont {Ollivier}\ \emph {et~al.}(2005)\citenamefont
  {Ollivier}, \citenamefont {Poulin},\ and\ \citenamefont
  {Zurek}}]{Ollivier:2004:im}%
  \BibitemOpen
  \bibfield  {author} {\bibinfo {author} {\bibfnamefont {H.}~\bibnamefont
  {Ollivier}}, \bibinfo {author} {\bibfnamefont {D.}~\bibnamefont {Poulin}}, 
  and\ \bibinfo {author} {\bibfnamefont {W.~H.}\ \bibnamefont {Zurek}},\
  }\href@noop {} {\bibfield  {journal} {\bibinfo  {journal} {Phys. Rev. A}\
  }\textbf {\bibinfo {volume} {72}},\ \bibinfo {pages} {042113} (\bibinfo
  {year} {2005})}\BibitemShut {NoStop}%
\bibitem [{\citenamefont {Blume-Kohout}\ and\ \citenamefont
  {Zurek}(2005)}]{Blume:2004:oo}%
  \BibitemOpen
  \bibfield  {author} {\bibinfo {author} {\bibfnamefont {R.}~\bibnamefont
  {Blume-Kohout}}\ and\ \bibinfo {author} {\bibfnamefont {W.~H.}\ \bibnamefont
  {Zurek}},\ }\href@noop {} {\bibfield  {journal} {\bibinfo  {journal} {Found.
  Phys.}\ }\textbf {\bibinfo {volume} {35}},\ \bibinfo {pages} {1857} (\bibinfo
  {year} {2005})}\BibitemShut {NoStop}%
\bibitem [{\citenamefont {Blume-Kohout}\ and\ \citenamefont
  {Zurek}(2006)}]{Blume:2005:oo}%
  \BibitemOpen
  \bibfield  {author} {\bibinfo {author} {\bibfnamefont {R.}~\bibnamefont
  {Blume-Kohout}}\ and\ \bibinfo {author} {\bibfnamefont {W.~H.}\ \bibnamefont
  {Zurek}},\ }\href@noop {} {\bibfield  {journal} {\bibinfo  {journal} {Phys.
  Rev. A}\ }\textbf {\bibinfo {volume} {73}},\ \bibinfo {pages} {062310}
  (\bibinfo {year} {2006})}\BibitemShut {NoStop}%
\bibitem [{\citenamefont {Riedel}\ and\ \citenamefont
  {Zurek}(2010)}]{Riedel:2010:un}%
  \BibitemOpen
  \bibfield  {author} {\bibinfo {author} {\bibfnamefont {C.~J.}\ \bibnamefont
  {Riedel}}\ and\ \bibinfo {author} {\bibfnamefont {W.~H.}\ \bibnamefont
  {Zurek}},\ }\href@noop {} {\bibfield  {journal} {\bibinfo  {journal} {Phys.
  Rev. Lett.}\ }\textbf {\bibinfo {volume} {105}},\ \bibinfo {pages} {020404}
  (\bibinfo {year} {2010})}\BibitemShut {NoStop}%
\bibitem [{\citenamefont {Riedel}\ and\ \citenamefont
  {Zurek}(2011)}]{Riedel:2011:un}%
  \BibitemOpen
  \bibfield  {author} {\bibinfo {author} {\bibfnamefont {C.~J.}\ \bibnamefont
  {Riedel}}\ and\ \bibinfo {author} {\bibfnamefont {W.~H.}\ \bibnamefont
  {Zurek}},\ }\href@noop {} {\bibfield  {journal} {\bibinfo  {journal} {New J.
  Phys.}\ }\textbf {\bibinfo {volume} {13}},\ \bibinfo {pages} {073038}
  (\bibinfo {year} {2011})}\BibitemShut {NoStop}%
\bibitem [{\citenamefont {Riedel}\ \emph {et~al.}(2012)\citenamefont {Riedel},
  \citenamefont {Zurek},\ and\ \citenamefont {Zwolak}}]{Riedel:2012:un}%
  \BibitemOpen
  \bibfield  {author} {\bibinfo {author} {\bibfnamefont {C.~J.}\ \bibnamefont
  {Riedel}}, \bibinfo {author} {\bibfnamefont {W.~H.}\ \bibnamefont {Zurek}}, 
  and\ \bibinfo {author} {\bibfnamefont {M.}~\bibnamefont {Zwolak}},\
  }\href@noop {} {\bibfield  {journal} {\bibinfo  {journal} {New J. Phys.}\
  }\textbf {\bibinfo {volume} {14}},\ \bibinfo {pages} {083010} (\bibinfo
  {year} {2012})}\BibitemShut {NoStop}%
\bibitem [{\citenamefont {Streltsov}\ and\ \citenamefont
  {Zurek}(2013)}]{Streltsov:2013:oo}%
  \BibitemOpen
  \bibfield  {author} {\bibinfo {author} {\bibfnamefont {A.}~\bibnamefont
  {Streltsov}}\ and\ \bibinfo {author} {\bibfnamefont {W.~H.}\ \bibnamefont
  {Zurek}},\ }\href@noop {} {\bibfield  {journal} {\bibinfo  {journal} {Phys.
  Rev. Lett.}\ }\textbf {\bibinfo {volume} {111}},\ \bibinfo {pages} {040401}
  (\bibinfo {year} {2013})}\BibitemShut {NoStop}%
\bibitem [{\citenamefont {Zurek}(2013)}]{Zurek:2013:xx}%
  \BibitemOpen
  \bibfield  {author} {\bibinfo {author} {\bibfnamefont {W.~H.}\ \bibnamefont
  {Zurek}},\ }\href@noop {} {\bibfield  {journal} {\bibinfo  {journal} {Phys.
  Rev. A}\ }\textbf {\bibinfo {volume} {87}},\ \bibinfo {pages} {052111}
  (\bibinfo {year} {2013})}\BibitemShut {NoStop}%
\bibitem [{\citenamefont {Zurek}(2018{\natexlab{b}})}]{Zurek:2018:om}%
  \BibitemOpen
  \bibfield  {author} {\bibinfo {author} {\bibfnamefont {W.~H.}\ \bibnamefont
  {Zurek}},\ }\href@noop {} {\bibfield  {journal} {\bibinfo  {journal} {Phil.
  Trans. R. Soc. A}\ }\textbf {\bibinfo {volume} {376}},\ \bibinfo {pages}
  {20170315} (\bibinfo {year} {2018}{\natexlab{b}})}\BibitemShut {NoStop}%
\bibitem [{\citenamefont {Fuchs}\ \emph {et~al.}(2014)\citenamefont {Fuchs},
  \citenamefont {Mermin},\ and\ \citenamefont {Schack}}]{Fuchs:2014:pp}%
  \BibitemOpen
  \bibfield  {author} {\bibinfo {author} {\bibfnamefont {C.~A.}\ \bibnamefont
  {Fuchs}}, \bibinfo {author} {\bibfnamefont {N.~D.}\ \bibnamefont {Mermin}}, \
  and\ \bibinfo {author} {\bibfnamefont {R.}~\bibnamefont {Schack}},\
  }\href@noop {} {\bibfield  {journal} {\bibinfo  {journal} {Am. J. Phys.}\
  }\textbf {\bibinfo {volume} {82}},\ \bibinfo {pages} {749} (\bibinfo {year}
  {2014})}\BibitemShut {NoStop}%
\bibitem [{\citenamefont {Paz}\ and\ \citenamefont
  {Zurek}(2001)}]{Paz:2001:aa}%
  \BibitemOpen
  \bibfield  {author} {\bibinfo {author} {\bibfnamefont {J.~P.}\ \bibnamefont
  {Paz}}\ and\ \bibinfo {author} {\bibfnamefont {W.~H.}\ \bibnamefont
  {Zurek}},\ }in\ \href@noop {} {\emph {\bibinfo {booktitle} {Coherent Atomic
  Matter Waves, Les Houches Session LXXII}}},\ \bibinfo {series} {Les Houches
  Summer School Series}, Vol.~\bibinfo {volume} {72},\ \bibinfo {editor}
  {edited by\ \bibinfo {editor} {\bibfnamefont {R.}~\bibnamefont {Kaiser}},
  \bibinfo {editor} {\bibfnamefont {C.}~\bibnamefont {Westbrook}}, and\
  \bibinfo {editor} {\bibfnamefont {F.}~\bibnamefont {David}}}\ (\bibinfo
  {publisher} {Springer},\ \bibinfo {address} {Berlin},\ \bibinfo {year}
  {2001}), pp.\ \bibinfo {pages} {533--614}\BibitemShut {NoStop}%
\bibitem [{\citenamefont {Breuer}\ and\ \citenamefont
  {Petruccione}(2002)}]{Breuer:2002:oq}%
  \BibitemOpen
  \bibfield  {author} {\bibinfo {author} {\bibfnamefont {H.-P.}\ \bibnamefont
  {Breuer}}\ and\ \bibinfo {author} {\bibfnamefont {F.}~\bibnamefont
  {Petruccione}},\ }\href@noop {} {\emph {\bibinfo {title} {The Theory of Open
  Quantum Systems}}}\ (\bibinfo  {publisher} {Oxford University Press},\
  \bibinfo {address} {Oxford},\ \bibinfo {year} {2002})\BibitemShut {NoStop}%
\bibitem [{\citenamefont {Joos}\ \emph {et~al.}(2003)\citenamefont {Joos},
  \citenamefont {Zeh}, \citenamefont {Kiefer}, \citenamefont {Giulini},
  \citenamefont {Kupsch},\ and\ \citenamefont {Stamatescu}}]{Joos:2003:jh}%
  \BibitemOpen
  \bibfield  {author} {\bibinfo {author} {\bibfnamefont {E.}~\bibnamefont
  {Joos}}, \bibinfo {author} {\bibfnamefont {H.~D.}\ \bibnamefont {Zeh}},
  \bibinfo {author} {\bibfnamefont {C.}~\bibnamefont {Kiefer}}, \bibinfo
  {author} {\bibfnamefont {D.}~\bibnamefont {Giulini}}, \bibinfo {author}
  {\bibfnamefont {J.}~\bibnamefont {Kupsch}}, and\ \bibinfo {author}
  {\bibfnamefont {I.-O.}\ \bibnamefont {Stamatescu}},\ }\href@noop {} {\emph
  {\bibinfo {title} {Decoherence and the Appearance of a Classical World in
  Quantum Theory}}},\ \bibinfo {edition} {2nd}\ ed.\ (\bibinfo  {publisher}
  {Springer},\ \bibinfo {address} {New York},\ \bibinfo {year}
  {2003})\BibitemShut {NoStop}%
\bibitem [{\citenamefont {Schlosshauer}(2007)}]{Schlosshauer:2007:un}%
  \BibitemOpen
  \bibfield  {author} {\bibinfo {author} {\bibfnamefont {M.}~\bibnamefont
  {Schlosshauer}},\ }\href@noop {} {\emph {\bibinfo {title} {Decoherence and
  the Quantum-to-Classical Transition}}}\ (\bibinfo  {publisher} {Springer},\
  \bibinfo {address} {Berlin/Heidelberg},\ \bibinfo {year} {2007})\BibitemShut
  {NoStop}%
\bibitem [{\citenamefont {Hornberger}(2009)}]{Hornberger:2009:aq}%
  \BibitemOpen
  \bibfield  {author} {\bibinfo {author} {\bibfnamefont {K.}~\bibnamefont
  {Hornberger}},\ }in\ \href@noop {} {\emph {\bibinfo {booktitle} {Entanglement
  and Decoherence: Foundations and Modern Trends}}},\ \bibinfo {series}
  {Lecture Notes in Physics}, Vol.\ \bibinfo {volume} {768},\ \bibinfo {editor}
  {edited by\ \bibinfo {editor} {\bibfnamefont {A.}~\bibnamefont
  {Buchleitner}}, \bibinfo {editor} {\bibfnamefont {C.}~\bibnamefont
  {Viviescas}}, and\ \bibinfo {editor} {\bibfnamefont {M.}~\bibnamefont
  {Tiersch}}}\ (\bibinfo  {publisher} {Springer},\ \bibinfo {address}
  {Berlin},\ \bibinfo {year} {2009}), pp.\ \bibinfo {pages}
  {221--276}\BibitemShut {NoStop}%
\bibitem [{\citenamefont {Schlosshauer}(2019)}]{Schlosshauer:2019:qd}%
  \BibitemOpen
  \bibfield  {author} {\bibinfo {author} {\bibfnamefont {M.}~\bibnamefont
  {Schlosshauer}},\ }\href {\doibase 10.1016/j.physrep.2019.10.001} {\bibfield
  {journal} {\bibinfo  {journal} {Phys. Rep.}\ }\textbf {\bibinfo {volume}
  {831}},\ \bibinfo {pages} {1} (\bibinfo {year} {2019})}\BibitemShut {NoStop}%
\bibitem [{\citenamefont {Lidar}\ and\ \citenamefont
  {Whaley}(2003)}]{Lidar:2003:aa}%
  \BibitemOpen
  \bibfield  {author} {\bibinfo {author} {\bibfnamefont {D.~A.}\ \bibnamefont
  {Lidar}}\ and\ \bibinfo {author} {\bibfnamefont {K.~B.}\ \bibnamefont
  {Whaley}},\ }in\ \href@noop {} {\emph {\bibinfo {booktitle} {Irreversible
  Quantum Dynamics}}},\ \bibinfo {series} {Springer Lecture Notes in Physics},
  Vol.\ \bibinfo {volume} {622},\ \bibinfo {editor} {edited by\ \bibinfo
  {editor} {\bibfnamefont {F.}~\bibnamefont {Benatti}}\ and\ \bibinfo {editor}
  {\bibfnamefont {R.}~\bibnamefont {Floreanini}}}\ (\bibinfo  {publisher}
  {Springer},\ \bibinfo {address} {Berlin},\ \bibinfo {year} {2003}), pp.\
  \bibinfo {pages} {83--120},\ \bibinfo {note} {also available as eprint
  quant-ph/0301032}\BibitemShut {NoStop}%
\bibitem [{\citenamefont {Lidar}(2014)}]{Lidar:2014:pp}%
  \BibitemOpen
  \bibfield  {author} {\bibinfo {author} {\bibfnamefont {D.~A.}\ \bibnamefont
  {Lidar}},\ }\href@noop {} {\bibfield  {journal} {\bibinfo  {journal} {Adv.
  Chem. Phys.}\ }\textbf {\bibinfo {volume} {154}},\ \bibinfo {pages} {295}
  (\bibinfo {year} {2014})}\BibitemShut {NoStop}%
\bibitem [{\citenamefont {Dalvit}\ \emph {et~al.}(2000)\citenamefont {Dalvit},
  \citenamefont {Dziarmaga},\ and\ \citenamefont {Zurek}}]{Dalvit:2000:bb}%
  \BibitemOpen
  \bibfield  {author} {\bibinfo {author} {\bibfnamefont {D.~A.~R.}\
  \bibnamefont {Dalvit}}, \bibinfo {author} {\bibfnamefont {J.}~\bibnamefont
  {Dziarmaga}}, and\ \bibinfo {author} {\bibfnamefont {W.~H.}\ \bibnamefont
  {Zurek}},\ }\href@noop {} {\bibfield  {journal} {\bibinfo  {journal} {Phys.
  Rev. A}\ }\textbf {\bibinfo {volume} {62}},\ \bibinfo {pages} {013607}
  (\bibinfo {year} {2000})}\BibitemShut {NoStop}%
\bibitem [{\citenamefont {Poyatos}\ \emph {et~al.}(1996)\citenamefont
  {Poyatos}, \citenamefont {Cirac},\ and\ \citenamefont
  {Zoller}}]{Poyatos:1996:um}%
  \BibitemOpen
  \bibfield  {author} {\bibinfo {author} {\bibfnamefont {J.~F.}\ \bibnamefont
  {Poyatos}}, \bibinfo {author} {\bibfnamefont {J.~I.}\ \bibnamefont {Cirac}},
   and\ \bibinfo {author} {\bibfnamefont {P.}~\bibnamefont {Zoller}},\
  }\href@noop {} {\bibfield  {journal} {\bibinfo  {journal} {Phys. Rev. Lett.}\
  }\textbf {\bibinfo {volume} {77}},\ \bibinfo {pages} {4728} (\bibinfo {year}
  {1996})}\BibitemShut {NoStop}%
\bibitem [{\citenamefont {Turchette}\ \emph {et~al.}(2000)\citenamefont
  {Turchette}, \citenamefont {Myatt}, \citenamefont {King}, \citenamefont
  {Sackett}, \citenamefont {Kielpinski}, \citenamefont {Itano}, \citenamefont
  {Monroe},\ and\ \citenamefont {Wineland}}]{Turchette:2000:aa}%
  \BibitemOpen
  \bibfield  {author} {\bibinfo {author} {\bibfnamefont {Q.~A.}\ \bibnamefont
  {Turchette}}, \bibinfo {author} {\bibfnamefont {C.~J.}\ \bibnamefont
  {Myatt}}, \bibinfo {author} {\bibfnamefont {B.~E.}\ \bibnamefont {King}},
  \bibinfo {author} {\bibfnamefont {C.~A.}\ \bibnamefont {Sackett}}, \bibinfo
  {author} {\bibfnamefont {D.}~\bibnamefont {Kielpinski}}, \bibinfo {author}
  {\bibfnamefont {W.~M.}\ \bibnamefont {Itano}}, \bibinfo {author}
  {\bibfnamefont {C.}~\bibnamefont {Monroe}}, and\ \bibinfo {author}
  {\bibfnamefont {D.~J.}\ \bibnamefont {Wineland}},\ }\href@noop {} {\bibfield
  {journal} {\bibinfo  {journal} {Phys. Rev. A}\ }\textbf {\bibinfo {volume}
  {62}},\ \bibinfo {pages} {053807} (\bibinfo {year} {2000})}\BibitemShut
  {NoStop}%
\bibitem [{\citenamefont {Carvalho}\ \emph {et~al.}(2001)\citenamefont
  {Carvalho}, \citenamefont {Milman}, \citenamefont {de~Matos~Filho},\ and\
  \citenamefont {Davidovich}}]{Carvalho:2001:ua}%
  \BibitemOpen
  \bibfield  {author} {\bibinfo {author} {\bibfnamefont {A.~R.~R.}\
  \bibnamefont {Carvalho}}, \bibinfo {author} {\bibfnamefont {P.}~\bibnamefont
  {Milman}}, \bibinfo {author} {\bibfnamefont {R.~L.}\ \bibnamefont
  {de~Matos~Filho}}, and\ \bibinfo {author} {\bibfnamefont {L.}~\bibnamefont
  {Davidovich}},\ }\href@noop {} {\bibfield  {journal} {\bibinfo  {journal}
  {Phys. Rev. Lett.}\ }\textbf {\bibinfo {volume} {86}},\ \bibinfo {pages}
  {4988} (\bibinfo {year} {2001})}\BibitemShut {NoStop}%
\bibitem [{\citenamefont {Viola}\ and\ \citenamefont
  {Lloyd}(1998)}]{Viola:1998:uu}%
  \BibitemOpen
  \bibfield  {author} {\bibinfo {author} {\bibfnamefont {L.}~\bibnamefont
  {Viola}}\ and\ \bibinfo {author} {\bibfnamefont {S.}~\bibnamefont {Lloyd}},\
  }\href@noop {} {\bibfield  {journal} {\bibinfo  {journal} {Phys. Rev. A}\
  }\textbf {\bibinfo {volume} {58}},\ \bibinfo {pages} {2733} (\bibinfo {year}
  {1998})}\BibitemShut {NoStop}%
\bibitem [{\citenamefont {Viola}\ \emph {et~al.}(1999)\citenamefont {Viola},
  \citenamefont {Knill},\ and\ \citenamefont {Lloyd}}]{Viola:1999:zp}%
  \BibitemOpen
  \bibfield  {author} {\bibinfo {author} {\bibfnamefont {L.}~\bibnamefont
  {Viola}}, \bibinfo {author} {\bibfnamefont {E.}~\bibnamefont {Knill}}, and\
  \bibinfo {author} {\bibfnamefont {S.}~\bibnamefont {Lloyd}},\ }\href@noop {}
  {\bibfield  {journal} {\bibinfo  {journal} {Phys. Rev. Lett.}\ }\textbf
  {\bibinfo {volume} {82}},\ \bibinfo {pages} {2417} (\bibinfo {year}
  {1999})}\BibitemShut {NoStop}%
\bibitem [{\citenamefont {Zanardi}(1999)}]{Zanardi:1999:oo}%
  \BibitemOpen
  \bibfield  {author} {\bibinfo {author} {\bibfnamefont {P.}~\bibnamefont
  {Zanardi}},\ }\href@noop {} {\bibfield  {journal} {\bibinfo  {journal} {Phys.
  Lett. A}\ }\textbf {\bibinfo {volume} {258}},\ \bibinfo {pages} {77}
  (\bibinfo {year} {1999})}\BibitemShut {NoStop}%
\bibitem [{\citenamefont {Viola}\ \emph {et~al.}(2000)\citenamefont {Viola},
  \citenamefont {Knill},\ and\ \citenamefont {Lloyd}}]{Viola:2000:pp}%
  \BibitemOpen
  \bibfield  {author} {\bibinfo {author} {\bibfnamefont {L.}~\bibnamefont
  {Viola}}, \bibinfo {author} {\bibfnamefont {E.}~\bibnamefont {Knill}}, and\
  \bibinfo {author} {\bibfnamefont {S.}~\bibnamefont {Lloyd}},\ }\href@noop {}
  {\bibfield  {journal} {\bibinfo  {journal} {Phys. Rev. Lett.}\ }\textbf
  {\bibinfo {volume} {85}},\ \bibinfo {pages} {3520} (\bibinfo {year}
  {2000})}\BibitemShut {NoStop}%
\bibitem [{\citenamefont {Wu}\ and\ \citenamefont {Lidar}(2002)}]{Wu:2002:aa}%
  \BibitemOpen
  \bibfield  {author} {\bibinfo {author} {\bibfnamefont {L.-A.}\ \bibnamefont
  {Wu}}\ and\ \bibinfo {author} {\bibfnamefont {D.~A.}\ \bibnamefont {Lidar}},\
  }\href@noop {} {\bibfield  {journal} {\bibinfo  {journal} {Phys. Rev. Lett.}\
  }\textbf {\bibinfo {volume} {88}},\ \bibinfo {pages} {207902} (\bibinfo
  {year} {2002})}\BibitemShut {NoStop}%
\bibitem [{\citenamefont {Wu}\ \emph {et~al.}(2002)\citenamefont {Wu},
  \citenamefont {Byrd},\ and\ \citenamefont {Lidar}}]{Wu:2002:bb}%
  \BibitemOpen
  \bibfield  {author} {\bibinfo {author} {\bibfnamefont {L.-A.}\ \bibnamefont
  {Wu}}, \bibinfo {author} {\bibfnamefont {M.~S.}\ \bibnamefont {Byrd}}, and\
  \bibinfo {author} {\bibfnamefont {D.~A.}\ \bibnamefont {Lidar}},\ }\href@noop
  {} {\bibfield  {journal} {\bibinfo  {journal} {Phys. Rev. Lett.}\ }\textbf
  {\bibinfo {volume} {89}},\ \bibinfo {pages} {127901} (\bibinfo {year}
  {2002})}\BibitemShut {NoStop}%
\bibitem [{\citenamefont {Steane}(1996)}]{Steane:1996:cd}%
  \BibitemOpen
  \bibfield  {author} {\bibinfo {author} {\bibfnamefont {A.~M.}\ \bibnamefont
  {Steane}},\ }\href@noop {} {\bibfield  {journal} {\bibinfo  {journal} {Phys.
  Rev. Lett.}\ }\textbf {\bibinfo {volume} {77}},\ \bibinfo {pages} {793}
  (\bibinfo {year} {1996})}\BibitemShut {NoStop}%
\bibitem [{\citenamefont {Shor}(1995)}]{Shor:1995:rx}%
  \BibitemOpen
  \bibfield  {author} {\bibinfo {author} {\bibfnamefont {P.~W.}\ \bibnamefont
  {Shor}},\ }\href@noop {} {\bibfield  {journal} {\bibinfo  {journal} {Phys.
  Rev. A}\ }\textbf {\bibinfo {volume} {52}},\ \bibinfo {pages} {R2493}
  (\bibinfo {year} {1995})}\BibitemShut {NoStop}%
\bibitem [{\citenamefont {Steane}(2001)}]{Steane:2001:dx}%
  \BibitemOpen
  \bibfield  {author} {\bibinfo {author} {\bibfnamefont {A.~M.}\ \bibnamefont
  {Steane}},\ }in\ \href@noop {} {\emph {\bibinfo {booktitle} {Decoherence and
  Its Implications in Quantum Computation and Information Transfer}}},\
  \bibinfo {editor} {edited by\ \bibinfo {editor} {\bibfnamefont
  {P.}~\bibnamefont {Turchi}}\ and\ \bibinfo {editor} {\bibfnamefont
  {A.}~\bibnamefont {Gonis}}}\ (\bibinfo  {publisher} {IOS Press},\ \bibinfo
  {address} {Amsterdam},\ \bibinfo {year} {2001}), pp.\ \bibinfo {pages}
  {284--298},\ \bibinfo {note} {also available as eprint
  quant-ph/0304016}\BibitemShut {NoStop}%
\bibitem [{\citenamefont {Gaitan}(2008)}]{Gaitan:2008:uu}%
  \BibitemOpen
  \bibfield  {author} {\bibinfo {author} {\bibfnamefont {F.}~\bibnamefont
  {Gaitan}},\ }\href@noop {} {\emph {\bibinfo {title} {Quantum Error Correction
  and Fault Tolerant Quantum Computing}}}\ (\bibinfo  {publisher} {CRC Press},\
  \bibinfo {address} {Boca Raton},\ \bibinfo {year} {2008})\BibitemShut
  {NoStop}%
\bibitem [{\citenamefont {Lidar}\ and\ \citenamefont
  {Brun}(2013)}]{Lidar:2013:pp}%
  \BibitemOpen
  \bibinfo {editor} {\bibfnamefont {D.~A.}\ \bibnamefont {Lidar}}\ and\
  \bibinfo {editor} {\bibfnamefont {T.~A.}\ \bibnamefont {Brun}},\ eds.,\
  \href@noop {} {\emph {\bibinfo {title} {Quantum Error Correction}}}\
  (\bibinfo  {publisher} {Cambridge University Press, Cambridge},\ \bibinfo {year}
  {2013})\BibitemShut {NoStop}%
\bibitem [{\citenamefont {Kwiat}\ \emph {et~al.}(2000)\citenamefont {Kwiat},
  \citenamefont {Berglund}, \citenamefont {Altepeter},\ and\ \citenamefont
  {White}}]{Kwiat:2000:kv}%
  \BibitemOpen
  \bibfield  {author} {\bibinfo {author} {\bibfnamefont {P.~G.}\ \bibnamefont
  {Kwiat}}, \bibinfo {author} {\bibfnamefont {A.~J.}\ \bibnamefont {Berglund}},
  \bibinfo {author} {\bibfnamefont {J.~B.}\ \bibnamefont {Altepeter}}, and\
  \bibinfo {author} {\bibfnamefont {A.~G.}\ \bibnamefont {White}},\ }\href@noop
  {} {\bibfield  {journal} {\bibinfo  {journal} {Science}\ }\textbf {\bibinfo
  {volume} {290}},\ \bibinfo {pages} {498} (\bibinfo {year}
  {2000})}\BibitemShut {NoStop}%
\bibitem [{\citenamefont {Altepeter}\ \emph {et~al.}(2004)\citenamefont
  {Altepeter}, \citenamefont {Hadley}, \citenamefont {Wendelken}, \citenamefont
  {Berglund},\ and\ \citenamefont {Kwiat}}]{Altepeter:2004:ll}%
  \BibitemOpen
  \bibfield  {author} {\bibinfo {author} {\bibfnamefont {J.~B.}\ \bibnamefont
  {Altepeter}}, \bibinfo {author} {\bibfnamefont {P.~G.}\ \bibnamefont
  {Hadley}}, \bibinfo {author} {\bibfnamefont {S.~M.}\ \bibnamefont
  {Wendelken}}, \bibinfo {author} {\bibfnamefont {A.~J.}\ \bibnamefont
  {Berglund}}, and\ \bibinfo {author} {\bibfnamefont {P.~G.}\ \bibnamefont
  {Kwiat}},\ }\href@noop {} {\bibfield  {journal} {\bibinfo  {journal} {Phys.
  Rev. Lett.}\ }\textbf {\bibinfo {volume} {92}},\ \bibinfo {pages} {147901}
  (\bibinfo {year} {2004})}\BibitemShut {NoStop}%
\bibitem [{\citenamefont {Kielpinski}\ \emph {et~al.}(2001)\citenamefont
  {Kielpinski}, \citenamefont {Meyer}, \citenamefont {Rowe}, \citenamefont
  {Sackett}, \citenamefont {Itano}, \citenamefont {Monroe},\ and\ \citenamefont
  {Wineland}}]{Kielpinski:2001:uu}%
  \BibitemOpen
  \bibfield  {author} {\bibinfo {author} {\bibfnamefont {D.}~\bibnamefont
  {Kielpinski}}, \bibinfo {author} {\bibfnamefont {V.}~\bibnamefont {Meyer}},
  \bibinfo {author} {\bibfnamefont {M.~A.}\ \bibnamefont {Rowe}}, \bibinfo
  {author} {\bibfnamefont {C.~A.}\ \bibnamefont {Sackett}}, \bibinfo {author}
  {\bibfnamefont {W.~M.}\ \bibnamefont {Itano}}, \bibinfo {author}
  {\bibfnamefont {C.}~\bibnamefont {Monroe}}, and\ \bibinfo {author}
  {\bibfnamefont {D.~J.}\ \bibnamefont {Wineland}},\ }\href@noop {} {\bibfield
  {journal} {\bibinfo  {journal} {Science}\ }\textbf {\bibinfo {volume}
  {291}},\ \bibinfo {pages} {1013} (\bibinfo {year} {2001})}\BibitemShut
  {NoStop}%
\bibitem [{\citenamefont {Roos}\ \emph {et~al.}(2004)\citenamefont {Roos},
  \citenamefont {Lancaster}, \citenamefont {Riebe}, \citenamefont {H\"affner},
  \citenamefont {H\"ansel}, \citenamefont {Gulde}, \citenamefont {Becher},
  \citenamefont {Eschner}, \citenamefont {Schmidt-Kaler},\ and\ \citenamefont
  {Blatt}}]{Roos:204:pp}%
  \BibitemOpen
  \bibfield  {author} {\bibinfo {author} {\bibfnamefont {C.~F.}\ \bibnamefont
  {Roos}}, \bibinfo {author} {\bibfnamefont {G.~P.~T.}\ \bibnamefont
  {Lancaster}}, \bibinfo {author} {\bibfnamefont {M.}~\bibnamefont {Riebe}},
  \bibinfo {author} {\bibfnamefont {H.}~\bibnamefont {H\"affner}}, \bibinfo
  {author} {\bibfnamefont {W.}~\bibnamefont {H\"ansel}}, \bibinfo {author}
  {\bibfnamefont {S.}~\bibnamefont {Gulde}}, \bibinfo {author} {\bibfnamefont
  {C.}~\bibnamefont {Becher}}, \bibinfo {author} {\bibfnamefont
  {J.}~\bibnamefont {Eschner}}, \bibinfo {author} {\bibfnamefont
  {F.}~\bibnamefont {Schmidt-Kaler}}, and\ \bibinfo {author} {\bibfnamefont
  {R.}~\bibnamefont {Blatt}},\ }\href {\doibase 10.1103/PhysRevLett.92.220402}
  {\bibfield  {journal} {\bibinfo  {journal} {Phys. Rev. Lett.}\ }\textbf
  {\bibinfo {volume} {92}},\ \bibinfo {pages} {220402} (\bibinfo {year}
  {2004})}\BibitemShut {NoStop}%
\bibitem [{\citenamefont {H{\"a}ffner}\ \emph {et~al.}(2005)\citenamefont
  {H{\"a}ffner}, \citenamefont {Schmidt-Kaler}, \citenamefont {H{\"a}nsel},
  \citenamefont {Roos}, \citenamefont {K{\"o}rber}, \citenamefont {Chwalla},
  \citenamefont {Riebe}, \citenamefont {Benhelm}, \citenamefont {Rapol},
  \citenamefont {Becher},\ and\ \citenamefont {Blatt}}]{Haffner:2005:zz}%
  \BibitemOpen
  \bibfield  {author} {\bibinfo {author} {\bibfnamefont {H.}~\bibnamefont
  {H{\"a}ffner}}, \bibinfo {author} {\bibfnamefont {F.}~\bibnamefont
  {Schmidt-Kaler}}, \bibinfo {author} {\bibfnamefont {W.}~\bibnamefont
  {H{\"a}nsel}}, \bibinfo {author} {\bibfnamefont {C.~F.}\ \bibnamefont
  {Roos}}, \bibinfo {author} {\bibfnamefont {T.}~\bibnamefont {K{\"o}rber}},
  \bibinfo {author} {\bibfnamefont {M.}~\bibnamefont {Chwalla}}, \bibinfo
  {author} {\bibfnamefont {M.}~\bibnamefont {Riebe}}, \bibinfo {author}
  {\bibfnamefont {J.}~\bibnamefont {Benhelm}}, \bibinfo {author} {\bibfnamefont
  {U.~D.}\ \bibnamefont {Rapol}}, \bibinfo {author} {\bibfnamefont
  {C.}~\bibnamefont {Becher}}, and\ \bibinfo {author} {\bibfnamefont
  {R.}~\bibnamefont {Blatt}},\ }\href@noop {} {\bibfield  {journal} {\bibinfo
  {journal} {Appl. Phys. B}\ }\textbf {\bibinfo {volume} {81}},\ \bibinfo
  {pages} {151} (\bibinfo {year} {2005})}\BibitemShut {NoStop}%
\bibitem [{\citenamefont {Langer}\ \emph {et~al.}(2005)\citenamefont {Langer},
  \citenamefont {Ozeri}, \citenamefont {Jost}, \citenamefont {Chiaverini},
  \citenamefont {DeMarco}, \citenamefont {Ben-Kish}, \citenamefont {Blakestad},
  \citenamefont {Britton}, \citenamefont {Hume}, \citenamefont {Itano},
  \citenamefont {Leibfried}, \citenamefont {Reichle}, \citenamefont
  {Rosenband}, \citenamefont {Schaetz}, \citenamefont {Schmidt},\ and\
  \citenamefont {Wineland}}]{Langer:2005:uu}%
  \BibitemOpen
  \bibfield  {author} {\bibinfo {author} {\bibfnamefont {C.}~\bibnamefont
  {Langer}}, \bibinfo {author} {\bibfnamefont {R.}~\bibnamefont {Ozeri}},
  \bibinfo {author} {\bibfnamefont {J.~D.}\ \bibnamefont {Jost}}, \bibinfo
  {author} {\bibfnamefont {J.}~\bibnamefont {Chiaverini}}, \bibinfo {author}
  {\bibfnamefont {B.}~\bibnamefont {DeMarco}}, \bibinfo {author} {\bibfnamefont
  {A.}~\bibnamefont {Ben-Kish}}, \bibinfo {author} {\bibfnamefont {R.~B.}\
  \bibnamefont {Blakestad}}, \bibinfo {author} {\bibfnamefont {J.}~\bibnamefont
  {Britton}}, \bibinfo {author} {\bibfnamefont {D.~B.}\ \bibnamefont {Hume}},
  \bibinfo {author} {\bibfnamefont {W.~M.}\ \bibnamefont {Itano}}, \bibinfo
  {author} {\bibfnamefont {D.}~\bibnamefont {Leibfried}}, \bibinfo {author}
  {\bibfnamefont {R.}~\bibnamefont {Reichle}}, \bibinfo {author} {\bibfnamefont
  {T.}~\bibnamefont {Rosenband}}, \bibinfo {author} {\bibfnamefont
  {T.}~\bibnamefont {Schaetz}}, \bibinfo {author} {\bibfnamefont {P.~O.}\
  \bibnamefont {Schmidt}}, and\ \bibinfo {author} {\bibfnamefont {D.~J.}\
  \bibnamefont {Wineland}},\ }\href {\doibase 10.1103/PhysRevLett.95.060502}
  {\bibfield  {journal} {\bibinfo  {journal} {Phys. Rev. Lett.}\ }\textbf
  {\bibinfo {volume} {95}},\ \bibinfo {pages} {060502} (\bibinfo {year}
  {2005})}\BibitemShut {NoStop}%
\bibitem [{\citenamefont {Viola}\ \emph {et~al.}(2001)\citenamefont {Viola},
  \citenamefont {Fortunato}, \citenamefont {Pravia}, \citenamefont {Knill},
  \citenamefont {Laflamme},\ and\ \citenamefont {Cory}}]{Viola:2001:ra}%
  \BibitemOpen
  \bibfield  {author} {\bibinfo {author} {\bibfnamefont {L.}~\bibnamefont
  {Viola}}, \bibinfo {author} {\bibfnamefont {E.~M.}\ \bibnamefont
  {Fortunato}}, \bibinfo {author} {\bibfnamefont {M.~A.}\ \bibnamefont
  {Pravia}}, \bibinfo {author} {\bibfnamefont {E.}~\bibnamefont {Knill}},
  \bibinfo {author} {\bibfnamefont {R.}~\bibnamefont {Laflamme}}, and\
  \bibinfo {author} {\bibfnamefont {D.~G.}\ \bibnamefont {Cory}},\ }\href@noop
  {} {\bibfield  {journal} {\bibinfo  {journal} {Science}\ }\textbf {\bibinfo
  {volume} {293}},\ \bibinfo {pages} {2059} (\bibinfo {year}
  {2001})}\BibitemShut {NoStop}%
\bibitem [{\citenamefont {Mohseni}\ \emph {et~al.}(2003)\citenamefont
  {Mohseni}, \citenamefont {Lundeen}, \citenamefont {Resch},\ and\
  \citenamefont {Steinberg}}]{Mohseni:2003:pp}%
  \BibitemOpen
  \bibfield  {author} {\bibinfo {author} {\bibfnamefont {M.}~\bibnamefont
  {Mohseni}}, \bibinfo {author} {\bibfnamefont {J.~S.}\ \bibnamefont
  {Lundeen}}, \bibinfo {author} {\bibfnamefont {K.~J.}\ \bibnamefont {Resch}},
   and\ \bibinfo {author} {\bibfnamefont {A.~M.}\ \bibnamefont {Steinberg}},\
  }\href@noop {} {\bibfield  {journal} {\bibinfo  {journal} {Phys. Rev. Lett.}\
  }\textbf {\bibinfo {volume} {91}},\ \bibinfo {pages} {187903} (\bibinfo
  {year} {2003})}\BibitemShut {NoStop}%
\bibitem [{\citenamefont {Zhang}\ \emph {et~al.}(2006)\citenamefont {Zhang},
  \citenamefont {Yin}, \citenamefont {Chen}, \citenamefont {Lu}, \citenamefont
  {Zhang}, \citenamefont {Li}, \citenamefont {Yang}, \citenamefont {Wang},\
  and\ \citenamefont {Pan}}]{Zhang:2006:zz}%
  \BibitemOpen
  \bibfield  {author} {\bibinfo {author} {\bibfnamefont {Q.}~\bibnamefont
  {Zhang}}, \bibinfo {author} {\bibfnamefont {J.}~\bibnamefont {Yin}}, \bibinfo
  {author} {\bibfnamefont {T.-Y.}\ \bibnamefont {Chen}}, \bibinfo {author}
  {\bibfnamefont {S.}~\bibnamefont {Lu}}, \bibinfo {author} {\bibfnamefont
  {J.}~\bibnamefont {Zhang}}, \bibinfo {author} {\bibfnamefont {X.-Q.}\
  \bibnamefont {Li}}, \bibinfo {author} {\bibfnamefont {T.}~\bibnamefont
  {Yang}}, \bibinfo {author} {\bibfnamefont {X.-B.}\ \bibnamefont {Wang}}, \
  and\ \bibinfo {author} {\bibfnamefont {J.-W.}\ \bibnamefont {Pan}},\
  }\href@noop {} {\bibfield  {journal} {\bibinfo  {journal} {Phys. Rev. A}\
  }\textbf {\bibinfo {volume} {73}},\ \bibinfo {pages} {020301(R)} (\bibinfo
  {year} {2006})}\BibitemShut {NoStop}%
\bibitem [{\citenamefont {Pushin}\ \emph {et~al.}(2011)\citenamefont {Pushin},
  \citenamefont {Huber}, \citenamefont {Arif},\ and\ \citenamefont
  {Cory}}]{Pushin:2011:zz}%
  \BibitemOpen
  \bibfield  {author} {\bibinfo {author} {\bibfnamefont {D.~A.}\ \bibnamefont
  {Pushin}}, \bibinfo {author} {\bibfnamefont {M.~G.}\ \bibnamefont {Huber}},
  \bibinfo {author} {\bibfnamefont {M.}~\bibnamefont {Arif}}, and\ \bibinfo
  {author} {\bibfnamefont {D.~G.}\ \bibnamefont {Cory}},\ }\href@noop {}
  {\bibfield  {journal} {\bibinfo  {journal} {Phys. Rev. Lett.}\ }\textbf
  {\bibinfo {volume} {107}},\ \bibinfo {pages} {150401} (\bibinfo {year}
  {2011})}\BibitemShut {NoStop}%
\bibitem [{\citenamefont {Raimond}\ \emph {et~al.}(2001)\citenamefont
  {Raimond}, \citenamefont {Brune},\ and\ \citenamefont
  {Haroche}}]{Raimond:2001:aa}%
  \BibitemOpen
  \bibfield  {author} {\bibinfo {author} {\bibfnamefont {J.~M.}\ \bibnamefont
  {Raimond}}, \bibinfo {author} {\bibfnamefont {M.}~\bibnamefont {Brune}}, 
  and\ \bibinfo {author} {\bibfnamefont {S.}~\bibnamefont {Haroche}},\
  }\href@noop {} {\bibfield  {journal} {\bibinfo  {journal} {Rev. Mod. Phys.}\
  }\textbf {\bibinfo {volume} {73}},\ \bibinfo {pages} {565} (\bibinfo {year}
  {2001})}\BibitemShut {NoStop}%
\bibitem [{\citenamefont {Kaiser}\ \emph {et~al.}(2001)\citenamefont {Kaiser},
  \citenamefont {Westbrook},\ and\ \citenamefont {David}}]{Kaiser:2001:tm}%
  \BibitemOpen
  \bibinfo {editor} {\bibfnamefont {R.}~\bibnamefont {Kaiser}}, \bibinfo
  {editor} {\bibfnamefont {C.}~\bibnamefont {Westbrook}}, and\ \bibinfo
  {editor} {\bibfnamefont {F.}~\bibnamefont {David}},\ eds.,\ \href@noop {}
  {\emph {\bibinfo {title} {Coherent Atomic Matter Waves, Les Houches Session
  LXXII}}},\ Les Houches Summer School Series\ (\bibinfo  {publisher}
  {Springer},\ \bibinfo {address} {Berlin},\ \bibinfo {year}
  {2001})\BibitemShut {NoStop}%
\bibitem [{\citenamefont {Haroche}\ and\ \citenamefont
  {Raimond}(2006)}]{Haroche:2006:hh}%
  \BibitemOpen
  \bibfield  {author} {\bibinfo {author} {\bibfnamefont {S.}~\bibnamefont
  {Haroche}}\ and\ \bibinfo {author} {\bibfnamefont {J.-M.}\ \bibnamefont
  {Raimond}},\ }\href@noop {} {\emph {\bibinfo {title} {Exploring the Quantum:
  Atoms, Cavities, and Photons}}}\ (\bibinfo  {publisher} {Oxford University
  Press},\ \bibinfo {address} {Oxford},\ \bibinfo {year} {2006})\BibitemShut
  {NoStop}%
\bibitem [{\citenamefont {Del{\'e}glise}\ \emph {et~al.}(2008)\citenamefont
  {Del{\'e}glise}, \citenamefont {Dotsenko}, \citenamefont {Sayrin},
  \citenamefont {Bernu}, \citenamefont {Brune}, \citenamefont {Raimond},\ and\
  \citenamefont {Haroche}}]{Deleglise:2008:oo}%
  \BibitemOpen
  \bibfield  {author} {\bibinfo {author} {\bibfnamefont {S.}~\bibnamefont
  {Del{\'e}glise}}, \bibinfo {author} {\bibfnamefont {I.}~\bibnamefont
  {Dotsenko}}, \bibinfo {author} {\bibfnamefont {C.}~\bibnamefont {Sayrin}},
  \bibinfo {author} {\bibfnamefont {J.}~\bibnamefont {Bernu}}, \bibinfo
  {author} {\bibfnamefont {M.}~\bibnamefont {Brune}}, \bibinfo {author}
  {\bibfnamefont {J.-M.}\ \bibnamefont {Raimond}}, and\ \bibinfo {author}
  {\bibfnamefont {S.}~\bibnamefont {Haroche}},\ }\href@noop {} {\bibfield
  {journal} {\bibinfo  {journal} {Nature}\ }\textbf {\bibinfo {volume} {455}},\
  \bibinfo {pages} {510} (\bibinfo {year} {2008})}\BibitemShut {NoStop}%
\bibitem [{\citenamefont {Arndt}\ \emph {et~al.}(1999)\citenamefont {Arndt},
  \citenamefont {Nairz}, \citenamefont {Vos-Andreae}, \citenamefont {Keller},
  \citenamefont {van~der Zouw},\ and\ \citenamefont
  {Zeilinger}}]{Arndt:1999:rc}%
  \BibitemOpen
  \bibfield  {author} {\bibinfo {author} {\bibfnamefont {M.}~\bibnamefont
  {Arndt}}, \bibinfo {author} {\bibfnamefont {O.}~\bibnamefont {Nairz}},
  \bibinfo {author} {\bibfnamefont {J.}~\bibnamefont {Vos-Andreae}}, \bibinfo
  {author} {\bibfnamefont {C.}~\bibnamefont {Keller}}, \bibinfo {author}
  {\bibfnamefont {G.}~\bibnamefont {van~der Zouw}}, and\ \bibinfo {author}
  {\bibfnamefont {A.}~\bibnamefont {Zeilinger}},\ }\href@noop {} {\bibfield
  {journal} {\bibinfo  {journal} {Nature}\ }\textbf {\bibinfo {volume} {401}},\
  \bibinfo {pages} {680} (\bibinfo {year} {1999})}\BibitemShut {NoStop}%
\bibitem [{\citenamefont {Hornberger}\ \emph {et~al.}(2003)\citenamefont
  {Hornberger}, \citenamefont {Uttenthaler}, \citenamefont {Brezger},
  \citenamefont {Hackerm{\"u}ller}, \citenamefont {Arndt},\ and\ \citenamefont
  {Zeilinger}}]{Hornberger:2003:tv}%
  \BibitemOpen
  \bibfield  {author} {\bibinfo {author} {\bibfnamefont {K.}~\bibnamefont
  {Hornberger}}, \bibinfo {author} {\bibfnamefont {S.}~\bibnamefont
  {Uttenthaler}}, \bibinfo {author} {\bibfnamefont {B.}~\bibnamefont
  {Brezger}}, \bibinfo {author} {\bibfnamefont {L.}~\bibnamefont
  {Hackerm{\"u}ller}}, \bibinfo {author} {\bibfnamefont {M.}~\bibnamefont
  {Arndt}}, and\ \bibinfo {author} {\bibfnamefont {A.}~\bibnamefont
  {Zeilinger}},\ }\href@noop {} {\bibfield  {journal} {\bibinfo  {journal}
  {Phys. Rev. Lett.}\ }\textbf {\bibinfo {volume} {90}},\ \bibinfo {pages}
  {160401} (\bibinfo {year} {2003})}\BibitemShut {NoStop}%
\bibitem [{\citenamefont {Hackerm{\"u}ller}\ \emph {et~al.}(2003)\citenamefont
  {Hackerm{\"u}ller}, \citenamefont {Hornberger}, \citenamefont {Brezger},
  \citenamefont {Zeilinger},\ and\ \citenamefont
  {Arndt}}]{Hackermuller:2003:uu}%
  \BibitemOpen
  \bibfield  {author} {\bibinfo {author} {\bibfnamefont {L.}~\bibnamefont
  {Hackerm{\"u}ller}}, \bibinfo {author} {\bibfnamefont {K.}~\bibnamefont
  {Hornberger}}, \bibinfo {author} {\bibfnamefont {B.}~\bibnamefont {Brezger}},
  \bibinfo {author} {\bibfnamefont {A.}~\bibnamefont {Zeilinger}}, and\
  \bibinfo {author} {\bibfnamefont {M.}~\bibnamefont {Arndt}},\ }\href@noop {}
  {\bibfield  {journal} {\bibinfo  {journal} {Appl. Phys. B}\ }\textbf
  {\bibinfo {volume} {77}},\ \bibinfo {pages} {781} (\bibinfo {year}
  {2003})}\BibitemShut {NoStop}%
\bibitem [{\citenamefont {Hackerm{\"u}ller}\ \emph {et~al.}(2004)\citenamefont
  {Hackerm{\"u}ller}, \citenamefont {Hornberger}, \citenamefont {Brezger},
  \citenamefont {Zeilinger},\ and\ \citenamefont
  {Arndt}}]{Hackermuller:2004:rd}%
  \BibitemOpen
  \bibfield  {author} {\bibinfo {author} {\bibfnamefont {L.}~\bibnamefont
  {Hackerm{\"u}ller}}, \bibinfo {author} {\bibfnamefont {K.}~\bibnamefont
  {Hornberger}}, \bibinfo {author} {\bibfnamefont {B.}~\bibnamefont {Brezger}},
  \bibinfo {author} {\bibfnamefont {A.}~\bibnamefont {Zeilinger}}, and\
  \bibinfo {author} {\bibfnamefont {M.}~\bibnamefont {Arndt}},\ }\href@noop {}
  {\bibfield  {journal} {\bibinfo  {journal} {Nature}\ }\textbf {\bibinfo
  {volume} {427}},\ \bibinfo {pages} {711} (\bibinfo {year}
  {2004})}\BibitemShut {NoStop}%
\bibitem [{\citenamefont {Fein}\ \emph {et~al.}(2019)\citenamefont {Fein},
  \citenamefont {Geyer}, \citenamefont {Zwick}, \citenamefont {Kia{\l}ka},
  \citenamefont {Pedalino}, \citenamefont {Mayor}, \citenamefont {Gerlich},\
  and\ \citenamefont {Arndt}}]{Fein:2019:aa}%
  \BibitemOpen
  \bibfield  {author} {\bibinfo {author} {\bibfnamefont {Y.~Y.}\ \bibnamefont
  {Fein}}, \bibinfo {author} {\bibfnamefont {P.}~\bibnamefont {Geyer}},
  \bibinfo {author} {\bibfnamefont {P.}~\bibnamefont {Zwick}}, \bibinfo
  {author} {\bibfnamefont {F.}~\bibnamefont {Kia{\l}ka}}, \bibinfo {author}
  {\bibfnamefont {S.}~\bibnamefont {Pedalino}}, \bibinfo {author}
  {\bibfnamefont {M.}~\bibnamefont {Mayor}}, \bibinfo {author} {\bibfnamefont
  {S.}~\bibnamefont {Gerlich}}, and\ \bibinfo {author} {\bibfnamefont
  {M.}~\bibnamefont {Arndt}},\ }\href {\doibase 10.1038/s41567-019-0663-9}
  {\bibfield  {journal} {\bibinfo  {journal} {Nature Phys.}\ }\textbf {\bibinfo
  {volume} {15}},\ \bibinfo {pages} {1242} (\bibinfo {year}
  {2019})}\BibitemShut {NoStop}%
\bibitem [{\citenamefont {Chiorescu}\ \emph {et~al.}(2003)\citenamefont
  {Chiorescu}, \citenamefont {Nakamura}, \citenamefont {Harmans},\ and\
  \citenamefont {Mooij}}]{Chiorescu:2003:ta}%
  \BibitemOpen
  \bibfield  {author} {\bibinfo {author} {\bibfnamefont {I.}~\bibnamefont
  {Chiorescu}}, \bibinfo {author} {\bibfnamefont {Y.}~\bibnamefont {Nakamura}},
  \bibinfo {author} {\bibfnamefont {C.~J. P.~M.}\ \bibnamefont {Harmans}}, \
  and\ \bibinfo {author} {\bibfnamefont {J.~E.}\ \bibnamefont {Mooij}},\
  }\href@noop {} {\bibfield  {journal} {\bibinfo  {journal} {Science}\ }\textbf
  {\bibinfo {volume} {21}},\ \bibinfo {pages} {1869} (\bibinfo {year}
  {2003})}\BibitemShut {NoStop}%
\bibitem [{\citenamefont {Vion}\ \emph {et~al.}(2002)\citenamefont {Vion},
  \citenamefont {Aassime}, \citenamefont {Cottet}, \citenamefont {Joyez},
  \citenamefont {Pothier}, \citenamefont {Urbina}, \citenamefont {Esteve},\
  and\ \citenamefont {Devoret}}]{Vion:2002:oo}%
  \BibitemOpen
  \bibfield  {author} {\bibinfo {author} {\bibfnamefont {D.}~\bibnamefont
  {Vion}}, \bibinfo {author} {\bibfnamefont {A.}~\bibnamefont {Aassime}},
  \bibinfo {author} {\bibfnamefont {A.}~\bibnamefont {Cottet}}, \bibinfo
  {author} {\bibfnamefont {P.}~\bibnamefont {Joyez}}, \bibinfo {author}
  {\bibfnamefont {H.}~\bibnamefont {Pothier}}, \bibinfo {author} {\bibfnamefont
  {C.}~\bibnamefont {Urbina}}, \bibinfo {author} {\bibfnamefont
  {D.}~\bibnamefont {Esteve}}, and\ \bibinfo {author} {\bibfnamefont {M.~H.}\
  \bibnamefont {Devoret}},\ }\href@noop {} {\bibfield  {journal} {\bibinfo
  {journal} {Science}\ }\textbf {\bibinfo {volume} {296}},\ \bibinfo {pages}
  {886} (\bibinfo {year} {2002})}\BibitemShut {NoStop}%
\bibitem [{\citenamefont {Yu}\ \emph {et~al.}(2002)\citenamefont {Yu},
  \citenamefont {Han}, \citenamefont {Chu}, \citenamefont {Chu},\ and\
  \citenamefont {Wang}}]{Yu:2002:yb}%
  \BibitemOpen
  \bibfield  {author} {\bibinfo {author} {\bibfnamefont {Y.}~\bibnamefont
  {Yu}}, \bibinfo {author} {\bibfnamefont {S.}~\bibnamefont {Han}}, \bibinfo
  {author} {\bibfnamefont {X.}~\bibnamefont {Chu}}, \bibinfo {author}
  {\bibfnamefont {S.-I.}\ \bibnamefont {Chu}}, and\ \bibinfo {author}
  {\bibfnamefont {Z.}~\bibnamefont {Wang}},\ }\href@noop {} {\bibfield
  {journal} {\bibinfo  {journal} {Science}\ }\textbf {\bibinfo {volume}
  {296}},\ \bibinfo {pages} {889} (\bibinfo {year} {2002})}\BibitemShut
  {NoStop}%
\bibitem [{\citenamefont {Martinis}\ \emph {et~al.}(2002)\citenamefont
  {Martinis}, \citenamefont {Nam}, \citenamefont {Aumentado},\ and\
  \citenamefont {Urbina}}]{Martinis:2002:qq}%
  \BibitemOpen
  \bibfield  {author} {\bibinfo {author} {\bibfnamefont {J.~M.}\ \bibnamefont
  {Martinis}}, \bibinfo {author} {\bibfnamefont {S.}~\bibnamefont {Nam}},
  \bibinfo {author} {\bibfnamefont {J.}~\bibnamefont {Aumentado}}, and\
  \bibinfo {author} {\bibfnamefont {C.}~\bibnamefont {Urbina}},\ }\href@noop {}
  {\bibfield  {journal} {\bibinfo  {journal} {Phys. Rev. Lett.}\ }\textbf
  {\bibinfo {volume} {89}},\ \bibinfo {pages} {117901} (\bibinfo {year}
  {2002})}\BibitemShut {NoStop}%
\bibitem [{\citenamefont {Schneider}\ and\ \citenamefont
  {Milburn}(1998)}]{Schneider:1998:yz}%
  \BibitemOpen
  \bibfield  {author} {\bibinfo {author} {\bibfnamefont {S.}~\bibnamefont
  {Schneider}}\ and\ \bibinfo {author} {\bibfnamefont {G.~J.}\ \bibnamefont
  {Milburn}},\ }\href@noop {} {\bibfield  {journal} {\bibinfo  {journal} {Phys.
  Rev. A}\ }\textbf {\bibinfo {volume} {57}},\ \bibinfo {pages} {3748}
  (\bibinfo {year} {1998})}\BibitemShut {NoStop}%
\bibitem [{\citenamefont {Miquel}\ \emph {et~al.}(1997)\citenamefont {Miquel},
  \citenamefont {Paz},\ and\ \citenamefont {Zurek}}]{Miquel:1997:zz}%
  \BibitemOpen
  \bibfield  {author} {\bibinfo {author} {\bibfnamefont {C.}~\bibnamefont
  {Miquel}}, \bibinfo {author} {\bibfnamefont {J.~P.}\ \bibnamefont {Paz}}, 
  and\ \bibinfo {author} {\bibfnamefont {W.~H.}\ \bibnamefont {Zurek}},\
  }\href@noop {} {\bibfield  {journal} {\bibinfo  {journal} {Phys. Rev. Lett.}\
  }\textbf {\bibinfo {volume} {78}},\ \bibinfo {pages} {3971} (\bibinfo {year}
  {1997})}\BibitemShut {NoStop}%
\bibitem [{\citenamefont {Steane}\ \emph {et~al.}(2000)\citenamefont {Steane},
  \citenamefont {Roos}, \citenamefont {Stevens}, \citenamefont {Mundt},
  \citenamefont {Leibfried}, \citenamefont {Schmidt-Kaler},\ and\ \citenamefont
  {Blatt}}]{Steane:2000:ii}%
  \BibitemOpen
  \bibfield  {author} {\bibinfo {author} {\bibfnamefont {A.}~\bibnamefont
  {Steane}}, \bibinfo {author} {\bibfnamefont {C.~F.}\ \bibnamefont {Roos}},
  \bibinfo {author} {\bibfnamefont {D.}~\bibnamefont {Stevens}}, \bibinfo
  {author} {\bibfnamefont {A.}~\bibnamefont {Mundt}}, \bibinfo {author}
  {\bibfnamefont {D.}~\bibnamefont {Leibfried}}, \bibinfo {author}
  {\bibfnamefont {F.}~\bibnamefont {Schmidt-Kaler}}, and\ \bibinfo {author}
  {\bibfnamefont {R.}~\bibnamefont {Blatt}},\ }\href {\doibase
  10.1103/PhysRevA.62.042305} {\bibfield  {journal} {\bibinfo  {journal} {Phys.
  Rev. A}\ }\textbf {\bibinfo {volume} {62}},\ \bibinfo {pages} {042305}
  (\bibinfo {year} {2000})}\BibitemShut {NoStop}%
\bibitem [{\citenamefont {H\"affner}\ \emph {et~al.}(2003)\citenamefont
  {H\"affner}, \citenamefont {Gulde}, \citenamefont {Riebe}, \citenamefont
  {Lancaster}, \citenamefont {Becher}, \citenamefont {Eschner}, \citenamefont
  {Schmidt-Kaler},\ and\ \citenamefont {Blatt}}]{Haffner:2003:oo}%
  \BibitemOpen
  \bibfield  {author} {\bibinfo {author} {\bibfnamefont {H.}~\bibnamefont
  {H\"affner}}, \bibinfo {author} {\bibfnamefont {S.}~\bibnamefont {Gulde}},
  \bibinfo {author} {\bibfnamefont {M.}~\bibnamefont {Riebe}}, \bibinfo
  {author} {\bibfnamefont {G.}~\bibnamefont {Lancaster}}, \bibinfo {author}
  {\bibfnamefont {C.}~\bibnamefont {Becher}}, \bibinfo {author} {\bibfnamefont
  {J.}~\bibnamefont {Eschner}}, \bibinfo {author} {\bibfnamefont
  {F.}~\bibnamefont {Schmidt-Kaler}}, and\ \bibinfo {author} {\bibfnamefont
  {R.}~\bibnamefont {Blatt}},\ }\href {\doibase 10.1103/PhysRevLett.90.143602}
  {\bibfield  {journal} {\bibinfo  {journal} {Phys. Rev. Lett.}\ }\textbf
  {\bibinfo {volume} {90}},\ \bibinfo {pages} {143602} (\bibinfo {year}
  {2003})}\BibitemShut {NoStop}%
\bibitem [{\citenamefont {Leibfried}\ \emph {et~al.}(2003)\citenamefont
  {Leibfried}, \citenamefont {DeMarco}, \citenamefont {Meyer}, \citenamefont
  {Lucas}, \citenamefont {Barrett}, \citenamefont {Britton}, \citenamefont
  {Itano}, \citenamefont {Jelenkovi{\'c}}, \citenamefont {Langer},
  \citenamefont {Rosenband},\ and\ \citenamefont
  {Wineland}}]{Leibfried:2003:mm}%
  \BibitemOpen
  \bibfield  {author} {\bibinfo {author} {\bibfnamefont {D.}~\bibnamefont
  {Leibfried}}, \bibinfo {author} {\bibfnamefont {B.}~\bibnamefont {DeMarco}},
  \bibinfo {author} {\bibfnamefont {V.}~\bibnamefont {Meyer}}, \bibinfo
  {author} {\bibfnamefont {D.}~\bibnamefont {Lucas}}, \bibinfo {author}
  {\bibfnamefont {M.}~\bibnamefont {Barrett}}, \bibinfo {author} {\bibfnamefont
  {J.}~\bibnamefont {Britton}}, \bibinfo {author} {\bibfnamefont {W.~M.}\
  \bibnamefont {Itano}}, \bibinfo {author} {\bibfnamefont {B.}~\bibnamefont
  {Jelenkovi{\'c}}}, \bibinfo {author} {\bibfnamefont {C.}~\bibnamefont
  {Langer}}, \bibinfo {author} {\bibfnamefont {T.}~\bibnamefont {Rosenband}}, 
  and\ \bibinfo {author} {\bibfnamefont {D.~J.}\ \bibnamefont {Wineland}},\
  }\href@noop {} {\bibfield  {journal} {\bibinfo  {journal} {Nature}\ }\textbf
  {\bibinfo {volume} {422}},\ \bibinfo {pages} {412} (\bibinfo {year}
  {2003})}\BibitemShut {NoStop}%
\bibitem [{\citenamefont {Schmidt-Kaler}\ \emph {et~al.}(2003)\citenamefont
  {Schmidt-Kaler}, \citenamefont {Gulde}, \citenamefont {Riebe}, \citenamefont
  {Deuschle}, \citenamefont {Kreuter}, \citenamefont {Lancaster}, \citenamefont
  {Becher}, \citenamefont {Eschner}, \citenamefont {H{\"a}ffner},\ and\
  \citenamefont {Blatt}}]{SchmidtKaler:2003:pp}%
  \BibitemOpen
  \bibfield  {author} {\bibinfo {author} {\bibfnamefont {F.}~\bibnamefont
  {Schmidt-Kaler}}, \bibinfo {author} {\bibfnamefont {S.}~\bibnamefont
  {Gulde}}, \bibinfo {author} {\bibfnamefont {M.}~\bibnamefont {Riebe}},
  \bibinfo {author} {\bibfnamefont {T.}~\bibnamefont {Deuschle}}, \bibinfo
  {author} {\bibfnamefont {A.}~\bibnamefont {Kreuter}}, \bibinfo {author}
  {\bibfnamefont {G.}~\bibnamefont {Lancaster}}, \bibinfo {author}
  {\bibfnamefont {C.}~\bibnamefont {Becher}}, \bibinfo {author} {\bibfnamefont
  {J.}~\bibnamefont {Eschner}}, \bibinfo {author} {\bibfnamefont
  {H.}~\bibnamefont {H{\"a}ffner}}, and\ \bibinfo {author} {\bibfnamefont
  {R.}~\bibnamefont {Blatt}},\ }\href@noop {} {\bibfield  {journal} {\bibinfo
  {journal} {J. Phys. B}\ }\textbf {\bibinfo {volume} {36}},\ \bibinfo {pages}
  {623} (\bibinfo {year} {2003})}\BibitemShut {NoStop}%
\bibitem [{\citenamefont {Brouard}\ and\ \citenamefont
  {Plata}(2004)}]{Brouard:2004:in}%
  \BibitemOpen
  \bibfield  {author} {\bibinfo {author} {\bibfnamefont {S.}~\bibnamefont
  {Brouard}}\ and\ \bibinfo {author} {\bibfnamefont {J.}~\bibnamefont
  {Plata}},\ }\href@noop {} {\bibfield  {journal} {\bibinfo  {journal} {Phys.
  Rev. A}\ }\textbf {\bibinfo {volume} {70}},\ \bibinfo {pages} {013413}
  (\bibinfo {year} {2004})}\BibitemShut {NoStop}%
\bibitem [{\citenamefont {Grotz}\ \emph {et~al.}(2006)\citenamefont {Grotz},
  \citenamefont {Heaney},\ and\ \citenamefont {Strunz}}]{Grotz:2006:km}%
  \BibitemOpen
  \bibfield  {author} {\bibinfo {author} {\bibfnamefont {T.}~\bibnamefont
  {Grotz}}, \bibinfo {author} {\bibfnamefont {L.}~\bibnamefont {Heaney}}, 
  and\ \bibinfo {author} {\bibfnamefont {W.~T.}\ \bibnamefont {Strunz}},\
  }\href@noop {} {\bibfield  {journal} {\bibinfo  {journal} {Phys. Rev. A}\
  }\textbf {\bibinfo {volume} {74}},\ \bibinfo {pages} {022102} (\bibinfo
  {year} {2006})}\BibitemShut {NoStop}%
\bibitem [{\citenamefont {H{\"a}ffner}\ \emph {et~al.}(2008)\citenamefont
  {H{\"a}ffner}, \citenamefont {Roos},\ and\ \citenamefont
  {Blatt}}]{Haffner:2008:pp}%
  \BibitemOpen
  \bibfield  {author} {\bibinfo {author} {\bibfnamefont {H.}~\bibnamefont
  {H{\"a}ffner}}, \bibinfo {author} {\bibfnamefont {C.~F.}\ \bibnamefont
  {Roos}}, and\ \bibinfo {author} {\bibfnamefont {R.}~\bibnamefont {Blatt}},\
  }\href {\doibase 10.1016/j.physrep.2008.09.003} {\bibfield  {journal}
  {\bibinfo  {journal} {Phys. Rep.}\ }\textbf {\bibinfo {volume} {469}},\
  \bibinfo {pages} {155} (\bibinfo {year} {2008})}\BibitemShut {NoStop}%
\bibitem [{\citenamefont {Kuhlmann}\ \emph {et~al.}(2013)\citenamefont
  {Kuhlmann}, \citenamefont {Houel}, \citenamefont {Ludwig}, \citenamefont
  {Greuter}, \citenamefont {Reuter}, \citenamefont {Wieck}, \citenamefont
  {Poggio},\ and\ \citenamefont {Warburton}}]{Kuhlmann:2013:aa}%
  \BibitemOpen
  \bibfield  {author} {\bibinfo {author} {\bibfnamefont {A.~V.}\ \bibnamefont
  {Kuhlmann}}, \bibinfo {author} {\bibfnamefont {J.}~\bibnamefont {Houel}},
  \bibinfo {author} {\bibfnamefont {A.}~\bibnamefont {Ludwig}}, \bibinfo
  {author} {\bibfnamefont {L.}~\bibnamefont {Greuter}}, \bibinfo {author}
  {\bibfnamefont {D.}~\bibnamefont {Reuter}}, \bibinfo {author} {\bibfnamefont
  {A.~D.}\ \bibnamefont {Wieck}}, \bibinfo {author} {\bibfnamefont
  {M.}~\bibnamefont {Poggio}}, and\ \bibinfo {author} {\bibfnamefont {R.~J.}\
  \bibnamefont {Warburton}},\ }\href@noop {} {\bibfield  {journal} {\bibinfo
  {journal} {Nature Phys.}\ }\textbf {\bibinfo {volume} {9}},\ \bibinfo {pages}
  {570} (\bibinfo {year} {2013})}\BibitemShut {NoStop}%
\bibitem [{\citenamefont {Arnold}\ \emph {et~al.}(2014)\citenamefont {Arnold},
  \citenamefont {Loo}, \citenamefont {Lema\^{\i}tre}, \citenamefont {Sagnes},
  \citenamefont {Krebs}, \citenamefont {Voisin}, \citenamefont {Senellart},\
  and\ \citenamefont {Lanco}}]{Arnold:2014:oo}%
  \BibitemOpen
  \bibfield  {author} {\bibinfo {author} {\bibfnamefont {C.}~\bibnamefont
  {Arnold}}, \bibinfo {author} {\bibfnamefont {V.}~\bibnamefont {Loo}},
  \bibinfo {author} {\bibfnamefont {A.}~\bibnamefont {Lema\^{\i}tre}}, \bibinfo
  {author} {\bibfnamefont {I.}~\bibnamefont {Sagnes}}, \bibinfo {author}
  {\bibfnamefont {O.}~\bibnamefont {Krebs}}, \bibinfo {author} {\bibfnamefont
  {P.}~\bibnamefont {Voisin}}, \bibinfo {author} {\bibfnamefont
  {P.}~\bibnamefont {Senellart}}, and\ \bibinfo {author} {\bibfnamefont
  {L.}~\bibnamefont {Lanco}},\ }\href {\doibase 10.1103/PhysRevX.4.021004}
  {\bibfield  {journal} {\bibinfo  {journal} {Phys. Rev. X}\ }\textbf {\bibinfo
  {volume} {4}},\ \bibinfo {pages} {021004} (\bibinfo {year}
  {2014})}\BibitemShut {NoStop}%
\bibitem [{\citenamefont {Fischer}\ \emph {et~al.}(2009)\citenamefont
  {Fischer}, \citenamefont {Trif}, \citenamefont {Coish},\ and\ \citenamefont
  {Loss}}]{Fischer:2009:ii}%
  \BibitemOpen
  \bibfield  {author} {\bibinfo {author} {\bibfnamefont {J.}~\bibnamefont
  {Fischer}}, \bibinfo {author} {\bibfnamefont {M.}~\bibnamefont {Trif}},
  \bibinfo {author} {\bibfnamefont {W.~A.}\ \bibnamefont {Coish}}, and\
  \bibinfo {author} {\bibfnamefont {D.}~\bibnamefont {Loss}},\ }\href@noop {}
  {\bibfield  {journal} {\bibinfo  {journal} {Solid State Comm.}\ }\textbf
  {\bibinfo {volume} {149}},\ \bibinfo {pages} {1443} (\bibinfo {year}
  {2009})}\BibitemShut {NoStop}%
\bibitem [{\citenamefont {Urbaszek}\ \emph {et~al.}(2013)\citenamefont
  {Urbaszek}, \citenamefont {Marie}, \citenamefont {Amand}, \citenamefont
  {Krebs}, \citenamefont {Voisin}, \citenamefont {Maletinsky}, \citenamefont
  {H\"ogele},\ and\ \citenamefont {Imamoglu}}]{Urbaszek:2013:pp}%
  \BibitemOpen
  \bibfield  {author} {\bibinfo {author} {\bibfnamefont {B.}~\bibnamefont
  {Urbaszek}}, \bibinfo {author} {\bibfnamefont {X.}~\bibnamefont {Marie}},
  \bibinfo {author} {\bibfnamefont {T.}~\bibnamefont {Amand}}, \bibinfo
  {author} {\bibfnamefont {O.}~\bibnamefont {Krebs}}, \bibinfo {author}
  {\bibfnamefont {P.}~\bibnamefont {Voisin}}, \bibinfo {author} {\bibfnamefont
  {P.}~\bibnamefont {Maletinsky}}, \bibinfo {author} {\bibfnamefont
  {A.}~\bibnamefont {H\"ogele}}, and\ \bibinfo {author} {\bibfnamefont
  {A.}~\bibnamefont {Imamoglu}},\ }\href@noop {} {\bibfield  {journal}
  {\bibinfo  {journal} {Rev. Mod. Phys.}\ }\textbf {\bibinfo {volume} {85}},\
  \bibinfo {pages} {79} (\bibinfo {year} {2013})}\BibitemShut {NoStop}%
\bibitem [{\citenamefont {Delteil}\ \emph {et~al.}(2014)\citenamefont
  {Delteil}, \citenamefont {Gao}, \citenamefont {Fallahi}, \citenamefont
  {Miguel-Sanchez},\ and\ \citenamefont {Imamo\ifmmode~\breve{g}\else
  \u{g}\fi{}lu}}]{Delteil:2014:aa}%
  \BibitemOpen
  \bibfield  {author} {\bibinfo {author} {\bibfnamefont {A.}~\bibnamefont
  {Delteil}}, \bibinfo {author} {\bibfnamefont {W.-b.}\ \bibnamefont {Gao}},
  \bibinfo {author} {\bibfnamefont {P.}~\bibnamefont {Fallahi}}, \bibinfo
  {author} {\bibfnamefont {J.}~\bibnamefont {Miguel-Sanchez}}, and\ \bibinfo
  {author} {\bibfnamefont {A.}~\bibnamefont {Imamo\ifmmode~\breve{g}\else
  \u{g}\fi{}lu}},\ }\href {\doibase 10.1103/PhysRevLett.112.116802} {\bibfield
  {journal} {\bibinfo  {journal} {Phys. Rev. Lett.}\ }\textbf {\bibinfo
  {volume} {112}},\ \bibinfo {pages} {116802} (\bibinfo {year}
  {2014})}\BibitemShut {NoStop}%
\bibitem [{\citenamefont {Tighineanu}\ \emph {et~al.}(2018)\citenamefont
  {Tighineanu}, \citenamefont {Dree\ss{}en}, \citenamefont {Flindt},
  \citenamefont {Lodahl},\ and\ \citenamefont
  {S\o{}rensen}}]{Tighineanu:2018:ii}%
  \BibitemOpen
  \bibfield  {author} {\bibinfo {author} {\bibfnamefont {P.}~\bibnamefont
  {Tighineanu}}, \bibinfo {author} {\bibfnamefont {C.~L.}\ \bibnamefont
  {Dree\ss{}en}}, \bibinfo {author} {\bibfnamefont {C.}~\bibnamefont {Flindt}},
  \bibinfo {author} {\bibfnamefont {P.}~\bibnamefont {Lodahl}}, and\ \bibinfo
  {author} {\bibfnamefont {A.~S.}\ \bibnamefont {S\o{}rensen}},\ }\href@noop {}
  {\bibfield  {journal} {\bibinfo  {journal} {Phys. Rev. Lett.}\ }\textbf
  {\bibinfo {volume} {120}},\ \bibinfo {pages} {257401} (\bibinfo {year}
  {2018})}\BibitemShut {NoStop}%
\bibitem [{\citenamefont {Fong}\ \emph {et~al.}(2012)\citenamefont {Fong},
  \citenamefont {Pernice},\ and\ \citenamefont {Tang}}]{Fong:2012:aa}%
  \BibitemOpen
  \bibfield  {author} {\bibinfo {author} {\bibfnamefont {K.~Y.}\ \bibnamefont
  {Fong}}, \bibinfo {author} {\bibfnamefont {W.~H.~P.}\ \bibnamefont
  {Pernice}}, and\ \bibinfo {author} {\bibfnamefont {H.~X.}\ \bibnamefont
  {Tang}},\ }\href {\doibase 10.1103/PhysRevB.85.161410} {\bibfield  {journal}
  {\bibinfo  {journal} {Phys. Rev. B}\ }\textbf {\bibinfo {volume} {85}},\
  \bibinfo {pages} {161410} (\bibinfo {year} {2012})}\BibitemShut {NoStop}%
\bibitem [{\citenamefont {Zhang}\ \emph {et~al.}(2014)\citenamefont {Zhang},
  \citenamefont {Moser}, \citenamefont {G\"uttinger}, \citenamefont
  {Bachtold},\ and\ \citenamefont {Dykman}}]{Zhang:2014:oo}%
  \BibitemOpen
  \bibfield  {author} {\bibinfo {author} {\bibfnamefont {Y.}~\bibnamefont
  {Zhang}}, \bibinfo {author} {\bibfnamefont {J.}~\bibnamefont {Moser}},
  \bibinfo {author} {\bibfnamefont {J.}~\bibnamefont {G\"uttinger}}, \bibinfo
  {author} {\bibfnamefont {A.}~\bibnamefont {Bachtold}}, and\ \bibinfo
  {author} {\bibfnamefont {M.~I.}\ \bibnamefont {Dykman}},\ }\href {\doibase
  10.1103/PhysRevLett.113.255502} {\bibfield  {journal} {\bibinfo  {journal}
  {Phys. Rev. Lett.}\ }\textbf {\bibinfo {volume} {113}},\ \bibinfo {pages}
  {255502} (\bibinfo {year} {2014})}\BibitemShut {NoStop}%
\bibitem [{\citenamefont {Miao}\ \emph {et~al.}(2014)\citenamefont {Miao},
  \citenamefont {Yeom}, \citenamefont {Wang}, \citenamefont {Standley},\ and\
  \citenamefont {Bockrath}}]{Miao:2014:ii}%
  \BibitemOpen
  \bibfield  {author} {\bibinfo {author} {\bibfnamefont {T.}~\bibnamefont
  {Miao}}, \bibinfo {author} {\bibfnamefont {S.}~\bibnamefont {Yeom}}, \bibinfo
  {author} {\bibfnamefont {P.}~\bibnamefont {Wang}}, \bibinfo {author}
  {\bibfnamefont {B.}~\bibnamefont {Standley}}, and\ \bibinfo {author}
  {\bibfnamefont {M.}~\bibnamefont {Bockrath}},\ }\href@noop {} {\bibfield
  {journal} {\bibinfo  {journal} {Nano Lett.}\ }\textbf {\bibinfo {volume}
  {14}},\ \bibinfo {pages} {2982} (\bibinfo {year} {2014})}\BibitemShut
  {NoStop}%
\bibitem [{\citenamefont {Moser}\ \emph {et~al.}(2014)\citenamefont {Moser},
  \citenamefont {Eichler}, \citenamefont {G{\"u}ttinger}, \citenamefont
  {Dykman},\ and\ \citenamefont {Bachtold}}]{Moser:2014:uu}%
  \BibitemOpen
  \bibfield  {author} {\bibinfo {author} {\bibfnamefont {J.}~\bibnamefont
  {Moser}}, \bibinfo {author} {\bibfnamefont {A.}~\bibnamefont {Eichler}},
  \bibinfo {author} {\bibfnamefont {J.}~\bibnamefont {G{\"u}ttinger}}, \bibinfo
  {author} {\bibfnamefont {M.~I.}\ \bibnamefont {Dykman}}, and\ \bibinfo
  {author} {\bibfnamefont {A.}~\bibnamefont {Bachtold}},\ }\href@noop {}
  {\bibfield  {journal} {\bibinfo  {journal} {Nature Nanotechnology}\ }\textbf
  {\bibinfo {volume} {9}},\ \bibinfo {pages} {1007} (\bibinfo {year}
  {2014})}\BibitemShut {NoStop}%
\bibitem [{\citenamefont {Maillet}\ \emph {et~al.}(2016)\citenamefont
  {Maillet}, \citenamefont {Vavrek}, \citenamefont {Fefferman}, \citenamefont
  {Bourgeois},\ and\ \citenamefont {Collin}}]{Maillet:2016:zz}%
  \BibitemOpen
  \bibfield  {author} {\bibinfo {author} {\bibfnamefont {O.}~\bibnamefont
  {Maillet}}, \bibinfo {author} {\bibfnamefont {F.}~\bibnamefont {Vavrek}},
  \bibinfo {author} {\bibfnamefont {A.~D.}\ \bibnamefont {Fefferman}}, \bibinfo
  {author} {\bibfnamefont {O.}~\bibnamefont {Bourgeois}}, and\ \bibinfo
  {author} {\bibfnamefont {E.}~\bibnamefont {Collin}},\ }\href@noop {}
  {\bibfield  {journal} {\bibinfo  {journal} {New J. Phys.}\ }\textbf {\bibinfo
  {volume} {18}},\ \bibinfo {pages} {073022} (\bibinfo {year}
  {2016})}\BibitemShut {NoStop}%
\bibitem [{\citenamefont {Hornberger}\ and\ \citenamefont
  {Sipe}(2003)}]{Hornberger:2003:un}%
  \BibitemOpen
  \bibfield  {author} {\bibinfo {author} {\bibfnamefont {K.}~\bibnamefont
  {Hornberger}}\ and\ \bibinfo {author} {\bibfnamefont {J.~E.}\ \bibnamefont
  {Sipe}},\ }\href@noop {} {\bibfield  {journal} {\bibinfo  {journal} {Phys.
  Rev. A}\ }\textbf {\bibinfo {volume} {68}},\ \bibinfo {pages} {012105}
  (\bibinfo {year} {2003})}\BibitemShut {NoStop}%
\bibitem [{\citenamefont {Hornberger}\ \emph {et~al.}(2004)\citenamefont
  {Hornberger}, \citenamefont {Sipe},\ and\ \citenamefont
  {Arndt}}]{Hornberger:2004:bb}%
  \BibitemOpen
  \bibfield  {author} {\bibinfo {author} {\bibfnamefont {K.}~\bibnamefont
  {Hornberger}}, \bibinfo {author} {\bibfnamefont {J.~E.}\ \bibnamefont
  {Sipe}}, and\ \bibinfo {author} {\bibfnamefont {M.}~\bibnamefont {Arndt}},\
  }\href@noop {} {\bibfield  {journal} {\bibinfo  {journal} {Phys. Rev. A}\
  }\textbf {\bibinfo {volume} {70}},\ \bibinfo {pages} {053608} (\bibinfo
  {year} {2004})}\BibitemShut {NoStop}%
\bibitem [{\citenamefont {Hornberger}\ \emph {et~al.}(2005)\citenamefont
  {Hornberger}, \citenamefont {Hackerm{\"u}ller},\ and\ \citenamefont
  {Arndt}}]{Hornberger:2005:mo}%
  \BibitemOpen
  \bibfield  {author} {\bibinfo {author} {\bibfnamefont {K.}~\bibnamefont
  {Hornberger}}, \bibinfo {author} {\bibfnamefont {L.}~\bibnamefont
  {Hackerm{\"u}ller}}, and\ \bibinfo {author} {\bibfnamefont
  {M.}~\bibnamefont {Arndt}},\ }\href@noop {} {\bibfield  {journal} {\bibinfo
  {journal} {Phys. Rev. A}\ }\textbf {\bibinfo {volume} {71}},\ \bibinfo
  {pages} {023601} (\bibinfo {year} {2005})}\BibitemShut {NoStop}%
\bibitem [{\citenamefont
  {Hornberger}(2006{\natexlab{a}})}]{Hornberger:2006:tx}%
  \BibitemOpen
  \bibfield  {author} {\bibinfo {author} {\bibfnamefont {K.}~\bibnamefont
  {Hornberger}},\ }\href@noop {} {\bibfield  {journal} {\bibinfo  {journal}
  {Phys. Rev. A}\ }\textbf {\bibinfo {volume} {73}},\ \bibinfo {pages} {052102}
  (\bibinfo {year} {2006}{\natexlab{a}})}\BibitemShut {NoStop}%
\bibitem [{\citenamefont {Hornberger}\ \emph {et~al.}(2012)\citenamefont
  {Hornberger}, \citenamefont {Gerlich}, \citenamefont {Nimmrichter},
  \citenamefont {Haslinger},\ and\ \citenamefont {Arndt}}]{Hornberger:2012:ii}%
  \BibitemOpen
  \bibfield  {author} {\bibinfo {author} {\bibfnamefont {K.}~\bibnamefont
  {Hornberger}}, \bibinfo {author} {\bibfnamefont {S.}~\bibnamefont {Gerlich}},
  \bibinfo {author} {\bibfnamefont {S.}~\bibnamefont {Nimmrichter}}, \bibinfo
  {author} {\bibfnamefont {P.}~\bibnamefont {Haslinger}}, and\ \bibinfo
  {author} {\bibfnamefont {M.}~\bibnamefont {Arndt}},\ }\href@noop {}
  {\bibfield  {journal} {\bibinfo  {journal} {Rev. Mod. Phys.}\ }\textbf
  {\bibinfo {volume} {84}},\ \bibinfo {pages} {157} (\bibinfo {year}
  {2012})}\BibitemShut {NoStop}%
\bibitem [{\citenamefont {Walls}\ and\ \citenamefont
  {Milburn}(1985)}]{Walls:1985:pp}%
  \BibitemOpen
  \bibfield  {author} {\bibinfo {author} {\bibfnamefont {D.~F.}\ \bibnamefont
  {Walls}}\ and\ \bibinfo {author} {\bibfnamefont {G.~J.}\ \bibnamefont
  {Milburn}},\ }\href@noop {} {\bibfield  {journal} {\bibinfo  {journal} {Phys.
  Rev. A}\ }\textbf {\bibinfo {volume} {31}},\ \bibinfo {pages} {2403}
  (\bibinfo {year} {1985})}\BibitemShut {NoStop}%
\bibitem [{\citenamefont {Brune}\ \emph {et~al.}(1992)\citenamefont {Brune},
  \citenamefont {Haroche}, \citenamefont {Raimond}, \citenamefont
  {Davidovich},\ and\ \citenamefont {Zagury}}]{Brune:1992:zz}%
  \BibitemOpen
  \bibfield  {author} {\bibinfo {author} {\bibfnamefont {M.}~\bibnamefont
  {Brune}}, \bibinfo {author} {\bibfnamefont {S.}~\bibnamefont {Haroche}},
  \bibinfo {author} {\bibfnamefont {J.~M.}\ \bibnamefont {Raimond}}, \bibinfo
  {author} {\bibfnamefont {L.}~\bibnamefont {Davidovich}}, and\ \bibinfo
  {author} {\bibfnamefont {N.}~\bibnamefont {Zagury}},\ }\href@noop {}
  {\bibfield  {journal} {\bibinfo  {journal} {Phys. Rev. A}\ }\textbf {\bibinfo
  {volume} {45}},\ \bibinfo {pages} {5193} (\bibinfo {year}
  {1992})}\BibitemShut {NoStop}%
\bibitem [{\citenamefont {Arndt}\ and\ \citenamefont
  {Hornberger}(2014)}]{Arndt:2014:oo}%
  \BibitemOpen
  \bibfield  {author} {\bibinfo {author} {\bibfnamefont {M.}~\bibnamefont
  {Arndt}}\ and\ \bibinfo {author} {\bibfnamefont {K.}~\bibnamefont
  {Hornberger}},\ }\href@noop {} {\bibfield  {journal} {\bibinfo  {journal}
  {Nature Phys.}\ }\textbf {\bibinfo {volume} {10}},\ \bibinfo {pages} {271}
  (\bibinfo {year} {2014})}\BibitemShut {NoStop}%
\bibitem [{\citenamefont {Auffeves}\ \emph {et~al.}(2003)\citenamefont
  {Auffeves}, \citenamefont {Maioli}, \citenamefont {Meunier}, \citenamefont
  {Gleyzes}, \citenamefont {Nogues}, \citenamefont {Brune}, \citenamefont
  {Raimond},\ and\ \citenamefont {Haroche}}]{Auffeves:2003:za}%
  \BibitemOpen
  \bibfield  {author} {\bibinfo {author} {\bibfnamefont {A.}~\bibnamefont
  {Auffeves}}, \bibinfo {author} {\bibfnamefont {P.}~\bibnamefont {Maioli}},
  \bibinfo {author} {\bibfnamefont {T.}~\bibnamefont {Meunier}}, \bibinfo
  {author} {\bibfnamefont {S.}~\bibnamefont {Gleyzes}}, \bibinfo {author}
  {\bibfnamefont {G.}~\bibnamefont {Nogues}}, \bibinfo {author} {\bibfnamefont
  {M.}~\bibnamefont {Brune}}, \bibinfo {author} {\bibfnamefont {J.~M.}\
  \bibnamefont {Raimond}}, and\ \bibinfo {author} {\bibfnamefont
  {S.}~\bibnamefont {Haroche}},\ }\href@noop {} {\bibfield  {journal} {\bibinfo
   {journal} {Phys. Rev. Lett.}\ }\textbf {\bibinfo {volume} {91}},\ \bibinfo
  {pages} {230405} (\bibinfo {year} {2003})}\BibitemShut {NoStop}%
\bibitem [{\citenamefont {Vlastakis}\ \emph {et~al.}(2013)\citenamefont
  {Vlastakis}, \citenamefont {Kirchmair}, \citenamefont {Leghtas},
  \citenamefont {Nigg}, \citenamefont {Frunzio}, \citenamefont {Girvin},
  \citenamefont {Mirrahimi}, \citenamefont {Devoret},\ and\ \citenamefont
  {Schoelkopf}}]{Vlastakis:2013:pp}%
  \BibitemOpen
  \bibfield  {author} {\bibinfo {author} {\bibfnamefont {B.}~\bibnamefont
  {Vlastakis}}, \bibinfo {author} {\bibfnamefont {G.}~\bibnamefont
  {Kirchmair}}, \bibinfo {author} {\bibfnamefont {Z.}~\bibnamefont {Leghtas}},
  \bibinfo {author} {\bibfnamefont {S.~E.}\ \bibnamefont {Nigg}}, \bibinfo
  {author} {\bibfnamefont {L.}~\bibnamefont {Frunzio}}, \bibinfo {author}
  {\bibfnamefont {S.~M.}\ \bibnamefont {Girvin}}, \bibinfo {author}
  {\bibfnamefont {M.}~\bibnamefont {Mirrahimi}}, \bibinfo {author}
  {\bibfnamefont {M.~H.}\ \bibnamefont {Devoret}}, and\ \bibinfo {author}
  {\bibfnamefont {R.~J.}\ \bibnamefont {Schoelkopf}},\ }\href {\doibase
  10.1126/science.1243289} {\bibfield  {journal} {\bibinfo  {journal}
  {Science}\ }\textbf {\bibinfo {volume} {342}},\ \bibinfo {pages} {607}
  (\bibinfo {year} {2013})}\BibitemShut {NoStop}%
\bibitem [{\citenamefont {Devoret}\ and\ \citenamefont
  {Schoelkopf}(2013)}]{Devoret:2013:pp}%
  \BibitemOpen
  \bibfield  {author} {\bibinfo {author} {\bibfnamefont {M.~H.}\ \bibnamefont
  {Devoret}}\ and\ \bibinfo {author} {\bibfnamefont {R.~J.}\ \bibnamefont
  {Schoelkopf}},\ }\href@noop {} {\bibfield  {journal} {\bibinfo  {journal}
  {Science}\ }\textbf {\bibinfo {volume} {339}},\ \bibinfo {pages} {1169}
  (\bibinfo {year} {2013})}\BibitemShut {NoStop}%
\bibitem [{\citenamefont {Rigetti}\ \emph {et~al.}(2012)\citenamefont
  {Rigetti}, \citenamefont {Gambetta}, \citenamefont {Poletto}, \citenamefont
  {Plourde}, \citenamefont {Chow}, \citenamefont {C\'orcoles}, \citenamefont
  {Smolin}, \citenamefont {Merkel}, \citenamefont {Rozen}, \citenamefont
  {Keefe}, \citenamefont {Rothwell}, \citenamefont {Ketchen},\ and\
  \citenamefont {Steffen}}]{Rigetti:2012:aa}%
  \BibitemOpen
  \bibfield  {author} {\bibinfo {author} {\bibfnamefont {C.}~\bibnamefont
  {Rigetti}}, \bibinfo {author} {\bibfnamefont {J.~M.}\ \bibnamefont
  {Gambetta}}, \bibinfo {author} {\bibfnamefont {S.}~\bibnamefont {Poletto}},
  \bibinfo {author} {\bibfnamefont {B.~L.~T.}\ \bibnamefont {Plourde}},
  \bibinfo {author} {\bibfnamefont {J.~M.}\ \bibnamefont {Chow}}, \bibinfo
  {author} {\bibfnamefont {A.~D.}\ \bibnamefont {C\'orcoles}}, \bibinfo
  {author} {\bibfnamefont {J.~A.}\ \bibnamefont {Smolin}}, \bibinfo {author}
  {\bibfnamefont {S.~T.}\ \bibnamefont {Merkel}}, \bibinfo {author}
  {\bibfnamefont {J.~R.}\ \bibnamefont {Rozen}}, \bibinfo {author}
  {\bibfnamefont {G.~A.}\ \bibnamefont {Keefe}}, \bibinfo {author}
  {\bibfnamefont {M.~B.}\ \bibnamefont {Rothwell}}, \bibinfo {author}
  {\bibfnamefont {M.~B.}\ \bibnamefont {Ketchen}}, and\ \bibinfo {author}
  {\bibfnamefont {M.}~\bibnamefont {Steffen}},\ }\href {\doibase
  10.1103/PhysRevB.86.100506} {\bibfield  {journal} {\bibinfo  {journal} {Phys.
  Rev. B}\ }\textbf {\bibinfo {volume} {86}},\ \bibinfo {pages} {100506}
  (\bibinfo {year} {2012})}\BibitemShut {NoStop}%
\bibitem [{\citenamefont {Sears}\ \emph {et~al.}(2012)\citenamefont {Sears},
  \citenamefont {Petrenko}, \citenamefont {Catelani}, \citenamefont {Sun},
  \citenamefont {Paik}, \citenamefont {Kirchmair}, \citenamefont {Frunzio},
  \citenamefont {Glazman}, \citenamefont {Girvin},\ and\ \citenamefont
  {Schoelkopf}}]{Sears:2012:ee}%
  \BibitemOpen
  \bibfield  {author} {\bibinfo {author} {\bibfnamefont {A.~P.}\ \bibnamefont
  {Sears}}, \bibinfo {author} {\bibfnamefont {A.}~\bibnamefont {Petrenko}},
  \bibinfo {author} {\bibfnamefont {G.}~\bibnamefont {Catelani}}, \bibinfo
  {author} {\bibfnamefont {L.}~\bibnamefont {Sun}}, \bibinfo {author}
  {\bibfnamefont {H.}~\bibnamefont {Paik}}, \bibinfo {author} {\bibfnamefont
  {G.}~\bibnamefont {Kirchmair}}, \bibinfo {author} {\bibfnamefont
  {L.}~\bibnamefont {Frunzio}}, \bibinfo {author} {\bibfnamefont {L.~I.}\
  \bibnamefont {Glazman}}, \bibinfo {author} {\bibfnamefont {S.~M.}\
  \bibnamefont {Girvin}}, and\ \bibinfo {author} {\bibfnamefont {R.~J.}\
  \bibnamefont {Schoelkopf}},\ }\href {\doibase 10.1103/PhysRevB.86.180504}
  {\bibfield  {journal} {\bibinfo  {journal} {Phys. Rev. B}\ }\textbf {\bibinfo
  {volume} {86}},\ \bibinfo {pages} {180504} (\bibinfo {year}
  {2012})}\BibitemShut {NoStop}%
\bibitem [{\citenamefont {Leggett}(2002)}]{Leggett:2002:uy}%
  \BibitemOpen
  \bibfield  {author} {\bibinfo {author} {\bibfnamefont {A.~J.}\ \bibnamefont
  {Leggett}},\ }\href@noop {} {\bibfield  {journal} {\bibinfo  {journal} {J.
  Phys.: Condens. Matter}\ }\textbf {\bibinfo {volume} {14}},\ \bibinfo {pages}
  {R415} (\bibinfo {year} {2002})}\BibitemShut {NoStop}%
\bibitem [{\citenamefont {Marshall}\ \emph {et~al.}(2003)\citenamefont
  {Marshall}, \citenamefont {Simon}, \citenamefont {Penrose},\ and\
  \citenamefont {Bouwmeester}}]{Marshall:2003:om}%
  \BibitemOpen
  \bibfield  {author} {\bibinfo {author} {\bibfnamefont {W.}~\bibnamefont
  {Marshall}}, \bibinfo {author} {\bibfnamefont {C.}~\bibnamefont {Simon}},
  \bibinfo {author} {\bibfnamefont {R.}~\bibnamefont {Penrose}}, and\
  \bibinfo {author} {\bibfnamefont {D.}~\bibnamefont {Bouwmeester}},\
  }\href@noop {} {\bibfield  {journal} {\bibinfo  {journal} {Phys. Rev. Lett}\
  }\textbf {\bibinfo {volume} {91}},\ \bibinfo {pages} {130401} (\bibinfo
  {year} {2003})}\BibitemShut {NoStop}%
\bibitem [{\citenamefont {Bassi}\ \emph {et~al.}(2005)\citenamefont {Bassi},
  \citenamefont {Ippoliti},\ and\ \citenamefont {Adler}}]{Bassi:2005:om}%
  \BibitemOpen
  \bibfield  {author} {\bibinfo {author} {\bibfnamefont {A.}~\bibnamefont
  {Bassi}}, \bibinfo {author} {\bibfnamefont {E.}~\bibnamefont {Ippoliti}}, 
  and\ \bibinfo {author} {\bibfnamefont {S.~L.}\ \bibnamefont {Adler}},\
  }\href@noop {} {\bibfield  {journal} {\bibinfo  {journal} {Phys. Rev. Lett.}\
  }\textbf {\bibinfo {volume} {94}},\ \bibinfo {pages} {030401} (\bibinfo
  {year} {2005})}\BibitemShut {NoStop}%
\bibitem [{\citenamefont {Pikovski}\ \emph {et~al.}(2012)\citenamefont
  {Pikovski}, \citenamefont {Vanner}, \citenamefont {Aspelmeyer}, \citenamefont
  {Kim},\ and\ \citenamefont {Brukner}}]{Pikovski:2012:aa}%
  \BibitemOpen
  \bibfield  {author} {\bibinfo {author} {\bibfnamefont {I.}~\bibnamefont
  {Pikovski}}, \bibinfo {author} {\bibfnamefont {M.~R.}\ \bibnamefont
  {Vanner}}, \bibinfo {author} {\bibfnamefont {M.}~\bibnamefont {Aspelmeyer}},
  \bibinfo {author} {\bibfnamefont {M.~S.}\ \bibnamefont {Kim}}, and\
  \bibinfo {author} {\bibfnamefont {{\v C}.}~\bibnamefont {Brukner}},\
  }\href@noop {} {\bibfield  {journal} {\bibinfo  {journal} {Nature Phys.}\
  }\textbf {\bibinfo {volume} {8}},\ \bibinfo {pages} {393} (\bibinfo {year}
  {2012})}\BibitemShut {NoStop}%
\bibitem [{\citenamefont {Wan}\ \emph {et~al.}(2016)\citenamefont {Wan},
  \citenamefont {Scala}, \citenamefont {Morley}, \citenamefont {Rahman},
  \citenamefont {Ulbricht}, \citenamefont {Bateman}, \citenamefont {Barker},
  \citenamefont {Bose},\ and\ \citenamefont {Kim}}]{Wan:2016:oo}%
  \BibitemOpen
  \bibfield  {author} {\bibinfo {author} {\bibfnamefont {C.}~\bibnamefont
  {Wan}}, \bibinfo {author} {\bibfnamefont {M.}~\bibnamefont {Scala}}, \bibinfo
  {author} {\bibfnamefont {G.~W.}\ \bibnamefont {Morley}}, \bibinfo {author}
  {\bibfnamefont {A.~A.}\ \bibnamefont {Rahman}}, \bibinfo {author}
  {\bibfnamefont {H.}~\bibnamefont {Ulbricht}}, \bibinfo {author}
  {\bibfnamefont {J.}~\bibnamefont {Bateman}}, \bibinfo {author} {\bibfnamefont
  {P.~F.}\ \bibnamefont {Barker}}, \bibinfo {author} {\bibfnamefont
  {S.}~\bibnamefont {Bose}}, and\ \bibinfo {author} {\bibfnamefont {M.~S.}\
  \bibnamefont {Kim}},\ }\href@noop {} {\bibfield  {journal} {\bibinfo
  {journal} {Phys. Rev. Lett.}\ }\textbf {\bibinfo {volume} {117}},\ \bibinfo
  {pages} {143003} (\bibinfo {year} {2016})}\BibitemShut {NoStop}%
\bibitem [{\citenamefont {Kaltenbaek}\ \emph {et~al.}(2016)\citenamefont
  {Kaltenbaek}, \citenamefont {Aspelmeyer}, \citenamefont {Barker},
  \citenamefont {Bassi}, \citenamefont {Bateman}, \citenamefont {Bongs},
  \citenamefont {Bose}, \citenamefont {Braxmaier}, \citenamefont {Brukner},
  \citenamefont {Christophe}, \citenamefont {Chwalla}, \citenamefont {Cohadon},
  \citenamefont {Cruise}, \citenamefont {Curceanu}, \citenamefont {Dholakia},
  \citenamefont {Di{\'o}si}, \citenamefont {D{\"o}ringshoff}, \citenamefont
  {Ertmer}, \citenamefont {Gieseler}, \citenamefont {G{\"u}rlebeck},
  \citenamefont {Hechenblaikner}, \citenamefont {Heidmann}, \citenamefont
  {Herrmann}, \citenamefont {Hossenfelder}, \citenamefont {Johann},
  \citenamefont {Kiesel}, \citenamefont {Kim}, \citenamefont {L{\"a}mmerzahl},
  \citenamefont {Lambrecht}, \citenamefont {Mazilu}, \citenamefont {Milburn},
  \citenamefont {M{\"u}ller}, \citenamefont {Novotny}, \citenamefont
  {Paternostro}, \citenamefont {Peters}, \citenamefont {Pikovski},
  \citenamefont {{Pilan Zanoni}}, \citenamefont {Rasel}, \citenamefont
  {Reynaud}, \citenamefont {Riedel}, \citenamefont {Rodrigues}, \citenamefont
  {Rondin}, \citenamefont {Roura}, \citenamefont {Schleich}, \citenamefont
  {Schmiedmayer}, \citenamefont {Schuldt}, \citenamefont {Schwab},
  \citenamefont {Tajmar}, \citenamefont {Tino}, \citenamefont {Ulbricht},
  \citenamefont {Ursin},\ and\ \citenamefont {Vedral}}]{Kaltenbaek:2016:pp}%
  \BibitemOpen
  \bibfield  {author} {\bibinfo {author} {\bibfnamefont {R.}~\bibnamefont
  {Kaltenbaek}}, \bibinfo {author} {\bibfnamefont {M.}~\bibnamefont
  {Aspelmeyer}}, \bibinfo {author} {\bibfnamefont {P.~F.}\ \bibnamefont
  {Barker}}, \bibinfo {author} {\bibfnamefont {A.}~\bibnamefont {Bassi}},
  \bibinfo {author} {\bibfnamefont {J.}~\bibnamefont {Bateman}}, \bibinfo
  {author} {\bibfnamefont {K.}~\bibnamefont {Bongs}}, \bibinfo {author}
  {\bibfnamefont {S.}~\bibnamefont {Bose}}, \bibinfo {author} {\bibfnamefont
  {C.}~\bibnamefont {Braxmaier}}, \bibinfo {author} {\bibfnamefont
  {{\v{C}}.}~\bibnamefont {Brukner}}, \bibinfo {author} {\bibfnamefont
  {B.}~\bibnamefont {Christophe}}, \bibinfo {author} {\bibfnamefont
  {M.}~\bibnamefont {Chwalla}}, \bibinfo {author} {\bibfnamefont {P.-F.}\
  \bibnamefont {Cohadon}}, \bibinfo {author} {\bibfnamefont {A.~M.}\
  \bibnamefont {Cruise}}, \bibinfo {author} {\bibfnamefont {C.}~\bibnamefont
  {Curceanu}}, \bibinfo {author} {\bibfnamefont {K.}~\bibnamefont {Dholakia}},
  \bibinfo {author} {\bibfnamefont {L.}~\bibnamefont {Di{\'o}si}}, \bibinfo
  {author} {\bibfnamefont {K.}~\bibnamefont {D{\"o}ringshoff}}, \bibinfo
  {author} {\bibfnamefont {W.}~\bibnamefont {Ertmer}}, \bibinfo {author}
  {\bibfnamefont {J.}~\bibnamefont {Gieseler}}, \bibinfo {author}
  {\bibfnamefont {N.}~\bibnamefont {G{\"u}rlebeck}}, \bibinfo {author}
  {\bibfnamefont {G.}~\bibnamefont {Hechenblaikner}}, \bibinfo {author}
  {\bibfnamefont {A.}~\bibnamefont {Heidmann}}, \bibinfo {author}
  {\bibfnamefont {S.}~\bibnamefont {Herrmann}}, \bibinfo {author}
  {\bibfnamefont {S.}~\bibnamefont {Hossenfelder}}, \bibinfo {author}
  {\bibfnamefont {U.}~\bibnamefont {Johann}}, \bibinfo {author} {\bibfnamefont
  {N.}~\bibnamefont {Kiesel}}, \bibinfo {author} {\bibfnamefont
  {M.}~\bibnamefont {Kim}}, \bibinfo {author} {\bibfnamefont {C.}~\bibnamefont
  {L{\"a}mmerzahl}}, \bibinfo {author} {\bibfnamefont {A.}~\bibnamefont
  {Lambrecht}}, \bibinfo {author} {\bibfnamefont {M.}~\bibnamefont {Mazilu}},
  \bibinfo {author} {\bibfnamefont {G.~J.}\ \bibnamefont {Milburn}}, \bibinfo
  {author} {\bibfnamefont {H.}~\bibnamefont {M{\"u}ller}}, \bibinfo {author}
  {\bibfnamefont {L.}~\bibnamefont {Novotny}}, \bibinfo {author} {\bibfnamefont
  {M.}~\bibnamefont {Paternostro}}, \bibinfo {author} {\bibfnamefont
  {A.}~\bibnamefont {Peters}}, \bibinfo {author} {\bibfnamefont
  {I.}~\bibnamefont {Pikovski}}, \bibinfo {author} {\bibfnamefont
  {A.}~\bibnamefont {{Pilan Zanoni}}}, \bibinfo {author} {\bibfnamefont
  {E.~M.}\ \bibnamefont {Rasel}}, \bibinfo {author} {\bibfnamefont
  {S.}~\bibnamefont {Reynaud}}, \bibinfo {author} {\bibfnamefont {C.~J.}\
  \bibnamefont {Riedel}}, \bibinfo {author} {\bibfnamefont {M.}~\bibnamefont
  {Rodrigues}}, \bibinfo {author} {\bibfnamefont {L.}~\bibnamefont {Rondin}},
  \bibinfo {author} {\bibfnamefont {A.}~\bibnamefont {Roura}}, \bibinfo
  {author} {\bibfnamefont {W.~P.}\ \bibnamefont {Schleich}}, \bibinfo {author}
  {\bibfnamefont {J.}~\bibnamefont {Schmiedmayer}}, \bibinfo {author}
  {\bibfnamefont {T.}~\bibnamefont {Schuldt}}, \bibinfo {author} {\bibfnamefont
  {K.~C.}\ \bibnamefont {Schwab}}, \bibinfo {author} {\bibfnamefont
  {M.}~\bibnamefont {Tajmar}}, \bibinfo {author} {\bibfnamefont {G.~M.}\
  \bibnamefont {Tino}}, \bibinfo {author} {\bibfnamefont {H.}~\bibnamefont
  {Ulbricht}}, \bibinfo {author} {\bibfnamefont {R.}~\bibnamefont {Ursin}}, 
  and\ \bibinfo {author} {\bibfnamefont {V.}~\bibnamefont {Vedral}},\ }\href
  {\doibase 10.1140/epjqt/s40507-016-0043-7} {\bibfield  {journal} {\bibinfo
  {journal} {EPJ Quantum Technology}\ }\textbf {\bibinfo {volume} {3}},\
  \bibinfo {pages} {5} (\bibinfo {year} {2016})}\BibitemShut {NoStop}%
\bibitem [{\citenamefont {Stickler}\ \emph
  {et~al.}(2018{\natexlab{a}})\citenamefont {Stickler}, \citenamefont
  {Papendell}, \citenamefont {Kuhn}, \citenamefont {Schrinski}, \citenamefont
  {Millen}, \citenamefont {Arndt},\ and\ \citenamefont
  {Hornberger}}]{Stickler:2018:ii}%
  \BibitemOpen
  \bibfield  {author} {\bibinfo {author} {\bibfnamefont {B.~A.}\ \bibnamefont
  {Stickler}}, \bibinfo {author} {\bibfnamefont {B.}~\bibnamefont {Papendell}},
  \bibinfo {author} {\bibfnamefont {S.}~\bibnamefont {Kuhn}}, \bibinfo {author}
  {\bibfnamefont {B.}~\bibnamefont {Schrinski}}, \bibinfo {author}
  {\bibfnamefont {J.}~\bibnamefont {Millen}}, \bibinfo {author} {\bibfnamefont
  {M.}~\bibnamefont {Arndt}}, and\ \bibinfo {author} {\bibfnamefont
  {K.}~\bibnamefont {Hornberger}},\ }\href@noop {} {\bibfield  {journal}
  {\bibinfo  {journal} {New J. Phys.}\ }\textbf {\bibinfo {volume} {20}},\
  \bibinfo {pages} {122001} (\bibinfo {year} {2018}{\natexlab{a}})}\BibitemShut
  {NoStop}%
\bibitem [{\citenamefont {Tegmark}(1993)}]{Tegmark:1993:uz}%
  \BibitemOpen
  \bibfield  {author} {\bibinfo {author} {\bibfnamefont {M.}~\bibnamefont
  {Tegmark}},\ }\href@noop {} {\bibfield  {journal} {\bibinfo  {journal}
  {Found. Phys. Lett.}\ }\textbf {\bibinfo {volume} {6}},\ \bibinfo {pages}
  {571} (\bibinfo {year} {1993})}\BibitemShut {NoStop}%
\bibitem [{\citenamefont {Nimmrichter}\ and\ \citenamefont
  {Hornberger}(2013)}]{Nimmrichter:2013:aa}%
  \BibitemOpen
  \bibfield  {author} {\bibinfo {author} {\bibfnamefont {S.}~\bibnamefont
  {Nimmrichter}}\ and\ \bibinfo {author} {\bibfnamefont {K.}~\bibnamefont
  {Hornberger}},\ }\href@noop {} {\bibfield  {journal} {\bibinfo  {journal}
  {Phys. Rev. Lett.}\ }\textbf {\bibinfo {volume} {110}},\ \bibinfo {pages}
  {160403} (\bibinfo {year} {2013})}\BibitemShut {NoStop}%
\bibitem [{\citenamefont {Nimmrichter}\ \emph {et~al.}(2011)\citenamefont
  {Nimmrichter}, \citenamefont {Hornberger}, \citenamefont {Haslinger},\ and\
  \citenamefont {Arndt}}]{Nimmrichter:2011:pr}%
  \BibitemOpen
  \bibfield  {author} {\bibinfo {author} {\bibfnamefont {S.}~\bibnamefont
  {Nimmrichter}}, \bibinfo {author} {\bibfnamefont {K.}~\bibnamefont
  {Hornberger}}, \bibinfo {author} {\bibfnamefont {P.}~\bibnamefont
  {Haslinger}}, and\ \bibinfo {author} {\bibfnamefont {M.}~\bibnamefont
  {Arndt}},\ }\href@noop {} {\bibfield  {journal} {\bibinfo  {journal} {Phys.
  Rev. A}\ }\textbf {\bibinfo {volume} {83}},\ \bibinfo {pages} {043621}
  (\bibinfo {year} {2011})}\BibitemShut {NoStop}%
\bibitem [{\citenamefont {Romero-Isart}\ \emph {et~al.}(2011)\citenamefont
  {Romero-Isart}, \citenamefont {Pflanzer}, \citenamefont {Blaser},
  \citenamefont {Kaltenbaek}, \citenamefont {Kiesel}, \citenamefont
  {Aspelmeyer},\ and\ \citenamefont {Cirac}}]{Romero:2011:aa}%
  \BibitemOpen
  \bibfield  {author} {\bibinfo {author} {\bibfnamefont {O.}~\bibnamefont
  {Romero-Isart}}, \bibinfo {author} {\bibfnamefont {A.~C.}\ \bibnamefont
  {Pflanzer}}, \bibinfo {author} {\bibfnamefont {F.}~\bibnamefont {Blaser}},
  \bibinfo {author} {\bibfnamefont {R.}~\bibnamefont {Kaltenbaek}}, \bibinfo
  {author} {\bibfnamefont {N.}~\bibnamefont {Kiesel}}, \bibinfo {author}
  {\bibfnamefont {M.}~\bibnamefont {Aspelmeyer}}, and\ \bibinfo {author}
  {\bibfnamefont {J.~I.}\ \bibnamefont {Cirac}},\ }\href@noop {} {\bibfield
  {journal} {\bibinfo  {journal} {Phys. Rev. Lett.}\ }\textbf {\bibinfo
  {volume} {107}},\ \bibinfo {pages} {020405} (\bibinfo {year}
  {2011})}\BibitemShut {NoStop}%
\bibitem [{\citenamefont {Pikovski}\ \emph {et~al.}(2015)\citenamefont
  {Pikovski}, \citenamefont {Zych}, \citenamefont {Costa},\ and\ \citenamefont
  {Brukner}}]{Pikovski:2015:oo}%
  \BibitemOpen
  \bibfield  {author} {\bibinfo {author} {\bibfnamefont {I.}~\bibnamefont
  {Pikovski}}, \bibinfo {author} {\bibfnamefont {M.}~\bibnamefont {Zych}},
  \bibinfo {author} {\bibfnamefont {F.}~\bibnamefont {Costa}}, and\ \bibinfo
  {author} {\bibfnamefont {{\v C}.}~\bibnamefont {Brukner}},\ }\href@noop {}
  {\bibfield  {journal} {\bibinfo  {journal} {Nature Phys.}\ ,\ \bibinfo
  {pages} {668}} (\bibinfo {year} {2015})}\BibitemShut {NoStop}%
\bibitem [{\citenamefont {Adler}(2007)}]{Adler:2007:um}%
  \BibitemOpen
  \bibfield  {author} {\bibinfo {author} {\bibfnamefont {S.~L.}\ \bibnamefont
  {Adler}},\ }\href@noop {} {\bibfield  {journal} {\bibinfo  {journal} {J.
  Phys. A}\ }\textbf {\bibinfo {volume} {40}},\ \bibinfo {pages} {2935}
  (\bibinfo {year} {2007})}\BibitemShut {NoStop}%
\bibitem [{\citenamefont {Bassi}\ \emph {et~al.}(2010)\citenamefont {Bassi},
  \citenamefont {Deckert},\ and\ \citenamefont {Ferialdi}}]{Bassi:2010:aa}%
  \BibitemOpen
  \bibfield  {author} {\bibinfo {author} {\bibfnamefont {A.}~\bibnamefont
  {Bassi}}, \bibinfo {author} {\bibfnamefont {D.-A.}\ \bibnamefont {Deckert}},
   and\ \bibinfo {author} {\bibfnamefont {L.}~\bibnamefont {Ferialdi}},\
  }\href@noop {} {\bibfield  {journal} {\bibinfo  {journal} {EPL}\ }\textbf
  {\bibinfo {volume} {92}},\ \bibinfo {pages} {50006} (\bibinfo {year}
  {2010})}\BibitemShut {NoStop}%
\bibitem [{\citenamefont {Harris}\ and\ \citenamefont
  {Stodolsky}(1981)}]{Harris:1981:rc}%
  \BibitemOpen
  \bibfield  {author} {\bibinfo {author} {\bibfnamefont {R.~A.}\ \bibnamefont
  {Harris}}\ and\ \bibinfo {author} {\bibfnamefont {L.}~\bibnamefont
  {Stodolsky}},\ }\href@noop {} {\bibfield  {journal} {\bibinfo  {journal} {J.
  Chem. Phys.}\ }\textbf {\bibinfo {volume} {74}},\ \bibinfo {pages} {2145}
  (\bibinfo {year} {1981})}\BibitemShut {NoStop}%
\bibitem [{\citenamefont {Trost}\ and\ \citenamefont
  {Hornberger}(2009)}]{Trost:2009:ll}%
  \BibitemOpen
  \bibfield  {author} {\bibinfo {author} {\bibfnamefont {J.}~\bibnamefont
  {Trost}}\ and\ \bibinfo {author} {\bibfnamefont {K.}~\bibnamefont
  {Hornberger}},\ }\href {\doibase 10.1103/PhysRevLett.103.023202} {\bibfield
  {journal} {\bibinfo  {journal} {Phys. Rev. Lett.}\ }\textbf {\bibinfo
  {volume} {103}},\ \bibinfo {pages} {023202} (\bibinfo {year}
  {2009})}\BibitemShut {NoStop}%
\bibitem [{\citenamefont {Bahrami}\ \emph {et~al.}(2012)\citenamefont
  {Bahrami}, \citenamefont {Shafiee},\ and\ \citenamefont
  {Bassi}}]{Bahrami:2012:oo}%
  \BibitemOpen
  \bibfield  {author} {\bibinfo {author} {\bibfnamefont {M.}~\bibnamefont
  {Bahrami}}, \bibinfo {author} {\bibfnamefont {A.}~\bibnamefont {Shafiee}}, 
  and\ \bibinfo {author} {\bibfnamefont {A.}~\bibnamefont {Bassi}},\
  }\href@noop {} {\bibfield  {journal} {\bibinfo  {journal} {Phys. Chem. Chem.
  Phys.}\ }\textbf {\bibinfo {volume} {14}},\ \bibinfo {pages} {9214} (\bibinfo
  {year} {2012})}\BibitemShut {NoStop}%
\bibitem [{\citenamefont {Joos}\ and\ \citenamefont
  {Zeh}(1985)}]{Joos:1985:iu}%
  \BibitemOpen
  \bibfield  {author} {\bibinfo {author} {\bibfnamefont {E.}~\bibnamefont
  {Joos}}\ and\ \bibinfo {author} {\bibfnamefont {H.~D.}\ \bibnamefont {Zeh}},\
  }\href@noop {} {\bibfield  {journal} {\bibinfo  {journal} {Z. Phys. B:
  Condens. Matter}\ }\textbf {\bibinfo {volume} {59}},\ \bibinfo {pages} {223}
  (\bibinfo {year} {1985})}\BibitemShut {NoStop}%
\bibitem [{\citenamefont {Joos}(1984)}]{Joos:1984:km}%
  \BibitemOpen
  \bibfield  {author} {\bibinfo {author} {\bibfnamefont {E.}~\bibnamefont
  {Joos}},\ }\href {\doibase 10.1103/PhysRevD.29.1626} {\bibfield  {journal}
  {\bibinfo  {journal} {Phys. Rev. D}\ }\textbf {\bibinfo {volume} {29}},\
  \bibinfo {pages} {1626} (\bibinfo {year} {1984})}\BibitemShut {NoStop}%
\bibitem [{\citenamefont {Gallis}\ and\ \citenamefont
  {Fleming}(1990)}]{Gallis:1990:un}%
  \BibitemOpen
  \bibfield  {author} {\bibinfo {author} {\bibfnamefont {M.~R.}\ \bibnamefont
  {Gallis}}\ and\ \bibinfo {author} {\bibfnamefont {G.~N.}\ \bibnamefont
  {Fleming}},\ }\href@noop {} {\bibfield  {journal} {\bibinfo  {journal} {Phys.
  Rev. A}\ }\textbf {\bibinfo {volume} {42}},\ \bibinfo {pages} {38} (\bibinfo
  {year} {1990})}\BibitemShut {NoStop}%
\bibitem [{\citenamefont {Di{\'o}si}(1995)}]{Diosi:1995:um}%
  \BibitemOpen
  \bibfield  {author} {\bibinfo {author} {\bibfnamefont {L.}~\bibnamefont
  {Di{\'o}si}},\ }\href@noop {} {\bibfield  {journal} {\bibinfo  {journal}
  {Europhys. Lett.}\ }\textbf {\bibinfo {volume} {30}},\ \bibinfo {pages} {63}
  (\bibinfo {year} {1995})}\BibitemShut {NoStop}%
\bibitem [{\citenamefont {Adler}(2006)}]{Adler:2006:yb}%
  \BibitemOpen
  \bibfield  {author} {\bibinfo {author} {\bibfnamefont {S.~L.}\ \bibnamefont
  {Adler}},\ }\href@noop {} {\bibfield  {journal} {\bibinfo  {journal} {J.
  Phys. A: Math. Gen.}\ }\textbf {\bibinfo {volume} {39}},\ \bibinfo {pages}
  {14067} (\bibinfo {year} {2006})}\BibitemShut {NoStop}%
\bibitem [{\citenamefont
  {Hornberger}(2006{\natexlab{b}})}]{Hornberger:2006:tb}%
  \BibitemOpen
  \bibfield  {author} {\bibinfo {author} {\bibfnamefont {K.}~\bibnamefont
  {Hornberger}},\ }\href@noop {} {\bibfield  {journal} {\bibinfo  {journal}
  {Phys. Rev. Lett.}\ }\textbf {\bibinfo {volume} {97}},\ \bibinfo {pages}
  {060601} (\bibinfo {year} {2006}{\natexlab{b}})}\BibitemShut {NoStop}%
\bibitem [{\citenamefont {Hornberger}(2007)}]{Hornberger:2006:tc}%
  \BibitemOpen
  \bibfield  {author} {\bibinfo {author} {\bibfnamefont {K.}~\bibnamefont
  {Hornberger}},\ }\href@noop {} {\bibfield  {journal} {\bibinfo  {journal}
  {EPL}\ }\textbf {\bibinfo {volume} {77}},\ \bibinfo {pages} {50007} (\bibinfo
  {year} {2007})}\BibitemShut {NoStop}%
\bibitem [{\citenamefont {Hornberger}\ and\ \citenamefont
  {Vacchini}(2008)}]{Hornberger:2008:ii}%
  \BibitemOpen
  \bibfield  {author} {\bibinfo {author} {\bibfnamefont {K.}~\bibnamefont
  {Hornberger}}\ and\ \bibinfo {author} {\bibfnamefont {B.}~\bibnamefont
  {Vacchini}},\ }\href@noop {} {\bibfield  {journal} {\bibinfo  {journal}
  {Phys. Rev. A}\ }\textbf {\bibinfo {volume} {77}},\ \bibinfo {pages} {022112}
  (\bibinfo {year} {2008})}\BibitemShut {NoStop}%
\bibitem [{\citenamefont {Busse}\ and\ \citenamefont
  {Hornberger}(2009)}]{Busse:2009:aa}%
  \BibitemOpen
  \bibfield  {author} {\bibinfo {author} {\bibfnamefont {M.}~\bibnamefont
  {Busse}}\ and\ \bibinfo {author} {\bibfnamefont {K.}~\bibnamefont
  {Hornberger}},\ }\href@noop {} {\bibfield  {journal} {\bibinfo  {journal} {J.
  Phys. A: Math. Theor.}\ }\textbf {\bibinfo {volume} {42}},\ \bibinfo {pages}
  {362001} (\bibinfo {year} {2009})}\BibitemShut {NoStop}%
\bibitem [{\citenamefont {Vacchini}\ and\ \citenamefont
  {Hornberger}(2009)}]{Vacchini:2009:pp}%
  \BibitemOpen
  \bibfield  {author} {\bibinfo {author} {\bibfnamefont {B.}~\bibnamefont
  {Vacchini}}\ and\ \bibinfo {author} {\bibfnamefont {K.}~\bibnamefont
  {Hornberger}},\ }\href@noop {} {\bibfield  {journal} {\bibinfo  {journal}
  {Phys. Rep.}\ }\textbf {\bibinfo {volume} {478}},\ \bibinfo {pages} {71}
  (\bibinfo {year} {2009})}\BibitemShut {NoStop}%
\bibitem [{\citenamefont {Busse}\ and\ \citenamefont
  {Hornberger}(2010)}]{Busse:2010:aa}%
  \BibitemOpen
  \bibfield  {author} {\bibinfo {author} {\bibfnamefont {M.}~\bibnamefont
  {Busse}}\ and\ \bibinfo {author} {\bibfnamefont {K.}~\bibnamefont
  {Hornberger}},\ }\href@noop {} {\bibfield  {journal} {\bibinfo  {journal} {J.
  Phys. A: Math. Theor.}\ }\textbf {\bibinfo {volume} {43}},\ \bibinfo {pages}
  {015303} (\bibinfo {year} {2010})}\BibitemShut {NoStop}%
\bibitem [{\citenamefont {Busse}\ \emph {et~al.}(2010)\citenamefont {Busse},
  \citenamefont {Pietrulewicz}, \citenamefont {Breuer},\ and\ \citenamefont
  {Hornberger}}]{Busse:2010:oo}%
  \BibitemOpen
  \bibfield  {author} {\bibinfo {author} {\bibfnamefont {M.}~\bibnamefont
  {Busse}}, \bibinfo {author} {\bibfnamefont {P.}~\bibnamefont {Pietrulewicz}},
  \bibinfo {author} {\bibfnamefont {H.-P.}\ \bibnamefont {Breuer}}, and\
  \bibinfo {author} {\bibfnamefont {K.}~\bibnamefont {Hornberger}},\
  }\href@noop {} {\bibfield  {journal} {\bibinfo  {journal} {Phys. Rev. E}\
  }\textbf {\bibinfo {volume} {82}},\ \bibinfo {pages} {026706} (\bibinfo
  {year} {2010})}\BibitemShut {NoStop}%
\bibitem [{\citenamefont {Gring}\ \emph {et~al.}(2010)\citenamefont {Gring},
  \citenamefont {Gerlich}, \citenamefont {Eibenberger}, \citenamefont
  {Nimmrichter}, \citenamefont {Berrada}, \citenamefont {Arndt}, \citenamefont
  {Ulbricht}, \citenamefont {Hornberger}, \citenamefont {M{\"u}ri},
  \citenamefont {Mayor}, \citenamefont {B{\"o}ckmann},  and\ \citenamefont
  {Doltsinis}}]{Gring:2010:aa}%
  \BibitemOpen
  \bibfield  {author} {\bibinfo {author} {\bibfnamefont {M.}~\bibnamefont
  {Gring}}, \bibinfo {author} {\bibfnamefont {S.}~\bibnamefont {Gerlich}},
  \bibinfo {author} {\bibfnamefont {S.}~\bibnamefont {Eibenberger}}, \bibinfo
  {author} {\bibfnamefont {S.}~\bibnamefont {Nimmrichter}}, \bibinfo {author}
  {\bibfnamefont {T.}~\bibnamefont {Berrada}}, \bibinfo {author} {\bibfnamefont
  {M.}~\bibnamefont {Arndt}}, \bibinfo {author} {\bibfnamefont
  {H.}~\bibnamefont {Ulbricht}}, \bibinfo {author} {\bibfnamefont
  {K.}~\bibnamefont {Hornberger}}, \bibinfo {author} {\bibfnamefont
  {M.}~\bibnamefont {M{\"u}ri}}, \bibinfo {author} {\bibfnamefont
  {M.}~\bibnamefont {Mayor}}, \bibinfo {author} {\bibfnamefont
  {M.}~\bibnamefont {B{\"o}ckmann}}, and\ \bibinfo {author} {\bibfnamefont
  {N.~L.}\ \bibnamefont {Doltsinis}},\ }\href@noop {} {\bibfield  {journal}
  {\bibinfo  {journal} {Phys. Rev. A}\ }\textbf {\bibinfo {volume} {81}},\
  \bibinfo {pages} {031604(R)} (\bibinfo {year} {2010})}\BibitemShut {NoStop}%
\bibitem [{\citenamefont {Stickler}\ and\ \citenamefont
  {Hornberger}(2015)}]{Stickler:2015:zz}%
  \BibitemOpen
  \bibfield  {author} {\bibinfo {author} {\bibfnamefont {B.~A.}\ \bibnamefont
  {Stickler}}\ and\ \bibinfo {author} {\bibfnamefont {K.}~\bibnamefont
  {Hornberger}},\ }\href@noop {} {\bibfield  {journal} {\bibinfo  {journal}
  {Phys. Rev. A}\ }\textbf {\bibinfo {volume} {92}},\ \bibinfo {pages} {023619}
  (\bibinfo {year} {2015})}\BibitemShut {NoStop}%
\bibitem [{\citenamefont {Walter}\ \emph {et~al.}(2016)\citenamefont {Walter},
  \citenamefont {Stickler},\ and\ \citenamefont {Hornberger}}]{Walter:2016:zz}%
  \BibitemOpen
  \bibfield  {author} {\bibinfo {author} {\bibfnamefont {K.}~\bibnamefont
  {Walter}}, \bibinfo {author} {\bibfnamefont {B.~A.}\ \bibnamefont
  {Stickler}},  and\ \bibinfo {author} {\bibfnamefont {K.}~\bibnamefont
  {Hornberger}},\ }\href@noop {} {\bibfield  {journal} {\bibinfo  {journal}
  {Phys. Rev. A}\ }\textbf {\bibinfo {volume} {93}},\ \bibinfo {pages} {063612}
  (\bibinfo {year} {2016})}\BibitemShut {NoStop}%
\bibitem [{\citenamefont {Stickler}\ \emph {et~al.}(2016)\citenamefont
  {Stickler}, \citenamefont {Papendell},\ and\ \citenamefont
  {Hornberger}}]{Stickler:2016:yy}%
  \BibitemOpen
  \bibfield  {author} {\bibinfo {author} {\bibfnamefont {B.~A.}\ \bibnamefont
  {Stickler}}, \bibinfo {author} {\bibfnamefont {B.}~\bibnamefont {Papendell}},
   and\ \bibinfo {author} {\bibfnamefont {K.}~\bibnamefont {Hornberger}},\
  }\href@noop {} {\bibfield  {journal} {\bibinfo  {journal} {Phys. Rev. A}\
  }\textbf {\bibinfo {volume} {94}},\ \bibinfo {pages} {033828} (\bibinfo
  {year} {2016})}\BibitemShut {NoStop}%
\bibitem [{\citenamefont {Papendell}\ \emph {et~al.}(2017)\citenamefont
  {Papendell}, \citenamefont {Stickler},\ and\ \citenamefont
  {Hornberger}}]{Papendell:2017:yy}%
  \BibitemOpen
  \bibfield  {author} {\bibinfo {author} {\bibfnamefont {B.}~\bibnamefont
  {Papendell}}, \bibinfo {author} {\bibfnamefont {B.~A.}\ \bibnamefont
  {Stickler}}, and\ \bibinfo {author} {\bibfnamefont {K.}~\bibnamefont
  {Hornberger}},\ }\href@noop {} {\bibfield  {journal} {\bibinfo  {journal}
  {New J. Phys.}\ }\textbf {\bibinfo {volume} {19}},\ \bibinfo {pages} {122001}
  (\bibinfo {year} {2017})}\BibitemShut {NoStop}%
\bibitem [{\citenamefont {Stickler}\ \emph
  {et~al.}(2018{\natexlab{b}})\citenamefont {Stickler}, \citenamefont
  {Ghahramani},\ and\ \citenamefont {Hornberger}}]{Stickler:2018:oo}%
  \BibitemOpen
  \bibfield  {author} {\bibinfo {author} {\bibfnamefont {B.~A.}\ \bibnamefont
  {Stickler}}, \bibinfo {author} {\bibfnamefont {F.~T.}\ \bibnamefont
  {Ghahramani}}, and\ \bibinfo {author} {\bibfnamefont {K.}~\bibnamefont
  {Hornberger}},\ }\href@noop {} {\bibfield  {journal} {\bibinfo  {journal}
  {Phys. Rev. Lett.}\ }\textbf {\bibinfo {volume} {121}},\ \bibinfo {pages}
  {243402} (\bibinfo {year} {2018}{\natexlab{b}})}\BibitemShut {NoStop}%
\bibitem [{\citenamefont {Stickler}\ \emph
  {et~al.}(2018{\natexlab{c}})\citenamefont {Stickler}, \citenamefont
  {Schrinski},\ and\ \citenamefont {Hornberger}}]{Stickler:2018:uu}%
  \BibitemOpen
  \bibfield  {author} {\bibinfo {author} {\bibfnamefont {B.~A.}\ \bibnamefont
  {Stickler}}, \bibinfo {author} {\bibfnamefont {B.}~\bibnamefont {Schrinski}},
   and\ \bibinfo {author} {\bibfnamefont {K.}~\bibnamefont {Hornberger}},\
  }\href@noop {} {\bibfield  {journal} {\bibinfo  {journal} {Phys. Rev. Lett.}\
  }\textbf {\bibinfo {volume} {121}},\ \bibinfo {pages} {040401} (\bibinfo
  {year} {2018}{\natexlab{c}})}\BibitemShut {NoStop}%
\bibitem [{\citenamefont {Schr{\"o}dinger}(1935)}]{Schrodinger:1935:jn}%
  \BibitemOpen
  \bibfield  {author} {\bibinfo {author} {\bibfnamefont {E.}~\bibnamefont
  {Schr{\"o}dinger}},\ }\href@noop {} {\bibfield  {journal} {\bibinfo
  {journal} {Proc. Cambridge Philos. Soc.}\ }\textbf {\bibinfo {volume} {31}},\
  \bibinfo {pages} {555} (\bibinfo {year} {1935})}\BibitemShut {NoStop}%
\bibitem [{\citenamefont {Zurek}(1986)}]{Zurek:1986:uz}%
  \BibitemOpen
  \bibfield  {author} {\bibinfo {author} {\bibfnamefont {W.~H.}\ \bibnamefont
  {Zurek}},\ }in\ \href@noop {} {\emph {\bibinfo {booktitle} {Frontiers of
  Nonequilibrium Statistical Mechanics}}},\ \bibinfo {editor} {edited by\
  \bibinfo {editor} {\bibfnamefont {G.~T.}\ \bibnamefont {Moore}}\ and\
  \bibinfo {editor} {\bibfnamefont {M.~O.}\ \bibnamefont {Scully}}}\ (\bibinfo
  {publisher} {Plenum Press},\ \bibinfo {address} {New York},\ \bibinfo {year}
  {1986}), pp.\ \bibinfo {pages} {145--149},\ \bibinfo {note} {first published
  in 1984 as Los Alamos report LAUR 84-2750}\BibitemShut {NoStop}%
\bibitem [{\citenamefont {Caldeira}\ and\ \citenamefont
  {Leggett}(1985)}]{Caldeira:1985:tt}%
  \BibitemOpen
  \bibfield  {author} {\bibinfo {author} {\bibfnamefont {A.~O.}\ \bibnamefont
  {Caldeira}}\ and\ \bibinfo {author} {\bibfnamefont {A.~J.}\ \bibnamefont
  {Leggett}},\ }\href@noop {} {\bibfield  {journal} {\bibinfo  {journal} {Phys.
  Rev. A}\ }\textbf {\bibinfo {volume} {31}},\ \bibinfo {pages} {1059}
  (\bibinfo {year} {1985})}\BibitemShut {NoStop}%
\bibitem [{\citenamefont {Caldeira}\ and\ \citenamefont
  {Leggett}(1983)}]{Caldeira:1983:on}%
  \BibitemOpen
  \bibfield  {author} {\bibinfo {author} {\bibfnamefont {A.~O.}\ \bibnamefont
  {Caldeira}}\ and\ \bibinfo {author} {\bibfnamefont {A.~J.}\ \bibnamefont
  {Leggett}},\ }\href@noop {} {\bibfield  {journal} {\bibinfo  {journal}
  {Physica A}\ }\textbf {\bibinfo {volume} {121}},\ \bibinfo {pages} {587}
  (\bibinfo {year} {1983})}\BibitemShut {NoStop}%
\bibitem [{\citenamefont {Gallis}(1992)}]{Gallis:1992:im}%
  \BibitemOpen
  \bibfield  {author} {\bibinfo {author} {\bibfnamefont {M.~R.}\ \bibnamefont
  {Gallis}},\ }\href@noop {} {\bibfield  {journal} {\bibinfo  {journal} {Phys.
  Rev. A}\ }\textbf {\bibinfo {volume} {45}},\ \bibinfo {pages} {47} (\bibinfo
  {year} {1992})}\BibitemShut {NoStop}%
\bibitem [{\citenamefont {Anglin}\ \emph {et~al.}(1997)\citenamefont {Anglin},
  \citenamefont {Paz},\ and\ \citenamefont {Zurek}}]{Anglin:1997:za}%
  \BibitemOpen
  \bibfield  {author} {\bibinfo {author} {\bibfnamefont {J.~R.}\ \bibnamefont
  {Anglin}}, \bibinfo {author} {\bibfnamefont {J.~P.}\ \bibnamefont {Paz}}, 
  and\ \bibinfo {author} {\bibfnamefont {W.~H.}\ \bibnamefont {Zurek}},\
  }\href@noop {} {\bibfield  {journal} {\bibinfo  {journal} {Phys. Rev. A}\
  }\textbf {\bibinfo {volume} {55}},\ \bibinfo {pages} {4041} (\bibinfo {year}
  {1997})}\BibitemShut {NoStop}%
\bibitem [{\citenamefont {Zurek}(1982)}]{Zurek:1982:tv}%
  \BibitemOpen
  \bibfield  {author} {\bibinfo {author} {\bibfnamefont {W.~H.}\ \bibnamefont
  {Zurek}},\ }\href {\doibase 10.1103/PhysRevD.26.1862} {\bibfield  {journal}
  {\bibinfo  {journal} {Phys. Rev. D}\ }\textbf {\bibinfo {volume} {26}},\
  \bibinfo {pages} {1862} (\bibinfo {year} {1982})}\BibitemShut {NoStop}%
\end{thebibliography}
% \bibliographystyle{apsrev4-1}

%merlin.mbs apsrev4-1.bst 2010-07-25 4.21a (PWD, AO, DPC) hacked
%Control: key (0)
%Control: author (72) initials jnrlst
%Control: editor formatted (1) identically to author
%Control: production of article title (-1) disabled
%Control: page (0) single
%Control: year (1) truncated
%Control: production of eprint (0) enabled
%

%TRANLSATED OFGDEN

% \bibitem [{Fre()}]{Freitas:2008:yy}%
%   \BibitemOpen
%   \href@noop {} {\bibinfo {title} {Interview with H.\ Dieter {Z}eh, conducted at Waldhilsbach, Germany, by
%   {F}{\'a}bio {F}reitas, {J}uly 25--26, 2008}.\ }\bibinfo {howpublished}
%   {Audio recording and transcript deposited at the American Institute of
%   Physics}\BibitemShut {NoStop}%

\end{document}